\documentclass[a4paper,11pt]{article}
\pdfoutput=1
\usepackage{jheppub}
\usepackage{amsmath, amsfonts, amssymb}
\usepackage{graphicx, hyperref, verbatim}
\usepackage{tikz}
\usepackage{tkz-euclide}
\usetikzlibrary{positioning}

\newcommand{\es}[2] {\begin{equation} \label{#1} \begin{split} #2 \end{split} \end{equation}}
\newcommand{\calE}{\mathcal{E}}
\newcommand{\calM}{\mathcal{M}}
\newcommand{\calMt}{\widetilde{\mathcal{M}}}
\newcommand{\calMh}{\widehat{\mathcal{M}}}
\newcommand{\calN}{\mathcal{N}}

\title{Mellin amplitudes for $AdS_3 \times S^3$}
\author{Connor Behan, Rodrigo S. Pitombo}
\affiliation{ICTP South American Institute for Fundamental Research, Instituto de F\'{i}sica Te\'{o}rica UNESP \\ Rua Dr. Bento Teobaldo Ferraz 271, 01140-070, S\~{a}o Paulo, SP, Brazil}
\emailAdd{connor.behan@gmail.com}
\emailAdd{rs.pitombo@unesp.br}

\abstract{There are holographic superconformal theories in all dimensions between two and six which allow arbitrary tree-level four-point functions to be fixed by basic consistency conditions. Although Mellin space is usually the most efficient setting for imposing these contraints, four-point functions in two dimensions have thus far been an exception due to their more intricate dependence on the conformal cross-ratios. In this paper, we introduce a simple fix which exploits the relation between a parity-odd conformal block in two dimensions and a parity-even conformal block in four dimensions. We then apply the resulting toolkit to a study of the paradigmatic holographic theory in two dimensions which is the D1-D5 CFT. For correlators involving Kaluza-Klein modes of the tensor multiplet, this analysis reproduces results which were previously obtained using hidden conformal symmetry. With four Kaluza-Klein modes of the graviton multiplet, it yields new results including a compact formula for the correlators of all pairwise identical operators,}

\begin{document}
\maketitle
\flushbottom

\section{Introduction}
\label{sec:intro}


Following early checks of AdS/CFT at the level of three-point functions \cite{lmms98}, precision holography has become increasingly concerned with the study of four-point functions. For the case of light operators, which are dual to single-particle states, results have progressed significantly since the first tree-level computation was done in \cite{dfmmr99}. In particular, it is now possible to compute infinite families of four-point functions which allow one to apply unitarity methods in AdS \cite{aabp16} or see the emergence of semiclassical strings \cite{av20}. The key idea, put forward in \cite{rz16,rz17,z17}, was to harness the power of superconformal Ward identities \cite{dgs04,no04,bllprv13} in Mellin space \cite{m09,p10,fkprv11} where the analytic structure of conformal correlators becomes manifest. Although this bootstrap technique has produced a complete library of tree-level four-point functions for a variety of holographic CFTs in dimension $3 \leq d \leq 6$ \cite{az20,abfz21}, it has not yet reached its full potential in $d = 2$. The goal of this paper is to bring the 2d case under better control with a focus on the holographic CFT which governs the near-horizon limit of a D1-D5 system.

The D1-D5 CFT is perhaps best known for its role in demonstrating that the Bekenstein-Hawking formula for the entropy of a black hole can be derived from string theory. In short, \cite{sv96} considered a large number of D1 and D5 branes on $\mathbb{R}^5 \times S^1 \times K3$
which T-dualize to a solitonic description of a 5d black hole when $\alpha'$ is small. On the other hand, the asymptotic density of states for this configuration can be determined from an open string calculation when $\alpha'$ is large using the fact that the elliptic genus is protected. Less than two years later, the first AdS/CFT paper \cite{m97} discussed the more general setup of a D1-D5 system on $\mathbb{R}^5 \times S^1 \times M_4$ where $M_4$ is either $K3$ or $T^4$. Taking $M_4$ to be small, type IIB string theory on the near-horizon geometry $AdS_3 \times S^3 \times M_4$ has a low energy limit given by supergravity on $AdS_3 \times S^3$. As reviewed in \cite{dmw02}, this sparked a sustained interest in the dual 2d CFT which is a symmetric orbifold for $N = Q_1 Q_5$ copies of the sigma model for $M_4$. Although the bulk dual of this orbifold has string scale curvature at weak coupling, non-trivial evidence that it matches supergravity at strong coupling was found in \cite{b98,b98b} which again made use of the elliptic genus.\footnote{At finite $N$, this calculation also revealed that many of the states allowed by symmetry had to be absent from the spectrum. This is an example of the stringy exclusion principle \cite{ms98} which has recently been discussed from a microscopic perspective in \cite{l23,cl24}.} The so called fuzzball program has since followed various heavy operators perturbatively in the coupling in order to distinguish which ones contribute to black hole microstates. Possible points of entry are \cite{hmz18,ghmm22} for purely left-moving excitations and \cite{lss20a,lss20b} for Ramond-Ramond ground states.

The present work, by contrast, is perturbative in $1/N$. As a blueprint for the set of results we would like to have, \cite{az20} studied the supergravity multiplet in $AdS_4 \times S^7$, $AdS_5 \times S^5$ and $AdS_7 \times S^4$ while \cite{abfz21} studied the super Yang-Mills multiplet in $AdS_5 \times S^3$, $AdS_6 \times S^3$ and $AdS_7 \times S^3$. These will be referred to as maximal supergraviton and supergluon type theories respectively.\footnote{Maximal means 16 supercharges for the graviton case and 8 supercharges for the gluon case. This is because superconformal algebras with 16 Poincar\'{e} supercharges cannot have flavor currents \cite{cdi16}. In a sense, the D1-D5 CFT is also maximal because a 2d superconformal algebra with more than 4 left-moving or right-moving supercharges can only be centrally extended if it is a W-algebra \cite{kv89}.} In both cases, it was possible to find Mellin space expressions for the tree-level four-point functions of arbitrary Kaluza-Klein modes. Applying this approach \textit{mutatis mutandis} to $AdS_3 \times S^3$ fails because of parity. In $d = 2p$ dimensions for instance, a $p$-form operator $\mathcal{O}_{\mu_1 \dots \mu_p}(x)$ should be split into its self-dual and anti-self-dual components. When these contribute to a correlation function through the OPE, there is no Mellin amplitude which can accommodate exchanged operators of the form $\varepsilon^{\mu_1 \dots \mu_p \nu_1 \dots \nu_p} \mathcal{O}_{\nu_1 \dots \nu_p}(x)$. This is rarely an issue for holographic theories in higher dimensions. Provided there is a sufficient amount of supersymmetry, single-trace operators are required to live in multiplets where the superconformal primary is a Lorentz scalar. When $p > 1$, this prevents any $p$-form from being exchanged. When $p = 1$ however, not only is the vector $\mathcal{O}_\mu(x)$ perfectly admissible in an OPE between two scalars but it is also easy to find single-trace operators which are super-descendants of a vector.\footnote{A simple example in the case of the small $\calN = 4$ superconformal algebra is the stress tensor \cite{ll19}.} Because of this, the partial results for the D1-D5 CFT in  \cite{ggr17,bggmr17,grw18,bg19,rrz19,grtw19,grtw20,wz21} have come from a combination of position space bootstrap methods and a hidden conformal symmetry first observed in \cite{ct18}. Despite the conceptually interesting nature of hidden conformal symmetry, using it to compute the remaining four-point functions appears to be a daunting task.

The following sections will detail an approach, only based on the manifest symmetries, which accomplishes the formerly daunting task by using the Mellin formalism after all. The logic is inspired by experience with spinning correlators in $d > 2$ which have multiple tensor structures and therefore multiple functions of the cross-ratios which need to be considered in tandem. As a result, $AdS_3/CFT_2$ is by no means beyond the reach of Mellin space bootstrap methods.\footnote{See \cite{rz24} for a discussion of $AdS_2/CFT_1$ which also has issues with Mellin space. These will not be addressed here.}
Even though a single-trace four-point function cannot be written as an integral transform of one Mellin amplitude, our analysis will make it clear that this problem can be cleanly circumvented by using \textit{two} Mellin amplitudes. With this proposal in hand, the non-trivial question is whether the algorithms of \cite{rz17,z17,az20,abfz21} can be adapted to still solve for these amplitudes in an efficient way.
We will show that the answer is yes and proceed to compute a large new class of four-point functions using single-trace cubic couplings from \cite{apt00} as the only dynamical input.

While our methods will use details about the D1-D5 CFT very sparingly, we would be remiss not to
discuss some of its remarkable structure.
At a generic point in moduli space, the bulk dual of this theory (which we will continue to refer to as the D1-D5 CFT) has both R-R and NS-NS flux. We can think of these as being sourced by a D1-D5 and NS1-NS5 system respectively \cite{gks98}. Since $AdS_3 \times S^3$ supergravity has an $SO(5)$ symmetry which rotates between the fluxes \cite{apt00}, it is possible to approach this regime along a path which only turns on NS-NS flux.
Strings in this background have a highly tractable worldsheet CFT which is a level $Q_5$ WZW model for either $SL(2; \mathbb{R}) \times SU(2)$ or $PSU(1, 1 | 2)$ depending on the formalism \cite{mo00,bvw99}. The latter (hybrid formalism) has been especially useful as it is well defined in the minimal NS-NS flux case of $Q_5 = 1$ which corresponds to a tensionless string. Building on previous evidence from \cite{gg14,gg18}, this allowed \cite{egg18} to show that the tensionless point is precisely dual to the free symmetric orbifold for $M_4 = T^4$.\footnote{Before this was realized, \cite{gk07} for instance had already used the pure NS-NS background to compute three-point functions in the D1-D5 CFT in order to enable comparisons to the free orbifold and supergravity. Agreement was found because half-BPS three-point functions are protected \cite{bbp12} but all three calculations were done at different points in moduli space.} In other words, this holographic CFT has a bulk dual which is solvable at both strong \textit{and} weak coupling making it the best candidate for deriving an AdS/CFT duality \cite{egg19,dggk20}.
In the planar limit, solvability extends beyond the pure NS-NS backgrounds owing to the integrability of type IIB string theory in $AdS_3 \times S^3 \times T^4$ \cite{bbz09}. An integrable bootstrap, reviewed in \cite{s14, ss24}, has been developed which allowed \cite{lsss14} to compute an exact S-matrix with R-R flux using worldsheet methods. A different calculation in \cite{ggn23}, based on deforming the free symmetric orbifold, has also been applied to compute the string S-matrix and the equivalence of the two approaches was shown in \cite{fs23}. Recent work in higher dimensions \cite{ahs22,ahs23,ah23} has shown that data from integrability can be leveraged to learn about correlators of light operators beyond the planar limit as well.

This paper is organized as follows. In section \ref{sec:setup}, we discuss single-trace superconformal primaries in the CFT and review how their four-point functions can be constrained by kinematics. The new technical developments for this problem are made in section \ref{sec:bootstrap} which culminates in an ansatz for two Mellin amplitudes together with an efficient algorithm for fixing their coefficients using superconformal Ward identities. Our ansatz fully incorporates the kinematic constraints but only requires the most basic data from the effective action in $AdS_3$. The correlators which can be obtained with this formalism are described in section \ref{sec:results} including the four-point functions of graviton Kaluza-Klein modes which are new. A discussion of future prospects is given in section \ref{sec:conc} and certain technical details are relegated to the appendices.

\section{Setup}
\label{sec:setup}

The symmetry most relevant for us will be two copies of the small $\calN = 4$ supergroup --- $PSU(1,1|2)_L \times PSU(1,1|2)_R$. This appears as a symmetry in the symmetric product orbifolds $\text{Sym}^N(M_4) / S_N$ with $M_4$ being the $T^4$ or $K3$ sigma model with appropriate fermion content. The maximal bosonic subgroup is $SL(2; \mathbb{R}) \times SU(2)$ in each factor (representing conformal symmetry and R-symmetry respectively) and these are indeed isometries of the $AdS_3 \times S^3$ background. In this section, we will review the small $\calN = 4$ multiplets which appear in the single-particle supergravity spectrum. We will then discuss how their correlation functions can be fixed in principle by imposing symmetries with a small amount of input from AdS/CFT. Strictly speaking, we will list all single-traces for $M_4 = K3$ in which case we are dealing with 6d $\calN = (2, 0)$ supergravity on $AdS_3 \times S^3$. The $M_4 = T^4$ case, which leads to $\calN = (2, 2)$ supergravity, has these same multiplets plus additional ones. The additional ones are fermionic and will therefore not affect the correlators of bosons at tree-level \cite{b98}.

Although we will not make use of it in this paper, it is worth mentioning that the superalgebra $\mathfrak{su}(1,1|2)_L \times \mathfrak{su}(1,1|2)_R$ enhances to a super-\textit{Virasoro} algebra with central charge $c = 6N$. Specifically the small $\calN = 4$ super-Virasoro algebra for $K3$ and a Wigner contraction of the large $\calN = 4$ super-Virasoro algebra for $T^4$. This symmetry will not help us solve for four-point functions of single-trace operators but it will tightly constrain more general correlation functions where the multi-trace operators involved are normal ordered products of spacetime symmetry currents and other single-trace operators.

\subsection{Supergravity spectrum}

The field content found by reducing type IIB supergravity on $AdS_3 \times S^3 \times K3$ was computed in \cite{dkss98,b98}. The resulting 6d $\calN = (2, 0)$ supergravity theory on $AdS_3 \times S^3$ is coupled to $h_{1,1} + 1$ tensor multiplets where $h_{1,1} = 20$. We will leave $h_{1,1}$ general since the $\calN = (2, 2)$ case for $T^4$ will lead to $h_{1,1} = 4$. The Kaluza-Klein modes of both tensor and graviton type were found to arrange themselves into multiplets of small $\calN = 4$ superconformal symmetry. What makes much of the subsequent analysis possible is that the superconformal primaries of these multiplets are annihilated by half of the $PSU(1,1|2)_L$ supercharges and half of the $PSU(1,1|2)_R$ supercharges. It would be interesting to find a structural explanation for this fact which currently appears to be a fortuitous output of the explicit Kaluza-Klein reduction.\footnote{The analogous statement for the maximal supergraviton and supergluon theories referenced in the introduction is inevitable. With 16 supercharges, half-BPS multiplets are the only ones small enough to be consistent with spin-2 gravity. With 8 supercharges, half-BPS multiplets are again the only ones small enough to be consistent with spin-1 Yang-Mills. Conversely, the $AdS_3/CFT_2$ setup here involves supergravity but only 8 supercharges. It is therefore harder to rule out long multiplets with primaries that are Lorentz scalars. A possible way to proceed would be finding an argument for why the hidden conformal symmetry observed for this model should be fundamental.}

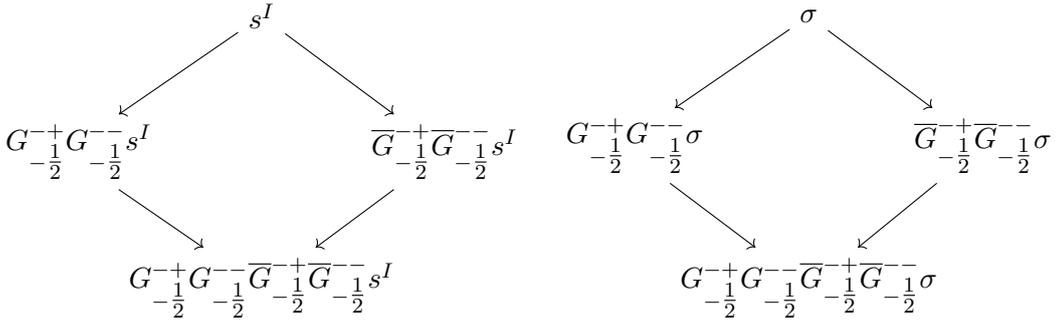
\begin{figure}
    \centering
    \begin{tikzpicture}[scale=0.6]
        \node[] at (-3,0) (s) {$s^I$};
        \node[] (bggs)[below right = of s] {$\overline{G}^{-+}_{-\tfrac{1}{2}} \overline{G}^{--}_{-\tfrac{1}{2}} s^I$};
        \node[] (ggs)[below left = of s] {$ \quad G^{-+}_{-\tfrac{1}{2}} G^{--}_{-\tfrac{1}{2}} s^I$};
        \node[] at (-3,-6) (ggggs) {$  G^{-+}_{-\tfrac{1}{2}} G^{--}_{-\tfrac{1}{2}} \overline{G}^{-+}_{-\tfrac{1}{2}} \overline{G}^{--}_{-\tfrac{1}{2}} s^I $};
        \draw [->] (s) -- (ggs);
        \draw [->] (s) -- (bggs);
        \draw [->] (ggs) -- (ggggs);
        \draw [->] (bggs) -- (ggggs);
        \node[] at (9,0) (sig) {$\sigma$};
        \node[] (bggsig)[below right = of sig] {$\overline{G}^{-+}_{-\tfrac{1}{2}} \overline{G}^{--}_{-\tfrac{1}{2}} \sigma$};
        \node[] (ggsig)[below left = of sig] {$ \quad G^{-+}_{-\tfrac{1}{2}} G^{--}_{-\tfrac{1}{2}} \sigma$};
        \node[] at (9,-6) (ggggsig) {$  G^{-+}_{-\tfrac{1}{2}} G^{--}_{-\tfrac{1}{2}} \overline{G}^{-+}_{-\tfrac{1}{2}} \overline{G}^{--}_{-\tfrac{1}{2}} \sigma $};
        \draw [->] (sig) -- (ggsig);
        \draw [->] (sig) -- (bggsig);
        \draw [->] (ggsig) -- (ggggsig);
        \draw [->] (bggsig) -- (ggggsig);
    \end{tikzpicture}
    \caption{Super-descendants of $s^I$ and $\sigma$ allowed to appear in bosonic OPEs.}
    \label{fig:superMultiplets1}
    \end{figure}

The following discussion will use the 2d notation $h, \bar{h} \equiv \frac{\Delta \pm \ell}{2}$. We will also abuse notation and use the same letter to refer to the superconformal primary and the multiplet as a whole. From the 6d tensor multiplet, we have $s^I_k$ where $k$ is an integer starting from $1$. It is a fundamental of $SO(h_{1,1} + 1)$ with $SL(2; \mathbb{R})_L \times SU(2)_L$ weights $h = j = \frac{k}{2}$ and $SL(2; \mathbb{R})_R \times SU(2)_R$ weights $\bar{h} = \bar{\jmath} = \frac{k}{2}$. The 6d graviton multiplet gives $\sigma_k$ where the integer $k$ this time starts from $2$. It is trivial under $SO(h_{1,1} + 1)$ but its $SL(2; \mathbb{R}) \times SU(2)$ weights are again $h = j = \frac{k}{2}$ and $\bar{h} = \bar{\jmath} = \frac{k}{2}$. The super-descendants of these operators which satisfy the right selection rules to appear in a bosonic OPE are in Figure \ref{fig:superMultiplets1}.
In the notation of \cite{ll19}, which writes $[j, \bar{\jmath}]_{h, \bar{h}}$ for a conformal primary, $s_k^I \sigma_k \in [\tfrac{k}{2}, \tfrac{k}{2}]_{\tfrac{k}{2}, \tfrac{k}{2}}$ while supercharges adjust the weights according to $G^{-\pm}_{-\tfrac{1}{2}} \in [-\tfrac{1}{2}, 0]_{\tfrac{1}{2}, 0}$ and $\overline{G}^{-\pm}_{-\tfrac{1}{2}} \in [0, -\tfrac{1}{2}]_{0, \tfrac{1}{2}}$.\footnote{In \cite{apt00,rrz19}, a much larger portion of the alphabet was used by assigning a different letter to each conformal primary. The choice to use just $s^I$ and $\sigma$ here, together with the action of supercharges, reflects the spirit of our bootstrap approach which pays greatest attention to the super-primaries because descendants have their contributions fixed by symmetry.}

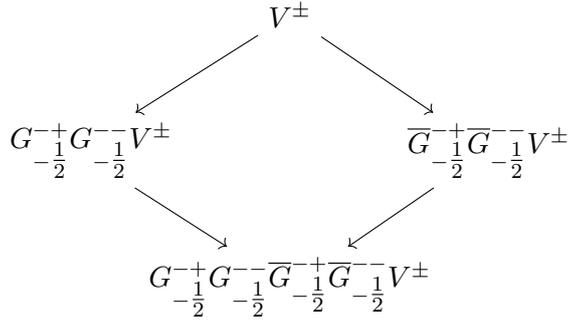
\begin{figure}
    \centering
    \begin{tikzpicture}[scale=0.6]
        \node[] at (0,0) (v) {$V^{\pm}$};
        \node[] (bggv)[below right = of v] {$\overline{G}^{-+}_{-\tfrac{1}{2}} \overline{G}^{--}_{-\tfrac{1}{2}} V^{\pm}$};
        \node[] (ggv)[below left = of v] {$ \quad G^{-+}_{-\tfrac{1}{2}} G^{--}_{-\tfrac{1}{2}} V^{\pm}$};
        \node[] at (0,-6) (ggggv) {$  G^{-+}_{-\tfrac{1}{2}} G^{--}_{-\tfrac{1}{2}} \overline{G}^{-+}_{-\tfrac{1}{2}} \overline{G}^{--}_{-\tfrac{1}{2}}V^{\pm} $};
        \draw [->] (v) -- (ggv);
        \draw [->] (v) -- (bggv);
        \draw [->] (ggv) -- (ggggv);
        \draw [->] (bggv) -- (ggggv);
    \end{tikzpicture}
    \caption{Bosonic super-descendants of $V^{\pm}.$}
    \label{fig:superMultiplets2}
    \end{figure}

In the multiplets above, the super-primary is a Lorentz scalar and the maximal spin among its super-descendants is $1$. In addition to these, the graviton gives us multiplets where the primary has spin $1$ and the maximal spin is $2$. These will be the Lorentz vectors $V^\pm_k$ which are both singlets under $SO(h_{1,1} + 1)$ and have $k$ starting at $1$. The quantum numbers of the self-dual $V^+_k$ are $h = j = \frac{k + 1}{2}$ and $\bar{h} = \bar{\jmath} = \frac{k - 1}{2}$. Those of the anti-self-dual $V^-_k$ are $h = j = \frac{k - 1}{2}$ and $\bar{h} = \bar{\jmath} = \frac{k + 1}{2}$. Referring to \cite{ll19} again, we can write $V^\pm_k \in [\tfrac{k \pm 1}{2}, \tfrac{k \mp 1}{2}]_{\tfrac{k \pm 1}{2}, \tfrac{k \mp 1}{2}}$ and the multiplet structure in Figure \ref{fig:superMultiplets2}.
Although half-BPS multiplets of $\mathfrak{psu}(1, 1 | 2)$ saturate a continuum of quantum numbers compatible with unitarity, the single-trace spectrum does not include pairs of operators which can recombine.

\subsection{Superconformal kinematics}

To study four-point functions of the form
\es{4pt-types}{
\left < s_{k_1}^{I_1} s_{k_2}^{I_2} s_{k_3}^{I_3} s_{k_4}^{I_4} \right >, \quad \left < s_{k_1}^{I_1} s_{k_2}^{I_2} \sigma_{k_3} \sigma_{k_4} \right >, \quad \left < \sigma_{k_1} \sigma_{k_2} \sigma_{k_3} \sigma_{k_4} \right >,
}
the first step is to package all R-symmetry components together by saturating $SU(2)_L$ and $SU(2)_R$ indices with commuting spinor polarizations.
\es{saturation}{
s_k^I(x, v, \bar{v}) &\equiv s_k^{I; \alpha_1 \dots \alpha_k \dot{\alpha}_1 \dots \dot{\alpha}_k}(x) v_{\alpha_1} \dots v_{\alpha_k} \bar{v}_{\dot{\alpha}_1} \dots \bar{v}_{\dot{\alpha}_k} \\
\sigma_k(x, v, \bar{v}) &\equiv \sigma_k^{\alpha_1 \dots \alpha_k \dot{\alpha}_1 \dots \dot{\alpha}_k}(x) v_{\alpha_1} \dots v_{\alpha_k} \bar{v}_{\dot{\alpha}_1} \dots \bar{v}_{\dot{\alpha}_k}
}
We can now write the four-point functions as
\es{4pt-form}{
\left < s_{k_1}^{I_1}(x_1, v_1, \bar{v}_1) s_{k_2}^{I_2}(x_2, v_2, \bar{v}_2) s_{k_3}^{I_3}(x_3, v_3, \bar{v}_3) s_{k_4}^{I_4}(x_4, v_4, \bar{v}_4) \right > &= \mathbf{K}(x_i, v_i, \bar{v}_i) G^{I_1 I_2 I_3 I_4}_{k_1 k_2 k_3 k_4}(z, \bar{z}, \alpha, \bar{\alpha}) \\
\left < s_{k_1}^{I_1}(x_1, v_1, \bar{v}_1) s_{k_2}^{I_2}(x_2, v_2, \bar{v}_2) \sigma_{k_3}(x_3, v_3, \bar{v}_3) \sigma_{k_4}(x_4, v_4, \bar{v}_4) \right > &= \mathbf{K}(x_i, v_i, \bar{v}_i) G^{I_1 I_2}_{k_1 k_2 k_3 k_4}(z, \bar{z}, \alpha, \bar{\alpha}) \\
\left < \sigma_{k_1}(x_1, v_1, \bar{v}_1) \sigma_{k_2}(x_2, v_2, \bar{v}_2) \sigma_{k_3}(x_3, v_3, \bar{v}_3) \sigma_{k_4}(x_4, v_4, \bar{v}_4) \right > &= \mathbf{K}(x_i, v_i, \bar{v}_i) G_{k_1 k_2 k_3 k_4}(z, \bar{z}, \alpha, \bar{\alpha})
}
where the kinematic prefactor $\mathbf{K}$ depends on
\es{differences}{
x^\mu_{ij} \equiv x_i^\mu - x_j^\mu, \quad v_{ij} \equiv \epsilon_{\alpha \beta} v_i^\alpha v_j^\beta, \quad \bar{v}_{ij} \equiv \epsilon_{\dot{\alpha} \dot{\beta}} \bar{v}_i^{\dot{\alpha}} \bar{v}_j^{\dot{\beta}}
}
in such a way that all bosonic symmetries are manifest. The dynamical function $G$ (suppressing Kaluza-Klein levels and possible $SO(h_{1,1} + 1)$ indices) depends on conformal cross-ratios $z, \bar{z}$ and R-symmetry cross-ratios $\alpha, \bar{\alpha}$. Writing $x_i^\mu \equiv [\text{Re}(z_i), \text{Im}(z_i)]$, these are defined as
\es{cross-ratios1}{
z \equiv \frac{z_{12} z_{34}}{z_{13} z_{24}}, \quad \bar{z} \equiv \frac{\bar{z}_{12} \bar{z}_{34}}{\bar{z}_{13} \bar{z}_{24}}, \quad \alpha \equiv \frac{v_{13} v_{24}}{v_{12} v_{34}}, \quad \bar{\alpha} \equiv \frac{\bar{v}_{13} \bar{v}_{24}}{\bar{v}_{12} \bar{v}_{34}}.
}
It is common to form the symmetric combinations
\es{cross-ratios2}{
U \equiv z\bar{z}, \quad V \equiv (1 - z)(1 - \bar{z}), \quad \sigma \equiv \alpha \bar{\alpha}, \quad \tau \equiv (1 - \alpha)(1 - \bar{\alpha})
}
but it is a crucial fact that not all functions $G(z, \bar{z}, \alpha, \bar{\alpha})$ can be written in terms of these combinations because there is not a unique way to invert the quadratic relations \eqref{cross-ratios2}. It is easy to see that for the cross-ratios in \eqref{cross-ratios2} to be enough, $G(z, \bar{z}, \alpha, \bar{\alpha})$ must be symmetric under $z \leftrightarrow \bar{z}$ and $\alpha \leftrightarrow \bar{\alpha}$. The dynamical functions appearing in \eqref{4pt-form} will not have this property. To see why, we can consider the expression for this function which follows from the OPE. One does not even need supersymmetry to see that exchanged operators have the factorized contributions
\es{toy-ope}{
g_h(z) g_{-j}(\alpha^{-1}) g_{\bar{h}}(\bar{z}) g_{-\bar{\jmath}}(\bar{\alpha}^{-1}) \subset G(z, \bar{z}, \alpha, \bar{\alpha})
}
in terms of $SL(2; \mathbb{R})$ blocks. This shows why we should not expect correlators to be symmetric functions of $z$ and $\bar{z}$. Such a symmetry would require every exchanged operator with $(h, j)$ on the left and $(\bar{h}, \bar{\jmath})$ on the right to come with another corresponding operator having $(\bar{h}, j)$ on the left and $(h, \bar{\jmath})$ on the right. A quick look at Figures \ref{fig:superMultiplets1} and \ref{fig:superMultiplets2} is enough to convince ourselves that the holographic dual of 6d supergravity on $AdS_3 \times S^3$ does not obey this constraint. Although $h \leftrightarrow \bar{h}$ alone does not leave the single-trace spectrum invariant, the simultaneous exchange $(h,j) \leftrightarrow (\bar{h},\bar{\jmath})$ does. If this property extends to the multi-trace spectrum and the OPE coefficients, it will follow that $G(z, \bar{z}, \alpha, \bar{\alpha})$ is invariant under $(z,\alpha) \leftrightarrow (\bar{z},\bar{\alpha})$. Parity symmetry guarantees that this is the case. 

Moving onto the prefactor $\mathbf{K}(x_i, v_i, \bar{v}_i)$, our choice for it is motivated by R-symmetry selection rules. These ensure that the number of distinct powers of $\alpha$ in $G(z, \bar{z}, \alpha, \bar{\alpha})$ (equal to the number of distinct powers of $\bar{\alpha}$) is finite. More precisely, it is $\calE + 1$ where the \textit{extremality} is defined as
\es{extremality}{
\calE = \text{min} \left [ \text{min}(k_i), \frac{1}{2} \sum_{i = 1}^4 k_i - \text{max}(k_i) \right ].
}
A convenient prefactor is one which makes the dynamical function a degree $\calE$ polynomial in $\alpha$ and $\bar{\alpha}$. To this end, we will set
\es{prefactor}{
\mathbf{K}(x_i,v_i,\bar{v}_i) = \left | \frac{v_{34}}{z_{34}} \right |^{k_4 + k_3 - k_2 - k_1} \left | \frac{v_{24}}{z_{24}} \right |^{k_2 + k_4 - k_1 - k_3} \left | \frac{v_{23}}{z_{23}} \right |^{k_1 + k_2 + k_3 - k_4 - 2\mathcal{E}} \left | \frac{v_{14}}{z_{14}} \right |^{2k_1 - 2\mathcal{E}} \left | \frac{v_{12} v_{34}}{z_{12} z_{34}} \right |^{2\mathcal{E}}
}
which was also used in \cite{az20,bfz21,abfz21}.\footnote{These papers referred to $\calE = \frac{1}{2} \sum_i k_i - \text{max}(k_i)$ as case I and $\calE = \text{min}(k_i)$ as case II. We will instead denote these two cases by $\calE \neq \text{min}(k_i)$ and $\calE = \text{min}(k_i)$ respectively.} This has the desired property when the weights satisfy one of the orderings
\es{3-orderings}{
& k_1 \leq k_2 \leq k_3 \leq k_4, \quad k_1 \leq k_3 \leq k_2 \leq k_4, \quad k_1 \leq k_4 \leq k_3 \leq k_2 \quad \text{for} \;\; \calE = \text{min}(k_i) \\
& k_1 \leq k_2 \leq k_3 \leq k_4, \quad k_1 \leq k_3 \leq k_2 \leq k_4, \quad k_3 \leq k_2 \leq k_1 \leq k_4 \quad \text{for} \;\; \calE \neq \text{min}(k_i).
}

Because our external operators are half-BPS, fermionic generators of $PSU(1,1|2)_L \times PSU(1,1|2)_R$ are not only useful for generating correlation functions of super-descendants. They also place non-trivial constraints on correlation functions of super-primaries which are captured in the superconformal Ward identities. When the prefactor is \eqref{prefactor}, these take the form \cite{grw18,rrz19}
\es{scwi}{
[\alpha \partial_\alpha - z \partial_z] G \bigl |_{\alpha = z^{-1}} = 0, \quad [\bar{\alpha} \partial_{\bar{\alpha}} - \bar{z} \partial_{\bar{z}}] G \bigl |_{\bar{\alpha} = \bar{z}^{-1}} = 0.
}
Using these identities, \cite{bfz21} computed superconformal blocks for exchanged operators that are also half-BPS. These have the factorized form
\es{bps-blocks}{
\frac{(\alpha z)^{h_{34}} (\bar{\alpha} \bar{z})^{\bar{h}_{34}} \mathcal{G}_h^{h_{12}, h_{34}}(z, \alpha) \mathcal{G}_{\bar{h}}^{\bar{h}_{12}, \bar{h}_{34}}(\bar{z}, \bar{\alpha})}{[(1 - z^{-1})/(1 - \alpha)]^{\calE - h_1 - h_2 - h_{34}} [(1 - \bar{z}^{-1})/(1 - \bar{\alpha})]^{\calE - \bar{h}_1 - \bar{h}_2 - \bar{h}_{34}}}
}
where the $PSU(1,1|2)$ block is given by
\es{psu112-block}{
\mathcal{G}_h^{h_{12}, h_{34}}(z, \alpha) &= g_h^{h_{12}, h_{34}}(z) g_{-h}^{-h_{12}, -h_{34}}(\alpha^{-1}) + \lambda_h^{h_{12}, h_{34}} g_{1 + h}^{h_{12}, h_{34}}(z) g_{1 - h}^{-h_{12}, -h_{34}}(\alpha^{-1}) \\
\lambda_h^{h_{12}, h_{34}} &\equiv \frac{(h^2 - h_{12}^2)(h^2 - h_{34}^2)}{4h^2(1 - 4h^2)}
}
and the $SL(2; \mathbb{R})$ block is given by
\es{sl2-block}{
g_h^{h_{12}, h_{34}}(z) = z^h {}_2F_1(h - h_{12}, h + h_{34}; 2h; z).
}
As in other even dimensions, there is a formal solution to the superconformal Ward identity which takes the form
\es{formal-sol}{
G(z, \bar{z}, \alpha, \bar{\alpha}) = \hat{G}(z, \bar{z}, \alpha, \bar{\alpha}) + (1 - z\alpha)(1 - \bar{z}\bar{\alpha}) H(z, \bar{z}, \alpha, \bar{\alpha}).
}
This splitting can be performed such that $\hat{G}(z, \bar{z}, \alpha, \bar{\alpha})$ is a rational function as shown by the cohomological argument in \cite{bllprv13,rrz19}. When $\alpha = z^{-1}$ it becomes independent of $z$ and when $\bar{\alpha} = \bar{z}^{-1}$ it becomes independent of $\bar{z}$. The non-rational part leftover denoted by $H(z, \bar{z}, \alpha, \bar{\alpha})$ is what we will refer to as the \textit{auxiliary correlator}.

\subsection{Review of previous approaches}

The first single-trace four-point function to be computed in the tree-level supergravity regime was $\left < s_1^1 s_1^1 s_1^1 s_1^1 \right >$ \cite{grw18}. The original calculation used results for $\left < s_1^1 s_1^1 \mathcal{O}_b \mathcal{O}_b \right >$ with $\mathcal{O}_b$ denoting a family of multi-trace operators having conformal dimensions that scale with $N$. This four-point function could be computed in \cite{ggr17,bggmr17} because the operators $\mathcal{O}_b$ were judiciously chosen to source asymptotically $AdS_3 \times S^3$ geometries that are exactly known. By varrying $b$, \cite{grw18} was able to learn enough about the $s_1^1 \times s_1^1$ OPE to recover the correlation function of four $s_1^1$ operators as a crossing symmetric completion. This method was developed further in \cite{bg19,grtw19}.

The first efforts to move away from this model dependent approach were made in \cite{rrz19}. In any holographic CFT, a four-point function can be written as a finite sum of Witten diagrams in all three channels instead of an infinite sum of conformal blocks in a single channel. Specializing to $\left < s_p^{I_1} s_p^{I_2} s_p^{I_3} s_p^{I_4} \right >$, \cite{rrz19} considered the ansatz
\begin{align}
G^{I_1 I_2 I_3 I_4}_{pppp}(z, \bar{z}, \alpha, \bar{\alpha}) &= \delta^{I_1 I_2} \delta^{I_3 I_4} G^{(s)}_{pppp}(z, \bar{z}, \alpha, \bar{\alpha}) + \text{crossed} \label{position-ansatz} \\
G^{(s)}_{pppp}(z, \bar{z}, \alpha, \bar{\alpha}) &= \sum_k C^{ssV}_{ppk} C^{Vss}_{kpp} \mathcal{S}_{V, k}(z, \bar{z}, \alpha, \bar{\alpha}) + C^{ss\sigma}_{ppk} C^{\sigma ss}_{kpp} \mathcal{S}_{\sigma, k}(z, \bar{z}, \alpha, \bar{\alpha}) + \mathcal{C}(z, \bar{z}, \alpha, \bar{\alpha}). \nonumber
\end{align}
From right to left, $\mathcal{C}(z, \bar{z}, \alpha, \bar{\alpha})$ is a linear combination of four-point contact diagrams with up to two derivatives whose coefficients are \textit{a priori} unknown degree $\calE = p$ polynomials in both $\alpha$ and $\bar{\alpha}$. The super-Witten diagrams $\mathcal{S}_{V, k}(z, \bar{z}, \alpha, \bar{\alpha})$ and $\mathcal{S}_{\sigma, k}(z, \bar{z}, \alpha, \bar{\alpha})$ are linear combinations of ordinary exchange Witten diagrams for all conformal primaries in the indicated multiplet weighted by $g^{0,0}_{-j}(\alpha^{-1}) g^{0,0}_{-\bar{\jmath}}(\bar{\alpha}^{-1})$ where $(j, \bar{\jmath})$ are the appropriate $SU(2)_L \times SU(2)_R$ spins.\footnote{These super-Witten diagrams were referred to as multiplet exchange amplitudes in \cite{az20,abfz21}. As with $\mathcal{C}(z, \bar{z}, \alpha, \bar{\alpha})$, \cite{rrz19} took the coefficients of the individual Witten diagrams comprising them to also be \textit{a priori} unknown but they could have been fixed (thereby speeding up the calculation) using the superconformal blocks \eqref{bps-blocks}.} Finally, $C^{ssV}_{ppk} C^{Vss}_{kpp}$ and $C^{ss\sigma}_{ppk} C^{\sigma ss}_{kpp}$ are the squared OPE coefficients of single-trace primaries.

On an ansatz like \eqref{position-ansatz}, three steps can be carried out to simplify the expressions for the Witten diagrams. First, the four-point diagrams are manifestly proportional to the so called $\bar{D}$-functions defined in \cite{do00}. Second, the spectrum is such that an algorithm inspired by \cite{dfr99} can be used to write exchange diagrams in terms of $\bar{D}$-functions as well. This requires modifications compared to \cite{dfr99} (to handle Levi-Civita symbols in the effective action for instance) and leads to expressions that grow with $p$. Third, even though they are more complicated in position space than Mellin space, the $\bar{D}$-functions can all be written as linear combinations over $\mathbb{R}(z, \bar{z})$ of $1$, $\log U$, $\log V$ and the scalar box diagram. Having found a suitable basis, \cite{rrz19} was able to apply \eqref{scwi} as a means of fixing unknown coefficients. When the squared OPE coefficients were computed from the supergravity analysis of \cite{apt00} and supplied as input, all remaining unknowns (the coefficients of contact terms) were able to be fixed. This allowed \cite{rrz19} to compute correlators explicitly up to $p = 4$. That the superconformal Ward identity has a unique solution is no accident. As explained in \cite{ct18,b22}, albeit for 4d theories, this is a consequence of the following facts.
\begin{enumerate}
\item Along the twisted locus $\alpha = z^{-1}$, half-BPS four-point functions can be bootstrapped using chiral symmetry \cite{bllprv13}.
\item As long as single-trace operators are half-BPS, their contributions to the full four-point function can be reconstructed from their contributions to the chiral algebra.
\item The single-trace part of the correlator fixes its double-discontinuity which in turn fixes the rest of the correlator following the logic in \cite{c17}.
\end{enumerate}
Taking this point of view, the cubic couplings from \cite{apt00} are needed simply because the mapping between operators in the chiral algebra and those in the original theory is ambiguous. Unlike in $\calN = 4$ SYM for instance, a single-trace superconformal primary is not uniquely determined by its left-moving quantum numbers $(h, j)$ due to the presence of $\sigma$ and $V$ type operators.

At the cost of once again solving the problem in a model dependent way, the patterns in the $p = 1,2,3,4$ results revealed a strategy for handling arbitrary $p$. This is because \cite{rrz19} noticed that the various four-point functions were related to each other through the \textit{hidden conformal symmetry} $SO(6, 2)$. This was the second example of hidden conformal symmetry to be observed after \cite{ct18} found an $SO(10, 2)$ symmetry underlying the tree-level four-point functions of $\calN = 4$ SYM. Similar results are now known to hold in another two theories as shown by \cite{abfz21,ahl21}.\footnote{This has been a significant benefit in certain subsequent calculations. Notably in \cite{hy21,dp22,hwyz23}, hidden conformal symmetry motivated a highly restrictive ansatz for the two-loop corrections to maximal supergluon and supergraviton scattering in $AdS_5$. The same consistency conditions used at tree-level were then enough to determine the full result up to contact terms. Ordinarily, two-loop amplitudes would lead to a mixing problem which can only be solved by computing four-point functions involving external long multiplets \cite{fm21,fm23a,fm23b}.} For a recent demonstration at the level of five-point functions, see \cite{hwyz24}.

Operationally, hidden conformal symmetry for the tensor Kaluza-Klein modes means that there is a correlation function depending on the 6d distances $x_{ij}^2 + 2v_{ij} \bar{v}_{ij}$ such that the 2d auxiliary correlators $H^{I_1 I_2 I_3 I_4}_{k_1 k_2 k_3 k_4}(z, \bar{z}, \alpha, \bar{\alpha})$ can all be obtained by Taylor expanding to collect powers of $v_{ij} \bar{v}_{ij}$ which correspond to the kinematic prefactor \eqref{prefactor}. The four-point function in question must be fixed by comparing it to low-lying results but this is easy to do because it is a four-point function of 6d scalars. This means that it only contains one function of cross-ratios and it is not surprising that this function is $H^{I_1 I_2 I_3 I_4}_{1111}(z, \bar{z})$. The task of obtaining 2d auxiliary correlators in this way was further simplified by the observation \cite{rrz19,grtw19} that Kaluza-Klein modes of the tensor multiplet enable the decomposition
\es{formal-tensor}{
G_{k_1 k_2 k_3 k_4}^{I_1 I_2 I_3 I_4}(z, \bar{z}, \alpha, \bar{\alpha}) = \hat{G}_{k_1 k_2 k_3 k_4}^{I_1 I_2 I_3 I_4}(z, \bar{z}, \alpha, \bar{\alpha}) + (1 - z\alpha) (1 - \bar{z}\bar{\alpha}) H_{k_1 k_2 k_3 k_4}^{I_1 I_2 I_3 I_4}(U, V, \sigma, \tau).
}
In other words, all the asymmetry of $G_{k_1 k_2 k_3 k_4}^{I_1 I_2 I_3 I_4}(z, \bar{z}, \alpha, \bar{\alpha})$ under exchanging $z \leftrightarrow \bar{z}$ and $\alpha \leftrightarrow \bar{\alpha}$ separately is accounted for by $(1 - z\alpha)(1 - \bar{z}\bar{\alpha})$. The non-trivial function in \eqref{formal-tensor} can then be obtained from an inverse Mellin transform according to
\begin{align}
H_{k_1 k_2 k_3 k_4}^{I_1 I_2 I_3 I_4}(U, V, \sigma, \tau) &= \int_{-i\infty}^{i\infty} \frac{ds dt}{(4\pi i)^2} U^{\frac{s + 2\calE - k_1 - k_2}{2}} V^{\frac{t - 2\calE + k_{14}}{2}} \calMt_{k_1 k_2 k_3 k_4}^{I_1 I_2 I_3 I_4}(s, t, \sigma, \tau) \label{mellin-tensor} \\
&\times \Gamma[\tfrac{k_1 + k_2 - s}{2}] \Gamma[\tfrac{k_3 + k_4 - s}{2}] \Gamma[\tfrac{k_1 + k_4 - t}{2}] \Gamma[\tfrac{k_2 + k_3 - t}{2}] \Gamma[\tfrac{k_1 + k_3 - \tilde{u}}{2}] \Gamma[\tfrac{k_2 + k_4 - \tilde{u}}{2}] \nonumber
\end{align}
where $s + t + u = \sum_{i = 1}^4 k_i$ and $\tilde{u} = u - 2$. Working out the Mellin amplitude, \cite{wz21} found
\begin{align}
\calMt_{k_1 k_2 k_3 k_4}^{I_1 I_2 I_3 I_4}(s, t, \sigma, \tau) &= \sum_{0 \leq m_1 + m_2 \leq \calE - 1} \frac{\sigma^{m_2} \tau^{\calE - m_1 - m_2 - 1}}{\prod_{i = 1}^6 m_i!} \frac{\delta^{I_1 I_2} \delta^{I_3 I_4}}{s + 2 + 2m_1 - k_1 - k_2} + \text{crossed} \nonumber \\
m_6 &= \frac{k_{21} + k_{43}}{2} + m_2, \quad m_5 = \frac{k_1 + k_2 + k_{34}}{2} - m_1 - m_2 - 1 \label{known-tensor} \\
m_4 &= \frac{k_{31} + k_{42}}{2} + m_1, \quad m_3 = k_1 - m_1 - m_2 - 1. \nonumber
\end{align}

As established by \cite{grtw20}, tree-level four-point functions from the graviton multiplet are also organized by a 6d hidden conformal symmetry but this time the external operators in the master correlator are not scalars. They are self-dual 3-forms which lead to functions of the cross-ratios appearing in several places. Even after a sufficient amount of low-lying data is used to fix them all, the Taylor expansion needed at the end of the day is harder to carry out in closed form. Nevertheless, \cite{wz21} was able to solve for the mixed tensor-graviton auxiliary correlator $H^{I_1 I_2}_{k_1 k_2 k_3 k_4}(z, \bar{z}, \alpha, \bar{\alpha})$. Since this is genuinely a function of $(z, \alpha)$ and $(\bar{z}, \bar{\alpha})$, results were presented as Mellin amplitudes for the individual functions in
\es{known-decomp}{
H^{I_1 I_2}_{k_1 k_2 k_3 k_4}(z, \bar{z}, \alpha, \bar{\alpha}) = \delta^{I_1 I_2} \left [ H^{(+)}_{k_1 k_2 k_3 k_4}(U, V, \sigma, \tau) + (z - \bar{z})(\alpha - \bar{\alpha})H^{(-)}_{k_1 k_2 k_3 k_4}(U, V, \sigma, \tau) \right ].
}
Defining $\Sigma = \sum_{i = 1}^4 k_i$, these are
\begin{align}
\calMt^{(-)}_{k_1 k_2 k_3 k_4}(s, t, \sigma, \tau) &= (-1)^{\frac{\Sigma}{2}} \sum_{0 \leq m_1 + m_2 \leq \calE - 2} \frac{2 \sigma^{m_2} \tau^{\calE - m_1 - m_2 - 2}}{3 \prod_{i = 1}^6 m_i!} \label{known-mixed-odd} \\
&\times \frac{(k_1 + k_3 - \tilde{u})(k_2 + k_4 - \tilde{u})}{(s + 2 + 2m_1 - k_1 - k_2)(t + 2 + 2m_5 - k_2 - k_3)(\tilde{u} + 2m_2 - k_1 - k_3)} \nonumber
\end{align}
and
\begin{align}
\calMt^{(+)}_{k_1 k_2 k_3 k_4}(s, t, \sigma, \tau) &= \calMt^{(0)}_{k_1 k_2 k_3 k_4}(s, t, \sigma, \tau) + \calMt^{(0)}_{k_2 k_1 k_3 k_4}(s, \tilde{u}, \tau, \sigma) \label{known-mixed-even} \\
&- (-1)^{\frac{\Sigma}{2}} \sum_{0 \leq m_1 + m_2 \leq \calE - 1} \frac{\sigma^{m_2} \tau^{\calE - m_1 - m_2 - 1}}{3\prod_{i = 1}^6 m_i!} \frac{(-1)^{k_{34}}}{s + 2 - 2m_1 - k_1 - k_2} \nonumber \\
&\times [4 m_3 m_5 \tau^{-1} - (k_1 + k_2 - 2m_1)(k_1 + k_2 - 2m_1 - 2) + k_3^2 + k_4^2 - 2] \nonumber
\end{align}
in the same conventions as \eqref{mellin-tensor}.\footnote{Compared to \cite{wz21}, we have changed notation slightly and fixed a relative sign between \eqref{known-mixed-odd} and \eqref{known-mixed-even}.} The function on the first line of \eqref{known-mixed-even} is
\begin{align}
\calMt^{(0)}_{k_1 k_2 k_3 k_4}(s, t, \sigma, \tau) &= (-1)^{\frac{\Sigma}{2}} \sum_{0 \leq m_1 + m_2 \leq \calE - 1} \frac{\sigma^{m_2} \tau^{\calE - m_1 - m_2 - 2}}{3 \prod_{i = 1}^6 m_i!} \label{known-mixed-other}  \\
&\times \frac{4 m_3 m_5}{(s + 2 + 2m_1 - k_1 - k_2)(t + 2m_5 - k_2 - k_3)} \nonumber \\
&\times \left [ (\tau - 1) (\tfrac{\tau - 2\sigma - 1}{\sigma} m_2 m_6 + m_{12} + m_{46}) - 2(2m_2 m_6 + m_2 + m_6 + 1) \right. \nonumber \\
&\left. - m_1 m_4 (\sigma - \tau + 1)^2 + (4 m_1 m_4 + m_2 m_6 + m_1 + m_2 + m_4 + m_6 + 2)\sigma \right ]. \nonumber
\end{align}
The following section will outline a method for going beyond these results in which the Mellin transforms of $G(z, \bar{z}, \alpha, \bar{\alpha})$ instead of $H(z, \bar{z}, \alpha, \bar{\alpha})$ take center stage.
\begin{figure}
    \centering
    \begin{tikzpicture}[scale=0.6]
    \newcommand{\cCy}{0}
    \newcommand{\cCx}{12}
    \newcommand{\sCx}{-6}
    \newcommand{\sCy}{0}
    \newcommand{\tCx}{0}
    \newcommand{\tCy}{0}
    \newcommand{\uCx}{6}
    \newcommand{\uCy}{0}
    \node at (3,0) {$+$};
    \node at (-3,0) {$+$};
    \node at (9,0) {$+$};
    \tkzDefPoint(0+\cCx,0+\cCy){a0};
    \tkzDefPoint(-1.41+\cCx,1.41+\cCy){a1};
    \tkzDefPoint(-1.41+\cCx,-1.41+\cCy){a2};
    \tkzDefPoint(1.41+\cCx,-1.41+\cCy){a3};
    \tkzDefPoint(1.41+\cCx,1.41+\cCy){a4};
    \draw  (a0) circle (2cm);
    \draw (a0) -- (a1);
    \draw (a0) -- (a2);
    \draw (a0) -- (a3);
    \draw (a0) -- (a4);
    \tkzDefPoint(\sCx,0+\sCy){b0};
    \tkzDefPoint(\sCx-1.41,1.41+\sCy){b1};
    \tkzDefPoint(\sCx-1.41,-1.41+\sCy){b2};
    \tkzDefPoint(\sCx+1.41,-1.41+\sCy){b3};
    \tkzDefPoint(\sCx+1.41,+1.41+\sCy){b4};
    \tkzDefPoint(\sCx-.8,0+\sCy){b5};
    \tkzDefPoint(\sCx+.8,0+\sCy){b6};
    \draw  (b0) circle (2cm);
    \draw (b5) -- (b1);
    \draw (b5) -- (b2);
    \draw (b6) -- (b3);
    \draw (b6) -- (b4);
    \draw (b0) -- (b6);
    \draw (b0) -- (b5);
    \tkzDefPoint(0+\tCx,0+\tCy){c0};
    \tkzDefPoint(1.41+\tCx,1.41+\tCy){c4};
    \tkzDefPoint(1.41+\tCx,-1.41+\tCy){c3};
    \tkzDefPoint(-1.41+\tCx,1.41+\tCy){c1};
    \tkzDefPoint(-1.41+\tCx,-1.41+\tCy){c2};
    \tkzDefPoint(0+\tCx,-.8+\tCy){c5};
    \tkzDefPoint(0+\tCx,.80+\tCy){c6};
    \draw  (c0) circle (2cm);
    \draw (c5) -- (c2);
    \draw (c5) -- (c3);
    \draw (c6) -- (c4);
    \draw (c6) -- (c1);
    \draw (c0) -- (c6);
    \draw (c0) -- (c5);
    \tkzDefPoint(0+\uCx,0+\uCy){c0};
    \tkzDefPoint(1.41+\uCx,1.41+\uCy){c4};
    \tkzDefPoint(1.41+\uCx,-1.41+\uCy){c3};
    \tkzDefPoint(-1.41+\uCx,1.41+\uCy){c1};
    \tkzDefPoint(-1.41+\uCx,-1.41+\uCy){c2};
    \tkzDefPoint(0+\uCx,-.8+\uCy){c5};
    \tkzDefPoint(0+\uCx,.80+\uCy){c6};
    \draw  (c0) circle (2cm);
    \draw (c5) -- (c2);
    \draw (c5) -- (c4);
    \draw (c6) -- (c3);
    \draw (c6) -- (c1);
    \draw (c0) -- (c6);
    \draw (c0) -- (c5);
    \end{tikzpicture}
    \caption{The correlators are sums of $s$, $t$ and $u$ exchange Witten diagrams plus a contact term}
    \label{fig:wittenDiag}
    \end{figure}
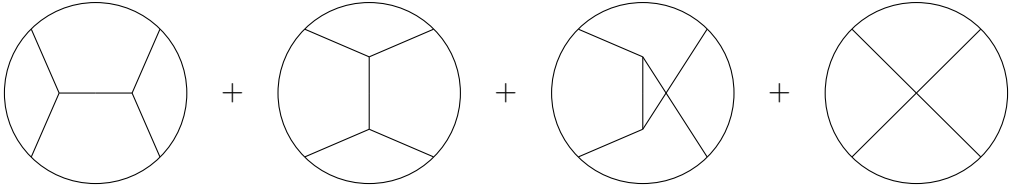
\section{A Mellin bootstrap in two dimensions}
\label{sec:bootstrap}

A worthwhile goal is to return to the first approach in \cite{rrz19}, which only assumes superconformal symmetry, and develop it to the point where correlators of large extremality no longer pose an obstacle. We will therefore advocate the use of Mellin space, not only as a tool for expressing final results, but also for deriving them in the first place. The Mellin representation is based on an integral transform which trades the cross-ratios $U$ and $V$ in \eqref{cross-ratios2} for Mandelstam-like variables $s$ and $t$. In this case, the spatial dimension $d$ is small enough compared to the external spin that four-point functions do not depend only on $U$ and $V$ but on $z$ and $\bar{z}$ individually. This can be traced to the fact that spin-1 operators in the spectrum (not only $V_k^\pm$ but also supercharge combinations like $G^{-+}_{-\frac{1}{2}} G^{--}_{-\frac{1}{2}}$ and $\overline{G}^{-+}_{-\frac{1}{2}} \overline{G}^{--}_{-\frac{1}{2}}$ acting on scalars) are self-dual and anti-self-dual.

This situation motivated \cite{wz21} to define a pair of Mellin amplitudes by splitting the auxiliary correlator into even and odd parts. Our starting point will consist of doing the same thing for the full correlator. The resulting even amplitude will admit an expansion into Witten diagrams which have the same poles and residues as the usual 2d conformal blocks. Interestingly, the most natural expansion for the odd amplitude will be almost the same except with 4d conformal blocks. Since these functions are just as well understood, it will be possible to formulate a bootstrap algorithm analogous to the one in \cite{z17} which solves the problem of going to high extremality.

\subsection{A tale of two Mellins}

Let us start by fixing conventions that differ slightly from those of \cite{wz21}. Holographic four-point functions at tree-level are typically decomposed into four parts, corresponding to the contact diagrams and three types of exchange diagrams. As suggested by \cite{az20}, the contact terms can often be associated with the three channels in a nice way, leading to
\es{g-stu}{
G(z, \bar{z}) = G^{(s)}(z, \bar{z}) + G^{(t)}(z, \bar{z}) + G^{(u)}(z, \bar{z}).
}
Splitting each of these functions into even and odd parts, we have
\es{g-stu-even-odd}{
G^{(s)}(z, \bar{z}) &= G^{(s,+)}(U, V) + \frac{z - \bar{z}}{U} G^{(s,-)}(U,V) \\
G^{(t)}(z, \bar{z}) &= G^{(t,+)}(U, V) - \frac{z - \bar{z}}{V} G^{(t,-)}(U,V) \\
G^{(u)}(z, \bar{z}) &= G^{(u,+)}(U, V) - (z - \bar{z}) G^{(u,-)}(U,V).
}
The prefactor $\frac{z - \bar{z}}{U}$, used for the $s$-channel, is mapped to the $t$-channel by a (13) or (24) permutation. Similarly, it is mapped to the $u$-channel by (14) or (23). These permutations generate the group of crossing transformations which is $S_3$. The Bose symmetry (12) or equivalently (34), which extends this group to $S_4$, is often not considered in CFT. This is because it is automatically respected by conformal blocks and exchange Witten diagrams in $d \geq 2$ as a consequence of the OPE being commutative. Despite this, it can still be helpful in AdS/CFT when formulating an ansatz for contact terms. It is reasonable to expect that the functions in \eqref{g-stu-even-odd} will be easier to bootstrap if all six of them transform simply under Bose symmetry. This agrees with our choice to extract $\frac{z - \bar{z}}{U}$, together with its orbits under $S_3$, since it has a nicer transformation under Bose symmetry than $z - \bar{z}$ alone. Nevertheless, $\frac{z - \bar{z}}{U}$ is not unique in this regard and it is fortunate that one does not need to make a more complicated choice in what follows.

We have not written any $\alpha$ or $\bar{\alpha}$ dependence yet because \eqref{g-stu-even-odd} is a valid decomposition in any holographic CFT. Even when we do have small $\calN = 4$ superconformal symmetry and parity symmetry, which guarantees that the odd correlators are anti-symmetric under $\alpha \leftrightarrow \bar{\alpha}$, we will not extract a factor of $\alpha - \bar{\alpha}$. For all six of the correlators just defined, we will relate them to Mellin amplitudes via
\begin{align}
G^{(*,\pm)}(U, V) &= \int_{-i\infty}^{i\infty} \frac{ds dt}{(4\pi i)^2} U^{\frac{s + 2\calE - \Delta_1 - \Delta_2}{2}} V^{\frac{t - 2\calE + \Delta_{14}}{2}} \calM^{(*,\pm)}(s, t) \Gamma_{\{ \Delta_i \}}(s, t) \label{g-mellin} \\
\Gamma_{\{ \Delta_i \}}(s, t) &\equiv \Gamma[\tfrac{\Delta_1 + \Delta_2 - s}{2}] \Gamma[\tfrac{\Delta_3 + \Delta_4 - s}{2}] \Gamma[\tfrac{\Delta_1 + \Delta_4 - t}{2}] \Gamma[\tfrac{\Delta_2 + \Delta_3 - t}{2}] \Gamma[\tfrac{\Delta_1 + \Delta_3 - u}{2}] \Gamma[\tfrac{\Delta_2 + \Delta_4 - u}{2}] \nonumber
\end{align}
where $s + t + u = \sum_{i = 1}^4 \Delta_i$. It is worth recalling that the standard definition \eqref{g-mellin} leads to a $\bar{z} \ll z \ll 1$ asymptotic of $z^{\frac{\Delta + \ell}{2}} \bar{z}^{\frac{\Delta - \ell}{2}}$ after extracting the pole $s = \Delta - \ell$ and projecting onto spin $\ell$. When also using \eqref{g-mellin} for the odd amplitude, as we have chosen to do, this fact can be used to gain some intuition for which special functions might appear. Since $\frac{z - \bar{z}}{U} \sim \frac{1}{\bar{z}}$ at leading order, one should be now looking for $z^{\frac{\Delta + \ell}{2}} \bar{z}^{\frac{\Delta - \ell + 2}{2}}$ which is the lightcone asymptotic behaviour of a conformal block with a scaling dimension and spin of $(\Delta + 1, \ell - 1)$.

Most of the technical manipulations that follow will be easiest to perform when there is a single even amplitude and a single odd amplitude instead of three of each. This means we need to choose one of the three prefactors in \eqref{g-stu-even-odd} to extract. Choosing $\frac{z - \bar{z}}{U}$ arbitrarily,
\es{g-even-odd}{
G^{(+)}(U, V) &= G^{(+,s)}(U, V) + G^{(+,t)}(U, V) + G^{(+,u)}(U, V) \\
G^{(-)}(U, V) &= G^{(-,s)}(U, V) - UV^{-1} G^{(-,t)}(U, V) - U G^{(-,u)}(U, V)
}
which implies
\es{m-even-odd}{
\calM^{(+)}(s, t) &= \calM^{(+,s)}(s, t) + \calM^{(+,t)}(s, t) + \calM^{(+,u)}(s, t) \\
\calM^{(-)}(s, t) &= \calM^{(-,s)}(s, t) - \widehat{UV^{-1}} \circ \calM^{(-,t)}(s, t) - \widehat{U} \circ \calM^{(-,u)}(s, t).
}
Here we have defined the difference operators
\es{diff-op}{
\widehat{U^m V^n} \circ \calM(s, t) \equiv \frac{\Gamma_{\{ \Delta_i \}}(s - 2m, t - 2n)}{\Gamma_{\{ \Delta_i \}}(s, t)} \calM(s - 2m, t - 2n).
}
Within a family of external operators, the various components are related by crossing transformations.
For convenient calculations, one can write down a Witten diagram ansatz for the $s$-channel and then use these relations to define the ansatz for the other two channels. Fortunately, it is possible to do this without leaving the set of favorable orderings as dictated by \eqref{3-orderings}. Therefore
\es{crossing-g}{
G^{(t,\pm)}_{1234}(U, V) &= \begin{cases}
\left ( \frac{U}{V} \right )^{\calE} G^{(s,\pm)}_{1432}(V, U), & \calE = \text{min}(\Delta_i) \\
\left ( \frac{U}{V} \right )^{\calE} G^{(s,\pm)}_{3214}(V, U), & \calE \neq \text{min}(\Delta_i)
\end{cases} \\
G^{(u,\pm)}_{1234}(U, V) &= U^{\calE} G^{(s,\pm)}_{1324} \left ( \frac{1}{U}, \frac{V}{U} \right )
}
which leads to
\es{crossing-m}{
\calM^{(t,\pm)}_{1234}(s, t) &= \begin{cases}
\calM^{(s,\pm)}_{1432}(t, s), & \calE = \text{min}(\Delta_i) \\
\calM^{(s,\pm)}_{3214}(t, s), & \calE \neq \text{min}(\Delta_i)
\end{cases} \\
\calM^{(u,\pm)}_{1234}(s, t) &= \calM^{(s,\pm)}_{1324}(u, t)
}
in Mellin space. When looking at full R-symmetry multiplets, the permutation of the labels will naturally be accompanied by an action on $\alpha$ and $\bar{\alpha}$.

We are now ready to discuss the even and odd versions of \eqref{position-ansatz}. The $s$-channel expressions, from which we assemble the full ansatz through \eqref{crossing-g} and \eqref{g-even-odd}, are
\es{non-susy-ansatz}{
G^{(s,\pm)}_{1234}(U, V) &= \sum_k C_{12k} C_{k34} \mathcal{W}^{(\pm)}_{\Delta_k, \ell_k}(U, V) + \mathcal{C}^{(\pm)}(U, V).
}
The sum runs over all single-trace operators allowed by selection rules. They have OPE coefficients $C_{12k}$ and $C_{k34}$ which may be inferred from cubic couplings in the effective action. Quartic couplings are much harder to compute in supergravity so these contributions, encoded in $\mathcal{C}^{(\pm)}(U, V)$, are taken as unknowns to be bootstrapped. Due to the separation of single-trace and double-trace data afforded by the measure in \eqref{g-mellin}, $\mathcal{C}^{(\pm)}(U, V)$ becomes a polynomial in $s$ and $t$ after a Mellin transform. Its degree is $1$ reflecting the fact that gravity is a two-derivative theory. Higher powers of $s$ and $t$ would give information about its UV completion in string theory. What becomes of $\mathcal{W}^{(\pm)}_{\Delta_k, \ell_k}(U, V)$ in Mellin space? We are mainly interested in their poles and residues since their regular terms can be absorbed into $\mathcal{C}^{(\pm)}(U, V)$. This means that, up to a kinematic prefactor, the original exchange Witten diagrams with $z$ and $\bar{z}$ can effectively be set equal to the blocks $g_h(z) g_{\bar{h}}(\bar{z})$ where $h, \bar{h} = \frac{\Delta_k \pm \ell_k}{2}$. Taking the even part $\mathcal{W}^{(+)}_{\Delta_k, \ell_k}(U, V)$ produces the symmetrized combination
\es{block-2d}{
g^{2d}_{h + \bar{h}, |h - \bar{h}|}(z, \bar{z}) = g_h^{\frac{\Delta_{12}}{2}, \frac{\Delta_{34}}{2}}(z) g_{\bar{h}}^{\frac{\Delta_{12}}{2}, \frac{\Delta_{34}}{2}}(\bar{z}) + (z \leftrightarrow \bar{z}).
}
These are nothing but the 2d global conformal blocks first computed in \cite{do00}. If $\mathcal{W}^{(-)}_{\Delta_k, \ell_k}(U, V)$ had been defined simply as an anti-symmetrization, the effect of this would be changing the sum in \eqref{block-2d} to a difference. But we have also extracted a factor of $\frac{z - \bar{z}}{U}$ which leads to an appearance of
\es{block-4d}{
g^{4d}_{h + \bar{h}, |h - \bar{h}|}(z, \bar{z}) = \frac{z\bar{z}}{z - \bar{z}} \left [ g_h^{\frac{\Delta_{12}}{2}, \frac{\Delta_{34}}{2}}(z) g_{\bar{h} - 1}^{\frac{\Delta_{12}}{2}, \frac{\Delta_{34}}{2}}(\bar{z}) - (z \leftrightarrow \bar{z}) \right ].
}
One therefore has 4d conformal blocks entering the problem as well. Keeping track of the kinematic prefactor \eqref{prefactor}, we can say that, up to contact terms,
\es{witten-to-block}{
\mathcal{W}^{(+)}_{\Delta, \ell}(U, V) &= V^{\frac{\Delta_{43}}{2}} \left ( \frac{V}{U} \right )^{\calE - \frac{\Delta_1 + \Delta_2}{2}} g^{2d}_{\Delta, \ell}(U, V) \\
\mathcal{W}^{(-)}_{\Delta, \ell}(U, V) &= V^{\frac{\Delta_{43}}{2}} \left ( \frac{V}{U} \right )^{\calE - \frac{\Delta_1 + \Delta_2}{2}} g^{4d}_{\Delta + 1, \ell - 1}(U, V).
}
The Mellin space poles and residues of the right hand sides are known and given in Appendix \ref{app:residues}. Note that it is crucial to write $g^{4d}_{\Delta + 1, \ell - 1}(U, V)$ in \eqref{witten-to-block} instead of $g^{4d}_{\Delta, \ell}(U, V)$ because $\bar{h}$ appears shifted by $1$ in the definition of a 4d conformal block \eqref{block-4d}. This shift means that a theory with 2d blocks up to $\ell = 2$ will only have 4d blocks up to $\ell = 1$.\footnote{Note also that \eqref{block-4d} vanishes at $\ell = -1$ which can be understood as a consequence of this representation going to minus itself under a Weyl reflection \cite{cdi16}.}

\subsection{The superconformal Ward identity}

The considerations in the previous subsection were completely general so let us now discuss the further ground that can be gained when there is small $\calN = 4$ superconformal symmetry. First off, the crossing transformations \eqref{crossing-g} and \eqref{crossing-m} become
\begin{align}
G^{(t,\pm)}_{k_1 k_2 k_3 k_4}(U, V, \alpha, \bar{\alpha}) &= \begin{cases}
\left [ \frac{U}{V} (\alpha - 1) (\bar{\alpha} - 1) \right ]^{\calE} G^{(s,\pm)}_{k_1 k_4 k_3 k_2} \left ( V, U, \frac{\alpha}{\alpha - 1}, \frac{\bar{\alpha}}{\bar{\alpha} - 1} \right ), & \calE = \text{min}(k_i) \\
\left [ \frac{U}{V} (\alpha - 1) (\bar{\alpha} - 1) \right ]^{\calE} G^{(s,\pm)}_{k_3 k_2 k_1 k_4} \left ( V, U, \frac{\alpha}{\alpha - 1}, \frac{\bar{\alpha}}{\bar{\alpha} - 1} \right ), & \calE \neq \text{min}(k_i)
\end{cases} \nonumber \\
G^{(u,\pm)}_{k_1 k_2 k_3 k_4}(U, V, \alpha, \bar{\alpha}) &= (U \alpha \bar{\alpha})^{\calE} G^{(s,\pm)}_{k_1 k_3 k_2 k_4} \left ( \frac{1}{U}, \frac{V}{U}, \frac{1}{\alpha}, \frac{1}{\bar{\alpha}} \right ) \label{super-crossing-g}
\end{align}
and
\begin{align}
\calM^{(t,\pm)}_{k_1 k_2 k_3 k_4}(s, t, \alpha, \bar{\alpha}) &= \begin{cases}
[(\alpha - 1)(\bar{\alpha} - 1)]^\calE \calM^{(s,\pm)}_{k_1 k_4 k_3 k_2} \left ( t, s, \frac{\alpha}{\alpha - 1}, \frac{\bar{\alpha}}{\bar{\alpha} - 1} \right ), & \calE = \text{min}(k_i) \\
[(\alpha - 1)(\bar{\alpha} - 1)]^\calE \calM^{(s,\pm)}_{k_3 k_2 k_1 k_4} \left ( t, s, \frac{\alpha}{\alpha - 1}, \frac{\bar{\alpha}}{\bar{\alpha} - 1} \right ), & \calE \neq \text{min}(k_i)
\end{cases} \nonumber \\
\calM^{(u,\pm)}_{k_1 k_2 k_3 k_4}(s, t, \alpha, \bar{\alpha}) &= (\alpha \bar{\alpha})^\calE \calM^{(s,\pm)}_{k_1 k_3 k_2 k_4} \left ( u, t, \frac{1}{\alpha}, \frac{1}{\bar{\alpha}} \right ) \label{super-crossing-m}
\end{align}
respectively once $\alpha$ and $\bar{\alpha}$ are included.\footnote{As a reminder, the $t$-channel part of $G_{k_1 k_2 k_3 k_4}(U, V, \alpha, \bar{\alpha})$ can always be obtained from the $s$-channel parts of either $G_{k_3 k_2 k_1 k_4}(U, V, \alpha, \bar{\alpha})$ or $G_{k_1 k_4 k_3 k_2}(U, V, \alpha, \bar{\alpha})$. It is simply convenient for our purposes to obtain it from a correlator whose $\alpha$ and $\bar{\alpha}$ dependence can still be a degree $\calE$ polynomial with the prefactor \eqref{prefactor}.} It is also straightforward to see that the ansatz
\es{before-super}{
\calM^{(s,+)}_{k_1 k_2 k_3 k_4}(s, t) &= \sum_k C_{k_1 k_2 k} C_{k k_3 k_4} \calM^{2d}_{\Delta_k, \ell_k}(s, t) + \calM^{(s,c,+)}_{k_1 k_2 k_3 k_4}(s, t) \\
\calM^{(s,-)}_{k_1 k_2 k_3 k_4}(s, t) &= \sum_k C_{k_1 k_2 k} C_{k k_3 k_4} \calM^{4d}_{\Delta_k + 1, \ell_k - 1}(s, t) + \calM^{(s,c,-)}_{k_1 k_2 k_3 k_4}(s, t)
}
should now be written as
\es{after-super}{
\calM^{(s,\pm)}_{k_1 k_2 k_3 k_4}(s, t, \alpha, \bar{\alpha}) &= \sum_k C_{k_1 k_2 k} C_{k k_3 k_4} \mathcal{S}^{(\pm)}_k(s, t, \alpha, \bar{\alpha}) + \calM^{(s,c,\pm)}_{k_1 k_2 k_3 k_4}(s, t, \alpha, \bar{\alpha}).
}
The even and odd super-Witten diagrams in Mellin space are found by taking symmetric and anti-symmetric combinations of the superconformal blocks \eqref{bps-blocks}. The 2d and 4d conformal blocks that appear are converted to the expressions in Appendix \ref{app:residues} while the $\alpha$ and $\bar{\alpha}$ dependence is left alone. For a half-BPS exchange with $j = h$ and $\bar{\jmath} = \bar{h}$, this produces
\begin{align}
\mathcal{S}^{(+)}_{h, \bar{h}}(s, t, \alpha, \bar{\alpha}) &= \left ( \frac{\tau}{\sigma} \right )^{\frac{k_{43}}{2}} \tau^{\calE - \frac{k_1 + k_2}{2}} \left [ \frac{1}{2} \calM^{2d}_{h + \bar{h}, |h - \bar{h}|}(s, t) \left [ g_{-h}(\alpha^{-1}) g_{-\bar{h}}(\bar{\alpha}^{-1}) + (\alpha \leftrightarrow \bar{\alpha}) \right ] \right. \nonumber \\
&+ \frac{\lambda_h}{2} \calM^{2d}_{h + \bar{h} + 1, |h - \bar{h} + 1|}(s, t) \left [ g_{1 - h}(\alpha^{-1}) g_{-\bar{h}}(\bar{\alpha}^{-1}) + (\alpha \leftrightarrow \bar{\alpha}) \right ] \nonumber \\
&+ \frac{\lambda_{\bar{h}}}{2} \calM^{2d}_{h + \bar{h} + 1, |h - \bar{h} - 1|}(s, t) \left [ g_{-h}(\alpha^{-1}) g_{1 - \bar{h}}(\bar{\alpha}^{-1}) + (\alpha \leftrightarrow \bar{\alpha}) \right ] \nonumber \\
&\left. + \frac{\lambda_h \lambda_{\bar{h}}}{2} \calM^{2d}_{h + \bar{h} + 2, |h - \bar{h}|}(s, t) \left [ g_{1 - h}(\alpha^{-1}) g_{1 - \bar{h}}(\bar{\alpha}^{-1}) + (\alpha \leftrightarrow \bar{\alpha}) \right ] \right ] \nonumber \\
\mathcal{S}^{(-)}_{h, \bar{h}}(s, t, \alpha, \bar{\alpha}) &= \left ( \frac{\tau}{\sigma} \right )^{\frac{k_{43}}{2}} \tau^{\calE - \frac{k_1 + k_2}{2}} \left [ \frac{1}{2} \calM^{4d}_{h + \bar{h} + 1, |h - \bar{h}| - 1}(s, t) \left [ g_{-h}(\alpha^{-1}) g_{-\bar{h}}(\bar{\alpha}^{-1}) - (\alpha \leftrightarrow \bar{\alpha}) \right ] \right. \nonumber \\
&+ \frac{\lambda_h}{2} \calM^{4d}_{h + \bar{h} + 2, |h - \bar{h} + 1| - 1}(s, t) \left [ g_{1 - h}(\alpha^{-1}) g_{-\bar{h}}(\bar{\alpha}^{-1}) - (\alpha \leftrightarrow \bar{\alpha}) \right ] \nonumber \\
&+ \frac{\lambda_{\bar{h}}}{2} \calM^{4d}_{h + \bar{h} + 2, |h - \bar{h} - 1| - 1}(s, t) \left [ g_{-h}(\alpha^{-1}) g_{1 - \bar{h}}(\bar{\alpha}^{-1}) - (\alpha \leftrightarrow \bar{\alpha}) \right ] \nonumber \\
&\left. + \frac{\lambda_h \lambda_{\bar{h}}}{2} \calM^{4d}_{h + \bar{h} + 3, |h - \bar{h}| - 1}(s, t) \left [ g_{1 - h}(\alpha^{-1}) g_{1 - \bar{h}}(\bar{\alpha}^{-1}) - (\alpha \leftrightarrow \bar{\alpha}) \right ] \right ]. \label{bps-witten}
\end{align}
The superscripts on $\lambda_h$, $\lambda_{\bar{h}}$ and the $SU(2)$ blocks have been suppressed but they are all $(-\frac{k_{12}}{2}, -\frac{k_{34}}{2})$.\footnote{To compute analogues of $\lambda_h$ and $\lambda_{\bar{h}}$ in higher dimensions, \cite{az20,abfz21} analyzed the so called maximal R-symmetry violating limit which is a method well adapted to Mellin space. This is not needed when the superconformal blocks have been found already using the Ward identity \cite{bfl19} as is the case here.} Along with the contact term, \eqref{bps-witten} is evidently a degree $\calE$ polynomial in both $\alpha$ and $\bar{\alpha}$. The even and odd functions which incorporate all three channels can therefore be written as
\es{full-monomials}{
G^{(\pm)}_{k_1 k_2 k_3 k_4}(U, V, \alpha, \bar{\alpha}) &= \sum_{i, j = 0}^{\calE} G^{(\pm)}_{i, j}(U, V) \alpha^i \bar{\alpha}^j \\
}
with $G^{(\pm)}_{j, i}(U, V) = \pm G^{(\pm)}_{i, j}(U, V)$.
It is these sums which must be fed into the Ward identity \eqref{scwi}.

As shown in \cite{z17}, formulating superconformal Ward identities in Mellin space is again a matter of symmetrizing and anti-symmetrizing in the cross-ratios. To see this, one should first act on $G_{k_1 k_2 k_3 k_4}(z, \bar{z}, \alpha, \bar{\alpha})$ with the two equations in \eqref{scwi} to produce
\es{ward-actions}{
\sum_{i, j = 0}^\calE \frac{\bar{\alpha}^j}{z^i} \left [ (z\partial_z - i) G^{(+)}_{i, j}(U, V) + \frac{z - \bar{z}}{U} (z\partial_z - i) G^{(-)}_{i, j}(U, V) + z^{-1} G^{(-)}_{i, j}(U, V) \right ] &= 0 \\
\sum_{i, j = 0}^\calE \frac{\alpha^i}{\bar{z}^j} \left [ (\bar{z}\partial_{\bar{z}} - j) G^{(+)}_{i, j}(U, V) + \frac{z - \bar{z}}{U} (\bar{z}\partial_{\bar{z}} - j) G^{(-)}_{i, j}(U, V) - \bar{z}^{-1} G^{(-)}_{i, j}(U, V) \right ] &= 0
}
respectively. Each power of the R-symmetry cross ratio that remains needs to have a vanishing coefficient so it is useful to write the $\bar{\alpha}^j$ term in the first line and the $\alpha^j$ term in the second line as sums over $i$. Using the symmetry and anti-symmetry of $G^{(\pm)}_{i, j}(U, V)$ leads to
\es{ward-monomials}{
\sum_{i = 0}^\calE z^{-i} \left [ (z\partial_z - i) G^{(+)}_{i, j}(U, V) + \frac{z - \bar{z}}{U} (z\partial_z - i) G^{(-)}_{i, j}(U, V) + z^{-1} G^{(-)}_{i, j}(U, V) \right ] &= 0 \\
\sum_{i = 0}^\calE \bar{z}^{-i} \left [ (\bar{z}\partial_{\bar{z}} - i) G^{(+)}_{i, j}(U, V) - \frac{z - \bar{z}}{U} (\bar{z}\partial_{\bar{z}} - i) G^{(-)}_{i, j}(U, V) + \bar{z}^{-1} G^{(-)}_{i, j}(U, V) \right ] &= 0.
}
We will now multiply the first equation by $U z^\calE (1 - z)$ and the second equation by $U \bar{z}^\calE (1 - \bar{z})$. This is not the minimal rescaling needed to eliminate inverse powers but the extra factors of $1 - z$ and $1 - \bar{z}$ are useful because of the identity
\es{multiplied-ward}{
z(1 - z)\partial_z = (1 - z) U \partial_U - z V \partial_V, \quad \bar{z}(1 - \bar{z})\partial_{\bar{z}} = (1 - \bar{z}) U \partial_U - \bar{z} V \partial_V.
}
All in all,
\es{ward-final}{
& \sum_{i = 0}^\calE \bar{z} z^{\calE + 1 - i} [(1 - z)U\partial_U - zV\partial_V - i(1 - z)] G^{(+)}_{i, j}(U, V) + \bar{z} z^{\calE - i}(1 - z)G^{(-)}_{i, j}(U, V) \\
&+ z^{\calE - i} (z - \bar{z}) [(1 - z)U\partial_U - zV\partial_V - i(1 - z)] G^{(-)}_{i, j}(U, V) \pm (z \leftrightarrow \bar{z}) = 0.
}
The translation to Mellin space is now possible since we have taken the sum and the difference of the two Ward identities. This allows the cross-ratio dependence in the sum to be expressed in terms of $U$ and $V$. The same works for the difference after factoring out $z - \bar{z}$. The resulting $U^m V^n$ monomials become the difference operators \eqref{diff-op} while the derivatives act multiplicatively according to
\es{deriv-op}{
U\partial_U \mapsto \left ( \frac{s - k_1 - k_2}{2} + \calE \right ) \times, \quad V\partial_V \mapsto \left ( \frac{t + k_{14}}{2} - \calE \right ) \times.
}
These equations labelled by $j$, which couple even and odd Mellin amplitudes, will be crucial for bootstrapping families of tree-level four-point functions.

\subsection{The auxiliary amplitude}\label{auxAmp}

After solving for the unknown parts of $\calM^{(\pm)}_{k_1 k_2 k_3 k_4}(s, t, \alpha, \bar{\alpha})$, it will sometimes be useful to recast them into even and odd amplitudes for the function $H_{k_1 k_2 k_3 k_4}(z, \bar{z}, \alpha, \bar{\alpha})$ appearing in \eqref{formal-sol}. This auxiliary correlator mixes information about the three channels in a non-trivial way so we will simply define
\es{h-even-odd}{
H(z, \bar{z}, \alpha, \bar{\alpha}) = H^{(+)}(U, V, \alpha, \bar{\alpha}) + (z - \bar{z}) H^{(-)}(U, V, \alpha, \bar{\alpha}).
}
Note that any one of the functions $G^{(\pm)}(U, V, \alpha, \bar{\alpha})$ will receive contributions from both $H^{(\pm)}(U, V, \alpha, \bar{\alpha})$. Using the fact that the crossing transformations of the full correlator are \eqref{super-crossing-g}, those of the auxiliary correlator are
\begin{align}
H^{(\pm)}_{k_1 k_2 k_3 k_4}(U, V, \alpha, \bar{\alpha}) &= \begin{cases}
\left ( \frac{U}{V} \right )^\calE [(\alpha - 1) (\bar{\alpha} - 1)]^{\calE - 1} H^{(\pm)}_{k_1 k_4 k_3 k_2} \left ( V, U, \frac{\alpha}{\alpha - 1}, \frac{\bar{\alpha}}{\bar{\alpha} - 1} \right ), & \calE = \text{min}(k_i) \\
\left ( \frac{U}{V} \right )^\calE [(\alpha - 1) (\bar{\alpha} - 1)]^{\calE - 1} H^{(\pm)}_{k_3 k_2 k_1 k_4} \left ( V, U, \frac{\alpha}{\alpha - 1}, \frac{\bar{\alpha}}{\bar{\alpha} - 1} \right ), & \calE \neq \text{min}(k_i)
\end{cases} \nonumber \\
H^{(+)}_{k_1 k_2 k_3 k_4}(U, V, \alpha, \bar{\alpha}) &= (U \alpha \bar{\alpha})^{\calE - 1} H^{(+)}_{k_1 k_3 k_2 k_4} \left ( \frac{1}{U}, \frac{V}{U}, \frac{1}{\alpha}, \frac{1}{\bar{\alpha}} \right ) \nonumber \\
H^{(-)}_{k_1 k_2 k_3 k_4}(U, V, \alpha, \bar{\alpha}) &= U^{\calE - 2} (\alpha \bar{\alpha})^{\calE - 1} H^{(-)}_{k_1 k_3 k_2 k_4} \left ( \frac{1}{U}, \frac{V}{U}, \frac{1}{\alpha}, \frac{1}{\bar{\alpha}} \right ). \label{crossing-h}
\end{align}
The R-symmetry cross-ratios enter in the expected way for a polynomial of degree $\calE - 1$. As for the conformal cross-ratios, only the first lines of \eqref{super-crossing-g} and \eqref{crossing-h} have the same form. The remaining transformation has an exponent on $U$ which is $\calE$ for $G^{(\pm)}(U, V, \alpha, \bar{\alpha})$, $\calE - 1$ for $H^{(+)}(U, V, \alpha, \bar{\alpha})$ and $\calE - 2$ for $H^{(-)}(U, V, \alpha, \bar{\alpha})$. The discussion in \cite{rz17} then implies that the Mellin amplitudes we associate to the latter two functions will only respect the usual permutation of Mandelstam variables if $u$ is suitably shifted. We will therefore define the Mellin representation of the auxiliary four-point functions by
\begin{align}
H^{(\pm)}(U, V, \alpha, \bar{\alpha}) &= \int_{-i\infty}^{i\infty} \frac{ds dt}{(4\pi i)^2} U^{\frac{s + 2\calE - k_1 - k_2}{2}} V^{\frac{t - 2\calE + k_{14}}{2}} \calMt^{(\pm)}(s, t, \alpha, \bar{\alpha}) \widetilde{\Gamma}_{\{ ki \}}(s, t) \label{h-mellin} \\
\widetilde{\Gamma}_{\{ k_i \}}(s, t) &\equiv \Gamma[\tfrac{k_1 + k_2 - s}{2}] \Gamma[\tfrac{k_3 + k_4 - s}{2}] \Gamma[\tfrac{k_1 + k_4 - t}{2}] \Gamma[\tfrac{k_2 + k_3 - t}{2}] \Gamma[\tfrac{k_1 + k_3 - \tilde{u}}{2}] \Gamma[\tfrac{k_2 + k_4 - \tilde{u}}{2}] \nonumber
\end{align}
with the understanding that $\tilde{u} = u - 2$ for the even amplitude and $\tilde{u} = u - 4$ for the odd amplitude.\footnote{When quoting results in the literature earlier, the equation \eqref{mellin-tensor} was defined with $\tilde{u} = u - 2$ in both cases because this was the convention used by \cite{wz21}. However, we can already see in \eqref{known-mixed-odd} that $\tilde{u} = u - 4$ for the odd amplitude would have been more natural there.}

Even though $\calMt^{(\pm)}(s, t, \alpha, \bar{\alpha})$ are nominally Mellin amplitudes for only the second term in \eqref{formal-sol}, they are actually enough to recover $\calM^{(\pm)}(s, t, \alpha, \bar{\alpha})$ which know about the full correlator.\footnote{How auxiliary Mellin amplitudes encode information about the rational function in \eqref{formal-sol} was a mystery until \cite{rz17} showed that such terms can arise when the Mellin-Barnes contour is pushed towards the edge of the critical strip where all gamma function arguments have positive real part.} Specifically, we can convert $(1 - z\alpha)(1 - \bar{z}\bar{\alpha})$ to a difference operator following \eqref{diff-op} to find
\es{aux-to-full}{
\calM^{(+)}(s, t, \alpha, \bar{\alpha}) &= \left [ 1 - \frac{\alpha + \bar{\alpha}}{2}(1 + \widehat{U} - \widehat{V}) + \alpha \bar{\alpha} \widehat{U} \right ] \circ \calMh^{(+)}(s, t, \alpha, \bar{\alpha}) \\
&- \frac{\alpha - \bar{\alpha}}{2} \left [ (\widehat{U} - \widehat{V})^2 - 2(\widehat{U} + \widehat{V}) + 1 \right ] \circ \calMh^{(-)}(s, t, \alpha, \bar{\alpha}) \\
\calM^{(-)}(s, t, \alpha, \bar{\alpha}) &= \widehat{U} \left [ 1 - \frac{\alpha + \bar{\alpha}}{2}(1 + \widehat{U} - \widehat{V}) + \alpha \bar{\alpha} \widehat{U} \right ] \circ \calMh^{(-)}(s, t, \alpha, \bar{\alpha}) \\
&- \frac{\alpha - \bar{\alpha}}{2} \widehat{U} \circ \calMh^{(+)}(s, t, \alpha, \bar{\alpha}).
}
The new amplitudes on the right hand side, defined by
\es{hack-amplitudes}{
\calMh^{(+)}_{k_1 k_2 k_3 k_4}(s, t, \alpha, \bar{\alpha}) &= \left ( \frac{k_1 + k_3 - u}{2} \right ) \left ( \frac{k_2 + k_4 - u}{2} \right ) \calMt^{(+)}_{k_1 k_2 k_3 k_4}(s, t, \alpha, \bar{\alpha}) \\
\calMh^{(-)}_{k_1 k_2 k_3 k_4}(s, t, \alpha, \bar{\alpha}) &= \left ( \frac{k_1 + k_3 - u}{2} \right )_2 \left ( \frac{k_2 + k_4 - u}{2} \right )_2 \calMt^{(-)}_{k_1 k_2 k_3 k_4}(s, t, \alpha, \bar{\alpha}),
}
are simply the ones that make $\calMt^{(\pm)}(s, t, \alpha, \bar{\alpha})$ appear with the original measure $\Gamma_{\{ k_i \}}(s, t)$. One could of course avoid this detour by instead defining difference operators that are analogous to \eqref{diff-op} but based on the measure $\widetilde{\Gamma}_{\{ k_i \}}(s, t)$. Since we will start by computing the full Mellin amplitude, it is more important to be able to go in the other direction. Once it is known that $\calM^{(\pm)}(s, t, \alpha, \bar{\alpha})$ satisfy the superconformal Ward identity, it is possible to invert the difference operators in \eqref{aux-to-full} and solve for $\calMt^{(\pm)}(s, t, \alpha, \bar{\alpha})$. One simply needs to follow the algorithm that \cite{rz17} used to show that auxiliary amplitudes in the $\mathcal{N} = 4$ SYM case were rational functions of $s$, $t$ and $\tilde{u}$. Changing notation slightly from \eqref{full-monomials}, we will expand amplitudes as
\es{hack-monomials}{
\calM^{(+)}_{k_1 k_2 k_3 k_4}(s, t, \alpha, \bar{\alpha}) &= \sum_{i, j = 0}^{\calE} \frac{1}{1 + \delta_{i,j}} \left ( \alpha^i \bar{\alpha}^j + \alpha^j \bar{\alpha}^i \right ) \calM^{(+)}_{i, j}(s, t) \\
\calM^{(-)}_{k_1 k_2 k_3 k_4}(s, t, \alpha, \bar{\alpha}) &= \sum_{i, j = 0}^{\calE} \left ( \alpha^i \bar{\alpha}^j - \alpha^j \bar{\alpha}^i \right ) \calM^{(-)}_{i, j}(s, t) \\
\calMh^{(+)}_{k_1 k_2 k_3 k_4}(s, t, \alpha, \bar{\alpha}) &= \sum_{i, j = 0}^{\calE - 1} \frac{1}{1 + \delta_{i,j}} \left ( \alpha^i \bar{\alpha}^j + \alpha^j \bar{\alpha}^i \right ) \calMh^{(+)}_{i, j}(s, t) \\
\calMh^{(-)}_{k_1 k_2 k_3 k_4}(s, t, \alpha, \bar{\alpha}) &= \sum_{i, j = 0}^{\calE - 1} \left ( \alpha^i \bar{\alpha}^j - \alpha^j \bar{\alpha}^i \right ) \calMh^{(-)}_{i, j}(s, t).
}
The procedure is then to recursively compute the left hand sides of
\begin{align}
\widehat{U} \circ \calMh^{(-)}_{i, j}(s, t) &= \calM^{(-)}_{i, j}(s, t) + \frac{1}{2} \widehat{U} \circ \left ( \calMh^{(+)}_{i - 1, j}(s, t) - \calMh^{(+)}_{i, j - 1}(s, t) \right ) \label{full-to-aux-minus} \\
&- \widehat{U^2} \circ \calMh^{(-)}_{i - 1, j - 1}(s, t) + \frac{1}{2} \widehat{U}(1 + \widehat{U} - \widehat{V}) \circ \left ( \calMh^{(-)}_{i - 1, j}(s, t) + \calMh^{(-)}_{i, j - 1}(s, t) \right ) \nonumber
\end{align}
and
\begin{align}
\calMh^{(+)}_{i,j}(s, t) &= \calM^{(+)}_{i, j}(s, t) + \frac{1}{2} (1 + \widehat{U} - \widehat{V}) \circ \left ( \calMh^{(+)}_{i - 1, j}(s, t) + \calMh^{(+)}_{i, j - 1}(s, t) \right ) - \widehat{U} \circ \calMh^{(+)}_{i - 1, j - 1}(s, t) \nonumber \\
&+ \frac{1}{2} \left [ (\widehat{U} - \widehat{V})^2 - 2(\widehat{U} + \widehat{V}) + 1 \right ] \circ \left ( \calMh^{(-)}_{i - 1, j}(s, t) - \calMh^{(-)}_{i, j - 1}(s, t) \right ) \label{full-to-aux-plus}
\end{align}
using lower order data on the right hand side together with the boundary condition that $\calMh^{(\pm)}_{i, j}(s, t) = 0$ when at least one of $i$ and $j$ is negative. After computing $\widehat{U} \circ \calMh^{(-)}_{i, j}(s, t)$ with \eqref{full-to-aux-minus}, one needs to invert the difference operator $\widehat{U}$. This is easy to do because it is a pure monomial.

\subsection{Input for the D1-D5 CFT}

At this point, all of the ingredients are in place for writing down super-Witten diagrams corresponding to half-BPS multiplets, determining the contact terms that must accompany them and analyzing the resulting correlators in two different forms. Let us therefore specialize these expressions to the operator content of Figures \ref{fig:superMultiplets1} and \ref{fig:superMultiplets2} in the $AdS_3 \times S^3$ description of the D1-D5 CFT. Analogously to \eqref{position-ansatz}, we will refer to the super-Witten diagrams $\mathcal{S}^{(\pm)}_{V, k}(s, t, \alpha, \bar{\alpha})$ and $\mathcal{S}^{(\pm)}_{\sigma, k}(s, t, \alpha, \bar{\alpha})$ which are needed to describe $\left < s^{I_1} s^{I_2} s^{I_3} s^{I_4} \right >$ and $\left < \sigma \sigma \sigma \sigma \right >$. In addition to these, a third pair of diagrams $\mathcal{S}^{(\pm)}_{s, k}(s, t, \alpha, \bar{\alpha})$ is needed for $\left < s^{I_1} s^{I_2} \sigma \sigma \right >$. For all six functions, it would be perfectly valid to define them by plugging the right quantum numbers into \eqref{bps-witten} but we will take this opportunity ot perform a slight modification.

We are interested in shifting the super-Witten diagrams by certain regular terms in hopes that this will cause the output of the superconformal Ward identity to take a simpler form which is more amenable to pattern recognition. A particularly successful prescription for doing this was found in \cite{az20}. Their idea starts with taking an $s$-channel super-Witten diagram for the quantum numbers $(\Delta, \ell)$ and manipulating its formal expression so that all poles manifestly appear at $s = \Delta - \ell + 2m$ for some integer $m \geq 0$. Although the multiplet contains a Witten diagram for the super-primary $\calM_{\Delta, \ell}(s, t)$ which has this property, it also contains at least one Witten diagrams for a super-descendant which has poles at $s = \Delta - \ell + 2 + 2m$. Accounting for this shift is a simple matter of re-indexing the sum over $m$. If we now look at the residues in Appendix \ref{app:residues}, this introduces a qualitative change. Before re-indexing, $m$ only appeared in the gamma functions, but afterwards the residues have a polynomial dependence on $m$. The prescription of \cite{az20} is to make the replacement
\es{m-replacement}{
m \mapsto \frac{1}{2} \left ( \sum_{i = 1}^4 k_i - t - u - \Delta + \ell \right )
}
in this polynomial. Since \eqref{m-replacement} is not an equality, this procedure does not amount to a simple rewriting of super-Witten diagrams. It changes them in such a way that they are able to absorb non-trivial contact terms. For the $d \geq 3$ holographic CFTs studied in \cite{az20,abfz21}, \eqref{m-replacement} was powerful enough that all contact terms were absorbed --- \textit{i.e.} the superconformal Ward identity was satisfied for the polar part of the ansatz alone. The D1-D5 results in the next section will show that contact terms still exist after using this prescription but are simpler than they would have been otherwise.

When applying this logic to the $s^I$, $\sigma$ and $V$ multiplets, there is a trick we can use in order to handle the $\calE = \text{min}(k_i)$ and $\calE \neq \text{min}(k_i)$ cases in a uniform way. This will lead to the appearsnce of the quantities
\es{weight-combos}{
\kappa_t \equiv |k_1 + k_4 - k_2 - k_3|, \quad \Sigma \equiv k_1 + k_2 + k_3 + k_4, \quad \kappa_u \equiv |k_1 + k_3 - k_2 - k_4|.
}
Specifically, we can take
\es{r-sym-block1}{
\alpha^{\frac{k_{34}}{2}} (1 - \alpha)^{\calE + k_4 - \frac{\Sigma}{2}} {}_2F_1 \left ( -\frac{k - k_{23}}{2}, -\frac{k + k_{34}}{2}; -k; \frac{1}{\alpha} \right ),
}
which is the $SU(2)$ block from \eqref{sl2-block} with the prefactor that follows from \eqref{prefactor}, and write it as
\es{r-sym-block2}{
\alpha^{-\frac{\kappa_t + \kappa_u}{4}} {}_2F_1 \left ( -\frac{2k - \kappa_t + \kappa_u}{4}, -\frac{2k - \kappa_t - \kappa_u}{4}; -k; \frac{1}{\alpha} \right ).
}
When $\calE \neq \text{min}(k_i)$, all three orderings in \eqref{3-orderings} satisfy $k_1 + k_4 \geq k_2 + k_3$ and $\calE = \frac{\Sigma}{2} - k_4$. This makes \eqref{r-sym-block2} manifestly identical to \eqref{r-sym-block1}. When $\calE = \text{min}(k_i)$, this equality is no longer manifest but it can be shown using an Euler transformation. Letting $\phi \in \{ s^I, \sigma \}$ (which have the same super-Witten diagrams), the residues of interest will now be written as
\es{bps-witten-residues}{
\underset{s = k + 1 \mp 1 + 2m}{\text{Res}} \mathcal{S}^{(\pm)}_{\phi, k}(s, t, \alpha, \bar{\alpha}) &= \sum_{i, j} K^{(\pm)}_{\phi, k, i, j}(t, u) H^{(\pm)}_{\phi, k, i, j}(m) \alpha^{\frac{2k - \kappa_t - \kappa_u}{4} - i} \bar{\alpha}^{\frac{2k - \kappa_t - \kappa_u}{4} - j} \\
\underset{s = k \mp 1 + 2m}{\text{Res}} \mathcal{S}^{(\pm)}_{V, k}(s, t, \alpha, \bar{\alpha}) &= \sum_{i, j} K^{(\pm)}_{V, k, i, j}(t, u) H^{(\pm)}_{V, k, i, j}(m) \alpha^{\frac{2k - \kappa_t - \kappa_u}{4} - i} \bar{\alpha}^{\frac{2k - \kappa_t - \kappa_u}{4} - j}
}
in terms of $(t_*, u_*) \equiv (t - \tfrac{\Sigma}{2}, u - \tfrac{\Sigma}{2})$.
The parts depending on $m$ can be cleanly stated as
\begin{align}
H^{(\pm)}_{\phi,k,i,j}(m) &= \frac{\zeta^{\pm}_{i,j} \Gamma[k]^2 (2k - \kappa_t - \kappa_u)^{-1} (2k - \kappa_t + \kappa_u)^{-1}}{m! i! j! \Gamma[k + 1 + m] (1 - k)_i (1 - k)_j \Gamma \left [ \frac{k_1 + k_2 - k - 1 \pm 1}{2} - m \right ] \Gamma \left [ \frac{k_3 + k_4 - k - 1 \pm 1}{2} - m \right ]} \nonumber \\
&\times \frac{\left ( \frac{\kappa_t + \kappa_u - 2k}{4} \right )_i \left ( \frac{\kappa_t - \kappa_u - 2k}{4} \right )_i \left ( \frac{\kappa_t + \kappa_u - 2k}{4} \right )_j \left ( \frac{\kappa_t - \kappa_u - 2k}{4} \right )_j}{\Gamma \left [ \frac{k + k_{12}}{2} \right ] \Gamma \left [ \frac{k + k_{34}}{2} \right ] \Gamma \left [ \frac{k - k_{12}}{2} \right ] \Gamma \left [ \frac{k - k_{34}}{2} \right ]} \nonumber \\
H^{(\pm)}_{V,k,i,j}(m) &= \frac{\zeta^{\pm}_{i,j} \Gamma[k]^2 (2k - \kappa_t - \kappa_u + 2)^{-2} (2k - \kappa_t + \kappa_u + 2)^{-2}}{2 i! j! m! \Gamma[k + 1 + m] (1 - k)_i (1 - k)_j \Gamma \left [ \frac{k_1 + k_2 - k \pm 1}{2} - m \right ] \Gamma \left [ \frac{k_3 + k_4 - k \pm 1}{2} - m \right ]} \nonumber \\
&\times \frac{\left ( \frac{\kappa_t + \kappa_u - 2k - 2}{4} \right )_i \left ( \frac{\kappa_t - \kappa_u - 2k - 2}{4} \right )_i \left ( \frac{\kappa_t + \kappa_u - 2k - 2}{4} \right )_j \left ( \frac{\kappa_t - \kappa_u - 2k - 2}{4} \right )_j}{\Gamma \left [ \frac{k + 1 + k_{12}}{2} \right ] \Gamma \left [ \frac{k + 1 + k_{34}}{2} \right ] \Gamma \left [ \frac{k + 1 - k_{12}}{2} \right ] \Gamma \left [ \frac{k + 1 - k_{34}}{2} \right ]} \nonumber \\
\zeta^-_{i,j} &= 16(i - j), \quad \zeta^+_{i,j} = 1. \label{residues-m}
\end{align}
The corresponding expressions with $t$ and $u$ are
\es{residues-tu}{
K^{(-)}_{\phi,k,i,j}(t, u) &= 1 \\
K^{(+)}_{\phi,k,i,j}(t, u) &= [\kappa_u^2 - \kappa_t^2 - 4k^2 - 4\kappa_t(i + j - k)](t_* + u_* + i + j - k) \\
&- 4(2ij - ik - jk)(2t_* + \kappa_t) + 4\kappa_t(i + j)(i + j - 2k) \\
K^{(-)}_{V,k,i,j}(t, u) &= 2ij[4(k^2 - \kappa_t + 1) + \kappa_t^2 - \kappa_u^2](t_* + u_* + k + 1) \\
&-k(i + j - k)[4(k + 1)(k + 1 - \kappa_t) + \kappa_t^2 - \kappa_u^2](t_* + u_* + k + 1) \\
&-2k(i - k)(j - k)(2k - \kappa_t - \kappa_u + 2)(2k - \kappa_t + \kappa_u + 2) \\
K^{(+)}_{V,k,i,j}(t, u) &= r_1(t_* + u_*)^2 + r_2(t_*^2 - u_*^2) + r_3t_*u_* + r_4(t_* + u_*) + r_5 t_* + r_6.
}
For the even part of the $V$ multiplet, which is the unique source of spin-2 Witten diagrams, the polynomial in \eqref{residues-tu} takes a somewhat longer form with the coefficients given by
\begin{align}
r_1 &= k(2k - \kappa_t - \kappa_u + 2)(2k - \kappa_t + \kappa_u + 2)(i + j)[\kappa_u^2 - \kappa_t^2 - 4(i + j - 1)(\kappa_t + k - 1)] \nonumber \\
&+ 2ij[(\kappa_t^2 - \kappa_u^2)^2 - 16(2k^2 + 1)\kappa_t^2 + 4(i + j - k - 2)\kappa_t(\kappa_t^2 - \kappa_u^2 - 4\kappa_t + 8k + 8)] \nonumber \\
&+ 8ij(5k^2 - 2ik - 2jk + 2ij + k - i - j + 2)(\kappa_t^2 - \kappa_u^2) + 64ij(2k^2 + 2k - ij + 1)\kappa_t \nonumber \\
&+ 32ij(k^2 - 1)(2ik + 2jk - 2ij + i + j - k - 1) \nonumber \\
r_2 &= 4k(k - 1)(2k - \kappa_t - \kappa_u + 2)(2k - \kappa_t + \kappa_u + 2)[i(k - i) + j(k - j)] \nonumber \\
&+ 8ij(2k^2 - 2ik - 2jk + 2ij + 2k - i - j)(\kappa_t^2 - \kappa_u^2 - 4k^2 + 4) \nonumber \\
&+ 32ij[2k(k^2 - k - 1) - (i + j)(k^2 - 2k - 1) - 2ij]\kappa_t \nonumber \\
r_3 &= 256ijk^2(k - i)(k - j) \nonumber \\
r_4 &= (2ij - ik - jk)(\kappa_t^2 - \kappa_u^2) \left [ (i + j - k - 2)(\kappa_t^2 - \kappa_u^2 - 4k\kappa_t + 8k^2 + 8) + 4(2ij - i - j + 1)\kappa_t \right ] \nonumber \\
& - 16[k(k + 1)(i^2 - i + j^2 - j) - 2ij(ik + jk - ij + i + j - 2k - 1)](2\kappa_t^2 - k^2\kappa_t) \nonumber \\
&+ 16(k - 1) \left [ k(k + 1)^2(k - 1)(i + j) + 2k(k + 1)(i^2 - i + j^2 - j) - 2ij(k + 1)^2(2k - i - j - 2) \right. \nonumber \\
&\left. - 2ij(ijk - ij + 3ik + 3jk + 3i + 3j - 3k + 2) \right ]\kappa_t - 16(k^2 - 1)^2(i + j - k - 2)(ik + jk - 2ij) \nonumber \\
r_5 &= 8 \left [ k(k^2 - 1)(i^2 + j^2 - ik - jk) + 4ij(k^2 - ij) - 2ij(i + j - 2k)(k^2 - 2k - 1) \right ] (\kappa_t^2 - \kappa_u^2) \nonumber \\
&- 32(k^2 - 1)(2ij - ik - jk)(2ij - ik - jk - i - j + k^2 + k)(\kappa_t - 1) \nonumber \\
&- 32k^2(k^2 - 1)[2ij(i + j + 2) - (k + 1)(i^2 - ik + j^2 - jk + 4ij)] \nonumber \\
r_6 &= [k(k + 1)(i^2 - i + j^2 - j - 2ij) - 2ij(i + j - 3k - 1)](\kappa_t^2 - \kappa_u^2)(\kappa_t^2 - \kappa_u^2 - 4\kappa_t + 8k^2 + 8) \nonumber \\
&- 4(2ij - ik - jk - i - j + k)(2ij - ik - jk)\kappa_t(\kappa_t^2 - \kappa_u^2 - 4\kappa_t - 8k^2 + 8) \nonumber \\
&+ 16[2ij(k^2 - 2k + i + j - 1) - k(k + 1)(i^2 - i + j^2 - j)][k^2\kappa_t^2 - (k^2 - 1)^2] \nonumber \\
&- 64ij(k^2 - 1)(k - i)(k - j)\kappa_t. \label{r-coeffs}
\end{align}

With these expressions for the residues in hand, the ansatz \eqref{after-super} becomes fully usable once single-trace OPE coefficients for the theory of interest are specified. These are all invariant under permutations of the three operators. As shown in Appendix \ref{app:3pt}, the non-zero OPE coefficients are given by
\es{ope-coeffs}{
C^{ss \sigma}_{k_1 k_2 k_3} &= \frac{1}{\sqrt{N}} \sqrt{\frac{2 k_1 k_2 k_3}{k_3^2 - 1}}, \quad C^{ss V}_{k_1 k_2 k_3} = i \sqrt{\frac{k_1 k_2}{N k_3}} \\
C^{\sigma \sigma \sigma}_{k_1 k_2 k_3} &= \frac{k_1^2 + k_2^2 + k_3^2 - 2}{\sqrt{N}} \sqrt{\frac{k_1 k_2 k_3}{2 (k_1^2 - 1) (k_2^2 - 1) (k_3^2 - 1)}} \\
C^{\sigma \sigma V}_{k_1 k_2 k_3} &= i \frac{k_1^2 + k_2^2 - k_3^2 - 1}{2\sqrt{N}} \sqrt{\frac{k_1 k_2}{(k_1^2 - 1)(k_2^2 - 1)k_3}}.
}
To identify the ones that vanish, we can use $SU(2)$ selection rules to see that $V^\pm_k$ is only exchanged when $k \in \{ |k_{12}| + 1, \dots, k_1 + k_2 - 1\}$. For the other three-point functions, the admissible range is $k \in \{ |k_{12}| + 2, \dots, k_1 + k_2 - 2\}$ because of the additional constraint that extremal couplings should vanish. The factors of $i$ in \eqref{ope-coeffs} are necessary because the standard conformal blocks for exchanged vectors assume that they are normalized as
\es{vector-norm}{
\left < V^\mu(x) V^\nu(0) \right > = \frac{x^2 \delta^{\mu\nu} - 2 x^\mu x^\nu}{|x|^{2\Delta + 2}}.
}
Defining $V^\pm_k = V^\parallel_k \pm i V^\perp_k$ in analogy with $z, \bar{z} = x^\parallel \pm i x^\perp$ gives the self-dual and anti-self-dual vectors a negative two-point function. There is a simple check based on the fact that the holomorphic stress tensor is a descendant of $V^+_1$ at level one. Using the relative coefficient in \eqref{bps-blocks}, which is negative, we see that \eqref{ope-coeffs} precisely matches the prediction of the Virasoro Ward identity for $c = 6N$.

\section{Results}
\label{sec:results}
The bootstrap procedure just described has allowed us to find formulae for tensor, graviton and mixed tensor-graviton correlators. The results we present below will apply to correlation functions with pairwise identical weights $(k_1, k_2, k_3, k_4) = (p,p,q,q)$ but it is clear that they come from methods which are completely general. Since the super-Witten diagram and OPE coefficient parts of the correlators have been completely specified by \eqref{bps-witten-residues} and \eqref{ope-coeffs} respectively, a solution boils down to an expression for the contact term which was denoted by $\calM^{(*, c, \pm)}(s, t, \alpha, \bar{\alpha})$ in \eqref{after-super}. One needs to choose a basis for expressing this degree $\calE$ polynomial in $\alpha$ and $\bar{\alpha}$. Our choices will be based on the R-symmetry polynomials that appear in the $PSU(1,1|2)_L\times PSU(1,1|2)_R$ superblocks, specifically the combinations
\es{contact-building-blocks}{
g^{k_{21}, k_{43}}_{[i, j]}(\alpha, \bar{\alpha}) &\equiv g^{k_{21}, k_{43}}_{-i}(\alpha^{-1}) g^{k_{21}, k_{43}}_{-j}(\bar{\alpha}^{-1}) - g^{k_{21}, k_{43}}_{-j}(\alpha^{-1}) g^{k_{21}, k_{43}}_{-i}(\bar{\alpha}^{-1}) \\
g^{k_{21}, k_{43}}_{\{ i, j \}}(\alpha, \bar{\alpha}) &\equiv g^{k_{21}, k_{43}}_{-i}(\alpha^{-1}) g^{k_{21}, k_{43}}_{-j}(\bar{\alpha}^{-1}) + g^{k_{21}, k_{43}}_{-j}(\alpha^{-1}) g^{k_{21}, k_{43}}_{-i}(\bar{\alpha}^{-1})
}
as this will ensure that many of the resulting coefficients are zero.

\subsection{Tensor correlators}
\label{4tensor}

Let us start with the tensor correlators $\left < s_p^{I_1} s_p^{I_2} s_q^{I_3} s_q^{I_4} \right >$. In this case, there are three exchanged $SO(h_{1, 1} + 1)$ structures: $\delta^{I_1 I_2} \delta^{I_3 I_4}$, $\delta^{I_1 I_4} \delta^{I_2 I_3}$ and $\delta^{I_1 I_3} \delta^{I_2 I_4}$. As long as the $s$-channel amplitude proportional to $\delta^{I_1 I_2} \delta^{I_3 I_4}$ is known for all weights obeying \eqref{3-orderings}, the other two channels follow from \eqref{super-crossing-m}. To get $\calM_{ppqq}^{(u, \pm)}(s, t, \alpha, \bar{\alpha})$, we can use $\calM_{pqpq}^{(u, \pm)}(s, t, \alpha, \bar{\alpha})$ and to get $\calM_{ppqq}^{(t, \pm)}(s, t, \alpha, \bar{\alpha})$, we can use either $\calM_{qppq}^{(u, \pm)}(s, t, \alpha, \bar{\alpha})$ or $\calM_{pqqp}^{(u, \pm)}(s, t, \alpha, \bar{\alpha})$. Which one is more convenient, given the prefactor \eqref{prefactor}, depends on which of $p$ and $q$ is larger. Below we will concentrate simply on $\calM_{ppqq}^{(s, \pm)}(s, t, \alpha, \bar{\alpha})$ or more precisely its contact term $\calM_{ppqq}^{(s, c, \pm)}(s, t, \alpha, \bar{\alpha})$ since the rest is known. We were able to find these for all $p, q$ and therefore directly fix the full amplitude without referring to auxiliary ones. The results agree with \eqref{known-tensor} but we will list them here for demonstration and also because they reveal a few previously unseen properties.

The ansatz is expressed in terms of \eqref{contact-building-blocks} with $k_{21} = k_{43} = 0$.\footnote{We could also have used Legendre polynomials as a basis in this case as $g^{0,0}_k(\alpha^{-1})=\frac{k!}{4^k \left(\frac{1}{2}\right)_k}P_k(2\alpha - 1)$.}
\es{EvenTensorAnsatz}{
    &\mathcal{M}_{ppqq}^{(s,c,+)}(s,t,\alpha,\bar{\alpha})=\sum_{j \leq i}^\mathcal{E} g^{0,0}_{\{i, j\}}(\alpha, \bar{\alpha})\big[a^{i,j}_0+(t+u)a_+^{i,j}+(t-u)a_-^{i,j}\big] \\
    &\mathcal{M}_{ppqq}^{(s,c,-)}(s,t,\alpha,\bar{\alpha})=\sum_{j \leq i}^\mathcal{E} g^{0,0}_{[i, j]}(\alpha, \bar{\alpha})\big[\tilde a^{i,j}_0+(t+u)\tilde a_+^{i,j}+(t-u)\tilde a_-^{i,j}\big].
}
It turns out that the only non-vanishing coefficients are $a_0^{i,i}$, $a_+^{i,i}$, $a_0^{i+1,i-1}$, $a_-^{i+1,i}$, $\tilde{a}_0^{i+1,i}$ and $\tilde{a}_+^{i+1,i}$. The first two coefficients in the even amplitude are more complicated. They are
\begin{align}
    a_0^{i,i}=& \frac{i!^2\left(-4\right)^{-2i - 2}pq\Gamma(q+p)}{\left(\frac{1}{2}\right)^2_i \left ( i - \frac{1}{2}\right )_3 \Gamma(q+2+i)\Gamma(p+2+i)\Gamma(p+1-i)\Gamma(q+1-i)} \times \label{tensor-even-complex1}
    \\&
    \big[-2 (i)_2^2 (-1 - 2 i + 2 i^2 + 8 i^3 + 4 i^4)
    -2 (1 + 2 i + 2 i^2) (-1 - 2 i + 2 i^2 + 8 i^3 + 4 i^4) p  q 
    \nonumber\\&
    +2 (1 + 2 i - 2 i^2 - 8 i^3 - 4 i^4)p^2q^2
    -(i)_2 (2 + 8 i + 7 i^2 - 22 i^3 - 61 i^4 - 60 i^5 - 20 i^6)(p + q)
   \nonumber\\&
   +(1 + 2 i)^2 (-2 - 3 i + 5 i^2 + 16 i^3 + 8 i^4)pq(p+q)
   + 16 (\tfrac{i}{2} - \tfrac{1}{4})_2 (1 + 3 i + 3 i^2)p^2  q^2  (p + q) 
   \nonumber\\&
   +2 (i)_2 (1 + 2 i - 2 i^2 - 8 i^3 - 4 i^4)
   (p+q)^2
   + 16 (\tfrac{i}{2} - \tfrac{1}{4})_2 (1 + 3 i + 3 i^2) p  q  (p + q)^2
   \nonumber\\&
   - 16 (i)_2 (\tfrac{i}{2} - \tfrac{1}{4})_2 (1 + 3 i + 3 i^2)(p + q)^3
    \big], \nonumber
\end{align}
and
\begin{align}
    a_+^{i,i}=& \frac{i!^2\left(-4\right)^{-2i-2}pq\Gamma(q+p)}{2\left(\frac{1}{2}\right)^2_i \left ( i - \frac{1}{2} \right )_3 \Gamma(q+2+i)\Gamma(p+2+i)\Gamma(p+1-i)\Gamma(q+1-i)} \times \label{tensor-even-complex2}
    \\&
    \big[2 (i)_2^2 (-1 - i + 3 i^2 + 8 i^3 + 4 i^4)
    +2 (1 + 2 i + 2 i^2) (-1 - i + 3 i^2 + 8 i^3 + 4 i^4)pq
    \nonumber\\&
    +2 (-1 - i + 3 i^2 + 8 i^3 + 4 i^4) p^2 q^2
    -(i)_2 (-2 - 3 i + i^2 + 8 i^3 + 4 i^4) (p + q)
    \nonumber\\&
    +(-1 + i + 5 i^2 + 8 i^3 + 4 i^4) p q (p + q)+ (i)_2 (p + q)^2
    \big]. \nonumber
\end{align}
The remaining coefficients in the even amplitudes are simpler and given by 
\es{tensor-even-simple}{
    &a_0^{i+1, i-1}=\frac{(i-1)!(i+1)!(-4)^{-2i - 1}pq\Gamma(q+p+1)i(i + 1)}{2\left(\frac{1}{2}\right)_{i-1}\left(\frac{1}{2}\right)_{i+1} \left ( i + \frac{1}{2} \right )_1 \Gamma(q+1+i)\Gamma(p+1+i)\Gamma(p-i)\Gamma(q-i)}\\
    &a_-^{i+1, i}=\frac{i!(i+1)!(-4)^{-2i-3}pq\Gamma(q+p)(i+1)^3 \left(2 + 4 i + 2 i^2 + p + q + 2 p q\right)}{\left(\frac{1}{2}\right)_{i}\left(\frac{1}{2}\right)_{i+1} \left ( i + \frac{1}{2} \right )_2 \Gamma(q+2+i)\Gamma(p+2+i)\Gamma(p-i)\Gamma(q-i)}.
}
We have two types of coefficients in the odd amplitude, the more complicated one is
\begin{align}
    \tilde{a}_0^{i + 1, i}=& \frac{i!(i+1)!(-4)^{-2i - 2}pq\Gamma(q+p-2)}{2 \left(\frac{1}{2}\right)_{i}\left(\frac{1}{2}\right)_{i+1} \left ( i + \frac{1}{2} \right )_2 \Gamma(q+2+i)\Gamma(p+2+i)\Gamma(p-i)\Gamma(q-i)} \times \label{tensor-odd-complex} \\
    & \big[(1 + i)^2(3 + 8i - 4i^2 - 8i^3 - 2i^4) - 2i (2 + i) (2 + 6 i + 3 i^2)pq \nonumber \\
    &+ (-5 - 12 i + 6 i^2) p^2q^2-p^3q^3 - (1 + i)^2 (8 + 22 i + 11 i^2)(p+q) \nonumber \\
    &+ (-7 - 20 i - 10 i^2)pq(p+q) + (1 + i)^2 (5 + 12 i + 6 i^2)(p+q)^2 \nonumber \\
    &+ (5 + 12 i + 6 i^2)pq(p+q)^2 \big]. \nonumber
\end{align}
Finally, the last coefficient is 
\begin{align}
    \tilde{a}_+^{i + 1, i}= -\frac{i!(i+1)!(-4)^{-2i-2}pq\Gamma(q+p+1)(i + 1)^2}{\left(\frac{1}{2}\right)_{i}\left(\frac{1}{2}\right)_{i+1}\Gamma(q+2+i)\Gamma(p+2+i)\Gamma(p-i)\Gamma(q-i)}. \label{tensor-odd-simple}
\end{align}
We checked this formula against the prediction \eqref{known-tensor} which \cite{wz21} found with hidden conformal symmetry and all results agreed. Moreover, it is possible to prove that $\tilde{a}_+^{i + 1, i}$ always agrees with \eqref{known-tensor}, which Appendix \ref{app:rigor} does for $q = p$. This establishes the correctness of one of our coefficients for arbitrarily high extremality without relying on extrapolating patterns.

It is interesting that when contact terms (made from double-trace operators) are decomposed into R-symmetry representations, the $SU(2)_L$ spin $j$ and the $SU(2)_R$ spin $\bar{\jmath}$ differ at most by $2$. As evident in Figures \ref{fig:superMultiplets1} and \ref{fig:superMultiplets2}, this is also the maximal difference one can find between $j$ and $\bar{\jmath}$ for \textit{single}-trace conformal primaries. When R-symmetry representations match, it is possible to absorb a contact Witten diagram into an exchange Witten diagram by shifting the residues of the latter. In other words, there exists a field redefinition that can be used to eliminate quartic couplings from the $AdS_3$ effective action. The difference compared to the maximal supergraviton and supergluon theories is that the redefinition needed here no longer coincides with \eqref{m-replacement}. Perhaps the most remarkable property of \eqref{m-replacement}, noticed in \cite{bfz21,abfz21}, is that it has a relation to Parisi-Sourlas supersymmetry from statistical physics \cite{ps79}. In particular, one can write a super-Witten diagram which uses this prescription as a differential operator acting on a Witten diagram for a single scalar conformal primary in a lower number of dimensions.\footnote{This was proven using an identity in \cite{z20} for the polar part of a Witten diagram which happens to be the full Witten diagram in the scalar case. It is a simple consequence of a similar identity for conformal blocks which was proven in \cite{krt19}.} Tree-level holographic correlators in the D1-D5 CFT can only have \textit{some} of their terms written in this way which means there is no strong reason to think that Parisi-Sourlas supersymmetry plays a role in this theory.

\subsection{Mixed tensor-graviton correlators}
Moving onto the tensor-graviton correlators $\left < s_p^{I_1} s_p^{I_2} \sigma_q \sigma_q \right >$, all terms in them now come with the same $SO(h_{1, 1} + 1)$ structure $\delta^{I_1 I_2}$. We have found that the most conveient basis to use for the ansatz is
\es{MixedAnsatz}{
    \calM_{ppqq}^{(c,+)}(s, t, \alpha, \bar{\alpha}) =& \sum_{i \leq j}^\calE g^{0,0}_{\{i,j\}}(\alpha, \bar{\alpha}) [b_s^{i, j} s + b_t^{i, j} t + (-1)^{i + j} b_u^{i, j} u] \\
    \calM_{ppqq}^{(s,c,-)}(s, t, \alpha, \bar{\alpha}) =& \sum_{i < j}^\calE g^{0,0}_{[i,j]}(\alpha, \bar{\alpha}) (b_{ss}^{i, j} s + b_{st}^{i, j} t + b_{su}^{i, j} u) \\
    \calM_{ppqq}^{(t,c,-)}(s, t, \alpha, \bar{\alpha}) =& \sum_{i < j}^\calE g^{\frac{p - q}{2},\frac{q - p}{2}}_{[i,j]} \left ( \frac{\alpha}{\alpha - 1}, \frac{\bar{\alpha}}{\bar{\alpha} - 1} \right ) |\alpha - 1|^{2\calE} (b_{ts}^{i, j} s + b_{tt}^{i, j} t + b_{tu}^{i, j} u) \\
    \calM_{ppqq}^{(u,c,-)}(s, t, \alpha, \bar{\alpha}) =& \sum_{i < j}^\calE g^{\frac{p - q}{2},\frac{p - q}{2}}_{[i,j]}(\alpha^{-1}, \bar{\alpha}^{-1}) |\alpha|^{2\calE} (-1)^{i + j + 1} (b_{us}^{i, j} s + b_{ut}^{i, j} t + b_{uu}^{i, j} u).
}
Due to \eqref{super-crossing-m}, the odd contact terms for the $t$- and $u$-channels should be viewed as crossing transformations of $s$-channel contact terms in a different correlator.
Moreover, Bose symmetry ensures that the coefficients are related by
\es{mixed-bose}{
    &b^{i,j}_t=b^{i,j}_u, \quad b^{i,j}_{st}=b^{i,j}_{su}, \quad b^{i,j}_{tt}=b^{i,j}_{uu}, \quad b^{i,j}_{ts}=b^{i,j}_{tu}=b^{i,j}_{us}=b^{i,j}_{ut}.
}
As in the four-tensor case, the coefficients in the ansatz truncate at a finite $|i-j|$. In particular, the non-zero ones are  $b_s^{i,i}$, $b_s^{i,i+2}$, $b_t^{i,i}$, $b_t^{i,i+1}$,  $b_t^{i,i+2}$, $b^{i,i+1}_{ss}$, $b^{i,i+1}_{st}$, $b^{i,i+1}_{ts}$, $b^{i,i+1}_{tt}$ and their images under \eqref{mixed-bose}. Moreover, the contact terms here are also given by the product of a simple prefactor and a symmetric polynomial in $p$ and $q$. While the polynomials were of third order at most in the last subsection, they have order up to six in this case. We will not write what they are because they all agree with \eqref{known-mixed-odd} and \eqref{known-mixed-even} from \cite{wz21}. Nevertheless, we would like to point out one property of this calculation which is manifest in a full but not auxiliary Mellin amplitude. Namely for $p = 1$, the vanishing of extremal correlators means that the $s^{I_1}_1 \times s^{I_2}_1$ OPE does not contain any $\sigma$ operators. By crossing, $s^{I_1}_1 \times \sigma_q$ cannot contain any $s^I$. The result is that the $\left < s_1^{I_1} s_1^{I_2} \sigma_q \sigma_q \right >$ correlator reduces to a function with only $s$-channel poles and the Ward identity then forces it to agree with $\left < s_1^{I_1} s_1^{I_2} s_q^{I_3} s_q^{I_4} \right >$ up to a trivial factor with $\delta^{I_3 I_4}$.

\subsection{Graviton correlators}
\label{gravAmp}

We now turn to the four-graviton correlators $\left < \sigma_p \sigma_p \sigma_q \sigma_q \right >$ which are not available in the literature. The simplest contact terms we have been able to find come about when we parameterize the even ones as
\es{GravAnsatz}{
    \calM_{ppqq}^{(s,c,+)}(s, t, \alpha, \bar{\alpha}) =& \sum_{i \leq j}^\calE g^{0,0}_{\{i,j\}}(\alpha, \bar{\alpha}) c_s^{i, j} s \\
    \calM_{ppqq}^{(t,c,+)}(s, t, \alpha, \bar{\alpha}) =& \sum_{i \leq j}^\calE g^{\frac{p - q}{2},\frac{q - p}{2}}_{\{i,j\}} \left ( \frac{\alpha}{\alpha - 1}, \frac{\bar{\alpha}}{\bar{\alpha} - 1} \right ) |\alpha - 1|^{2\calE} c_t^{i, j} t \\
    \calM_{ppqq}^{(u,c,+)}(s, t, \alpha, \bar{\alpha}) =& \sum_{i \leq j}^\calE g^{\frac{p - q}{2},\frac{p - q}{2}}_{\{i,j\}}(\alpha^{-1}, \bar{\alpha}^{-1}) |\alpha|^{2\calE} (-1)^{i + j} c_u^{i, j} u
}
and the odd ones as in \eqref{MixedAnsatz}. The Bose symmetry relations are given by \eqref{mixed-bose} again (with $b_*^{i, j}$ replaced by $c_*^{i, j}$).
Unfortunately, while the coefficients of the odd amplitudes truncate at finite $|i-j|$, all coefficients are generically non-zero in the even contact term. This does not have to be the case if we depart from \eqref{GravAnsatz}, for instance by including $g_{\{ i, j \}}(\alpha, \bar{\alpha})$ and $g_{\{ i, j \}}(\alpha^{-1}, \bar{\alpha}^{-1}) |\alpha|^{2\calE}$ in the same channel. When this is done however, there is a large redundancy in the solutions and we have not found a way to fix this without spoiling truncation of the basis. A more pragmatic problem is that the symmetric polynomials in $p$ and $q$ that appear when we content ourselves with \eqref{GravAnsatz} are of order thirteen. Due to these complications, the families of full amplitudes we were able to infer using the method of subsection \ref{4tensor} all had fixed extremality. Nevertheless, we have overcome this limitation by using a different method. By feeding our case-by-case full amplitudes $
\calM^{(\pm)}_{ppqq}(s, t, \alpha, \bar{\alpha})$ into \eqref{full-to-aux-minus} and \eqref{full-to-aux-plus}, we have been able to fix a formula for the \textit{auxiliary} amplitudes $\widetilde{\calM}^{(\pm)}_{ppqq}(s, t, \alpha, \bar{\alpha})$ which holds for all $p$ and $q$. One can now easily go back to full amplitudes by using \eqref{aux-to-full}.\footnote{Based on the tensor case, it should not be surprising that patterns which are hard to notice in the full amplitude become easier in the auxiliary amplitude. The expressions \eqref{known-tensor} and \eqref{EvenTensorAnsatz} are equivalent but the former can be written down in much less space.}

The two types of monomials to consider when it comes to auxiliary Mellin amplitudes are $\alpha^i \bar{\alpha}^j$ and $\sigma^i \tau^j$. While running the algorithm in subsection \ref{auxAmp}, we should use $\alpha^i \bar{\alpha}^j$ so that the remark under \eqref{full-to-aux-minus} about only needing to invert $\widehat{U}$ holds true. After obtaining an expression however, we should change the basis because 
because each $\sigma^i\tau^j$ coefficient of the auxiliary amplitude has a finite (independent of $p$ and $q$) number of poles.\footnote{This feature is almost certainly related to the existence of hidden 6d conformal symmetry at tree-level in the D1-D5 CFT.} This simplicity is especially apparent in the odd auxiliary Mellin amplitude
\es{grav-aux-odd}{
    \calMt^{(-)}_{ppqq}(s, t, \alpha, \bar{\alpha}) &= \frac{8p^2q^2(p^2 + q^2 - 2)(\alpha - \bar{\alpha})}{(p + 1)!(q + 1)!} \sum_{0 \leq i + j \leq \calE - 2} \frac{\sigma^i \tau^j}{(i! j!)^2} \\
    &\times \frac{(p - i - j - 1)_{i + j} (q - i - j - 1)_{i + j}}{(s - 2i - 2j - 2)(t - p - q + 2j + 2)(\tilde{u} - p - q + 2i + 2)}
}
where we recall that $\tilde{u} = u - 4$.
The even amplitude, which has $\tilde{u} = u - 2$, is the more complicated one and its analytic structure is 
\begin{align}
    \calMt^{(+)}_{ppqq}(s, t, \alpha, \bar{\alpha}) &= \frac{p^2 q^2}{(p + 1)! (q + 1)!} \sum_{0 \leq i + j \leq \calE - 1} \frac{\sigma^i \tau^j}{(i! j!)^2} (p - i - j)_{i + j - 1} (q - i - j)_{i + j - 1} \nonumber \\
&\left [ \sum_{m = 0}^2 \sum_{n = m - 1}^1 \frac{4(p^2 + q^2 - 2)}{s - 2i - 2j - 2m} \left ( \frac{d_{st}^{i,j}(m,n)}{t - p - q + 2n + 2j} + \frac{d_{su}^{i,j}(m,n)}{\tilde{u} - p - q + 2n + 2i} \right ) \right . \nonumber \\
&+ \sum_{m = -1}^1 \left ( \frac{d_s^{i,j}(m)}{s - 2i - 2j - 2m - 2} + \frac{d_t^{i,j}(m)}{t - p - q + 2m + 2j} + \frac{d_u^{i,j}(m)}{\tilde{u} - p - q + 2m + 2i} \right ) \nonumber \\
&\left. + \sum_{n = -1}^1 \frac{d_{tu}^{i,j}(n) (p^2 + q^2 - 2)}{(t - p - q + 2n + 2j)(\tilde{u} - p - q - 2n + 2i)} \right ] \label{grav-aux-even}
\end{align}
where the solution is now specified by listing the coefficients. For the single poles, we have
\begin{align}
d_t^{i,j}(-1) &= -8j^2(j - 1)^2 \nonumber \\
d_t^{i,j}(0) &= 8j^2[2j^2 - 2j(p + q) + pq + 1] \nonumber \\
d_t^{i,j}(1) &= -2[2j^2 - 2j(p + q - 1) - p + pq - q + 1]^2 \nonumber \\
d_s^{i,j}(-1) &= -2[-1 - 2 (i + j) - 2 (i + j)^2 + p^2][-1 - 2 (i + j) - 2 (i + j)^2 + q^2] \nonumber \\
d_s^{i,j}(0) &= -4[-6 - 8 (i + j) - 4 (i + j)^2 + p^2 + q^2](p - i - j - 1)(q - i - j - 1) \nonumber \\
d_s^{i,j}(1) &= -8(p - i - j - 2)_2(q - i - j - 2)_2. \label{grav-single-poles}
\end{align}
The other single poles follow from ($i \leftrightarrow j$, $t \leftrightarrow \tilde{u}$) so let us move onto simultaneous poles. The ones involving both $t$ and $\tilde{u}$ are simple because the non-factorized polynomial in $p$ and $q$ turns out to be independent of $i$ and $j$. It is also the same no matter which pair of poles we are looking at --- always $ p^2 + q^2-2 $.  The full result is
\es{grav-tu-poles}{
d_{tu}^{i,j}(-1) = 4j^2(j - 1)^2, \quad d_{tu}^{i,j}(0) = -8i^2j^2, \quad d_{tu}^{i,j}(1) = 4i^2(i - 1)^2
}
which is clearly Bose symmetric. The simultaneous poles in $s$ and $t$ are more complicated and given by
\begin{align}
d_{st}^{i,j}(2,1) &= (p - i - j - 2)_2(q - i - j - 2)_2 \nonumber \\
d_{st}^{i,j}(1,1) &= -(p - i - j - 1)(q - i - j - 1)[2(i)_2 + (i + j)(i + j - p - q + 2) - p + pq - q + 2] \nonumber \\
d_{st}^{i,j}(1,0) &= -2j^2(p - i - j - 1)(q - i - j - 1) \nonumber \\
d_{st}^{i,j}(0,1) &= i^2[3i^2 + 2j^2 + 4ij - 2(i + j)(p + q - 1) - p + 2pq - q + 1] \nonumber \\
d_{st}^{i,j}(0,0) &= j^2[2j^2 + 4ij + 2j - 2(i + j)(p + q) - p + 2pq - q] \nonumber \\
d_{st}^{i,j}(0,-1) &= j^2(j - 1)^2. \label{grav-st-poles}
\end{align}
The coefficients we have not written follow from
\es{grav-bse}{
d_{u}^{i,j}(m) = d_t^{j,i}(m), \quad d^{i,j}_{su}(m,n) = d^{j,i}_{st}(m,n)
}
due to Bose symmetry.

\subsection{The flat-space limit}

With expressions for the $\left < s_p^{I_1} s_p^{I_1} s_q^{I_1} s_q^{I_1} \right >$, $\left < s_p^{I_1} s_p^{I_2} \sigma_q \sigma_q \right >$  and $\left < \sigma_p \sigma_p \sigma_q \sigma_q \right >$ correlators, one can extract the flat-space amplitudes from the asymptotic behaviour of the Mellin amplitudes as $s,t\to \infty$. Another way of getting these amplitudes is to impose the Ward identities directly on a flat-space version of our ansätze. Considering the two Ward identities in \eqref{ward-final}, \cite{az20} found that to get non-trivial flat-space constraints in higher dimensions, one should use the Ward identity with the $-$ sign instead of the $+$ sign. For the even amplitude, we have found the same thing here. The $+$ Ward identity, which was enough to find all AdS results in the previous subsections, is automatically satisfied whille the $-$ Ward identity fixes the flat space limit which is proportional to $(s + t - \alpha s)(s + t - \bar{\alpha} s)$. This can also be seen from the function of cross-ratios acting on the auxiliary correlator in \eqref{formal-sol}.

The flat-space limits of \eqref{known-tensor} and \eqref{known-mixed-even} were studied in \cite{wz21}. Since the tensor correlator is manifestly made from single poles, the linear growth condition as $s,t \to \infty$ fixes the non-trivial part of the flat-space tensor amplitude to be a polynomial in $\sigma$ and $\tau$. Defining $m_i$ as in \eqref{known-tensor},
\es{tensor-flat}{
\calM^{(+, s)}_{\mathrm{flat}}(s, t, \alpha, \bar{\alpha}) \propto \frac{(s + t - \alpha s)(s + t - \bar{\alpha} s)}{s} \sum_{0 \leq m_1 + m_2 \leq \calE - 1} \frac{\sigma^{m_2} \tau^{\calE - m_1 - m_2 - 1}}{\prod_{i = 1}^6 m_i!}.
}
The mixed auxiliary amplitude has simultaneous poles but they drop out of the flat-space limit leaving
\es{mixed-flat}{
\calM^{(+)}_{\mathrm{flat}}(s, t, \alpha, \bar{\alpha}) &\propto \frac{(s + t - \alpha s)(s + t - \bar{\alpha} s)}{s} \sum_{0 \leq m_1 + m_2 \leq \calE - 1} \frac{\sigma^{m_2} \tau^{\calE - m_1 - m_2 - 1}}{3 \prod_{i = 1}^6 m_i} \\
&\times [2 m_3 m_5 \tau^{-1} - 2(p - m_1)(p - m_1 - 1) + q^2 - 1]
}
from the single poles. Note that in \cite{abfz21}, the non-trivial term in flat space depended only on $\sigma$ and $\tau$ for a different reason. The amplitudes had simultaneous poles but a \textit{constant} growth condition as $s,t \to \infty$ characteristic of gluons.

One could similarly study the graviton amplitudes of subsection \ref{gravAmp} as a function of $\sigma$ and $\tau$ but we have found it more convenient to express the flat-space limit in terms of $\alpha$ and $\bar{\alpha}$. The result is of the form
\begin{align}
    \mathcal{M}^{(+)}_{\mathrm{flat}}(s, t, \alpha, \bar{\alpha}) \propto \frac{(s + t - \alpha s) (s + t - \bar\alpha s)}{stu}\sum_{i,j=0}^{\calE - 1}\alpha^i\bar\alpha^j \mathcal{P}^{i,j}_{p,q}(s,t) \label{flatEvenGeneral}
\end{align}
where the denominator is a typical signature of supergravity amplitudes. Moreover, $\mathcal{P}^{i,j}_{p,q}(s,t)$ is a homogeneous polynomial of order two in $s$ and $t$ given by
\begin{align}
    \mathcal{P}^{i,j}_{p,q}(s,t)=\frac{(-1)^{i+j}p^2q^2 \Gamma(i+j-1) [(s^2+t^2) P^{i,j}_1(p,q)+st(i+j-1) P^{i,j}_2(p,q)]}{2(i!j!)^2(p^2-1) (q^2-1) \Gamma (p-i) \Gamma (q-i) \Gamma (p-j) \Gamma (q-j) \Gamma (p+q-i-j-3)^{-1}}
\end{align}
where $P^{i,j}_1(p,q)$ and $P^{i,j}_2(p,q)$ are polynomials, symmetric under $p\leftrightarrow q$ and $i\leftrightarrow j$. They are given by 
\begin{align}
    P^{i,j}_1(p,q)&=-8 i^2 j^2 (p + q - 3) (p + q - 2) (p + q - 1)
   \nonumber\\
   &-(i+j)^3 (p - 1) (q - 1) (5 - 7 p + 2 p^2 - 7 q + 5 p q + 2 q^2)
   \nonumber\\
   &-(i+j) (p - 1)^2 (q - 1)^2 (-2 + 3 p + p^2 + 3 q + 2 p q + q^2)
   \nonumber\\
   &+(i+j)^2(p - 1) (q - 1) (7 - 11 p + 2 p^2 + 2 p^3 - 11 q + 7 p q + 2 q^2 + 
   2 q^3)
   \nonumber\\
   &-2 i j (i+j)(p + q - 2) (p + q - 1) (-10 + 7 p + p^2 + 7 q - 6 p q + q^2)
   \nonumber\\
   &+2 i j (i+j)^2 (p+q+2)^2 (p+q+3),
\end{align}
and 
\begin{align}
    P^{i,j}_2(p,q)&=-(i + j)^4 (p-1)^2 ( q-1)^2 + (i + j)(p - 1)^2 (q - 1)^2 (2 - p + p^2 - q + 2 p q + q^2)
    \nonumber\\
    & 
- 2 (i + j)^3(p - 1) (q - 1) (-4 + 7 p - 3 p^2 + 7 q - 6 p q + p^2 q - 3 q^2 + 
p q^2)  
    \nonumber\\&
 -4 i j(p - 1) (q - 1) (p + q - 3) (p + q - 2) (p + q - 1)
    \nonumber\\&
 +4 i j (i + j)^2(p + q - 2) (p + q - 1) (3 - 2 p - 2 q + p q) 
    \nonumber\\&
-(i + j)^2(p - 1) (q - 1) [9 - p(p - 4)^2 - q(q - 4)^2 + 17 p q + pq(p + q)(p + q - 6)]
    \nonumber\\&
 +4 i j (i + j)(p + q - 2) (p + q - 1) (-6 + 6 p - p^2 + 6 q - 6 p q + p^2 q - 
 q^2 + p q^2)
    \nonumber\\&
 -4 i^2 j^2[24 - 50 (p + q) + 35 (p + q)^2 - 10 (p + q)^3 + (p + q)^4].
\end{align}

The odd amplitudes decouple from the flat-space Ward identities. Then, to directly derive their $s,t\to\infty$ behaviour we would need to consider subleading terms in the identities. This implies, in particular, that there is no analogous nice prefactor of $(s + t - \alpha s)(s + t - \bar{\alpha} s)$ for the flat-space limit of the odd amplitudes. We verified this by explicitly performing the limit on finite radius results.

\section{Conclusion}
\label{sec:conc}

In this work, we were able to push the holographic correlator bootstrap in $AdS_3 \times S^3$ to previously inaccessible Kaluza-Klein levels by developing a formalism based on two Mellin amplitudes. One of these Mellin amplitudes can be specified in a convenient way with the help of 4d conformal blocks which is surprising for a purely 2d problem.\footnote{It is tempting to speculate that a relation to a 4d theory can be revealed by the chiral algebra \cite{bllprv13,rrz19}. For the D1-D5 CFT, correlators in the twisted configuration are invariant under a single copy of $PSU(1, 1 | 2)$ and this property is shared by $\mathcal{N} = 4$ SYM.} The all-graviton four-point functions, which are presented here for the first time, display a number of properties which will make them interesting to study further. In particular, the presence of a spin-2 multiplet with only 8 supercharges causes the flat-space limit to take a form not seen in either the maximal supergraviton or supergluon amplitudes of \cite{az20,abfz21}. The factor growing with the external weights is not simply an overlap of wavefunctions on $S^3$ but contains a non-trivial dependence on the Mandelstam variables as well. Exploring the double-trace spectrum could also be fruitful since the contact terms were special enough that they could all be absorbed into exchange Witten diagrams. A difference compared to four-point functions involving tensor Kaluza-Klein modes is that we were not able to find a most natural candidate among the different ways to partition contact terms between the three channels.

The most drastic simplifying feature of the auxiliary amplitudes is an avatar of hidden 6d conformal symmetry \cite{ct18,rrz19,grtw20}. Although the set of poles possessed by such an amplitude grows without bound as the extremality is increased, this is no longer true if one restricts to a fixed R-symmetry monomial in the right basis. This in fact plays a significant role in how the double-trace spectrum is organized as can be seen in \cite{as21,dgs22}. Reproducing the graviton-type four-point functions using hidden conformal symmetry should be possible by extending the analysis in \cite{wz21} but we find the direct approach used here more appealing. This is partly because it avoids the technical pitfalls associated with 6d tensor structures but the main reason is that it gives us a tool for more general $AdS_3/CFT_2$ backgrounds. One of these which has received much interest is $AdS_3 \times S^3 \times S^3 \times S^1$ whose dual CFT must be invariant under the large $\mathcal{N} = 4$ super-Virasoro algebra.
Although the considerations in \cite{gmms04} presented a paradox for many years, this was based on an error in the supergravity spectrum of \cite{bps99}. After correcting it, \cite{eggl17,egl17,eg19} found evidence that the holographic dual is a deformed symmetric orbifold of the $S^3 \times S^1$ sigma model for a minimal value of one of the fluxes. Recently, \cite{w24} argued that the seed theory should more generally be thought of as a sigma model for the moduli space of instantons on $S^3 \times S^1$. Tihs provides a concrete description of the IR fixed-point for the flow devised in \cite{t14}.

When exploring other backgrounds, it can be expected that they will similarly require single-trace three-point functions like \eqref{ope-coeffs} to be supplied as input before superconformal symmetry can be used to fix the four-point function. Our calculation was possible because of cubic couplings for 6d $\calN = (2, 0)$ supergravity on $AdS_3 \times S^3$ that have been available since the work of \cite{apt00}. This is the exception rather than the rule. For the majority of otherwise well studied $AdS_{d + 1}$ backgrounds with 8 supercharges, the calculation of these cubic couplings has not yet been done.\footnote{It is possible that such calculations can be made less cumbersome in the future due to the development of new approaches to Kaluza-Kelin effective actions based on exceptional field theory \cite{dms23}.} It would be especially advantageous to know what they are for the orbifold backgrounds with a singular locus studied in \cite{bcf23,cpwy23,bcf24}. This would permit the evaluation of gluon scattering amplitudes with tree-level graviton exchange which in turn make it possible to fix ambiguities in gluon loops confined to the singularity. Even in the $AdS_3 \times S^3$ case, results in \cite{apt00} are limited to three-point functions in which two of the external operators are scalars. Those involving two or more vectors should be worked out as well so that the methods in this paper can be used to solve for an even wider class of half-BPS four-point functions. These 2d four-point functions with external vectors should be among the simplest examples of spinning correlators in AdS/CFT which \cite{fsv17,ckk17} have considered in the Mellin formalism. An interesting application of them would be computing the one-loop correction to the simplest scalar correlators encountered here using the AdS unitarity method \cite{aabp16}.

In addition to loops, which introduce new singularities, holographic correlators also have stringy corrections which are polynomial in the Mandelstam variables. In $\mathcal{N} = 4$ SYM for instance, these were first computed in \cite{bcpw19,c19,cp20} using supersymmetric localization. Although this method does not have a direct analogue in the D1-D5 CFT, it was shown in \cite{ahs22,ahs23,ah23} that much of the same data can be obtained from single-valuedness which is more powerful than was previously thought.
It would be very satisfying if this also places strong constraints on amplitudes of tensor and graviton vertex operators in $AdS_3 \times S^3 \times M_4$ supported by R-R flux.\footnote{For the tensor multiplet, it will be important that the $SO(h_{1, 1} + 1)$ global symmetry seen here is broken to $SO(h_{1, 1})$ in the full string theory \cite{t07}.}
Previous efforts to turn on R-R flux beyond the planar limit have started with the free symmetric orbifolds of $K3$ and $T^4$ sigma models.
Along with the works \cite{hmz18,lss20a,lss20b,ghmm22} studying heavy operators, \cite{gpz15,bkz21} showed that the Hagedorn density of states and low-lying higher-spin currents at the orbifold point are both resolved in conformal perturbation theory.\footnote{There is by now a similar story for symmetric orbifolds of $\mathcal{N} = 2$ super-Virasoro minimal models which are conjectured to be holographic \cite{bbchk20,abbck22,bbch22}. Explicit checks were performed up to spin-3 but it is possible that the algorithms in \cite{ab22} which probed higher spins can be adapted to this supersymmetric case.} A complementary expansion using holographic correlators as a starting point will bring us closer to the goal of being able to follow the spectrum of this rich theory from weak to strong coupling.

\acknowledgments
We are grateful for conversations with Francesco Aprile, Pinaki Banerjee, Alexandre Belin, Nathan Berkovits, Pietro Ferrero, Ji Hoon Lee, Jo\~{a}o Penedones, Eric Perlmutter, Leonardo Rastelli, Michele Santagata, Alessandro Sfondrini and Ida Zadeh. This project received funding from the S\~{a}o Paulo Research Foundation (FAPESP) grants 2019/24277-8, 2021/14335-0, 2022/05236-1 and 2023/03825-2. CB thanks the Pascal Institute and the organizers of ``Speakable and Unspeakable in Quantum Gravity'' for providing a stimulating work environment.

\appendix
\section{Exchange Witten diagrams}
\label{app:residues}

Exchange Witten diagrams in Mellin space are most conveniently expressed as a sum over poles with the contact term set to zero.\footnote{Conformal blocks on the other hand, even though they have the same poles and residues as Witten diagrams, have an unambiguous regular term which grows faster than any polynomial \cite{fk11}.} This is allowed because we are including contact terms separately in our ansatz for a four-point function anyway. Equivalently, the degree $\ell - 1$ contact term which appears in a spin-$\ell$ Witten diagram defined using AdS propagators may be changed at will with a field redefinition. The explicit expression, which can be fixed using the conformal Casimir, is
\es{exchange-amp}{
\calM_{\Delta,\ell} = \sum_{m = 0}^\infty \frac{f^m_{\Delta,\ell} Q_{\Delta,\ell}(t,u)}{s - \Delta + \ell - 2m}
}
where the $m = 0$ pole is due to the exchanged conformal primary and $m > 0$ poles are due to its descendants. The numerator factors $f^m_{\Delta, \ell}$ and $Q_{\Delta, \ell}(t, u)$ both depend on the spatial dimension $d$. As explained in the main text, we will need to consider $\ell = 0,1,2$ for $d = 2$ and $\ell = 0,1$ for $d = 4$. Using the notation $\Gamma(x \pm y) \equiv \Gamma(x + y) \Gamma(x - y)$,
\es{exchange-f}{
f^m_{\Delta, \ell} = \frac{-2^{2 - 2\ell} \Gamma[\Delta + \ell]}{m! \left ( \Delta - \frac{d - 2}{2} \right )_m \Gamma \left [ \frac{\Delta_1 + \Delta_2 - \Delta + \ell}{2} - m \right ] \Gamma \left [ \frac{\Delta_3 + \Delta_4 - \Delta + \ell}{2} - m \right ] \Gamma \left [ \frac{\Delta + \ell \pm \Delta_{12}}{2} \right ] \Gamma \left [ \frac{\Delta + \ell \pm \Delta_{34}}{2} \right ]}
}
in the conventions of \cite{az20,abfz21}.\footnote{Note that \cite{cgp12,gkss16,dgs17} use different conventions. Their normalization of Witten diagrams, namely $U^{\tau / 2}(1 - V)^\ell$ after taking $U \to 0$ first and $V \to 1$ second, is the same as ours. The splitting between $f^m_{\Delta, \ell}$ and $Q_{\Delta, \ell}(t, u)$ is what differs.} The Mack polynomials are then
\es{exchange-q}{
Q_{\Delta,0}(t, u) &= 1 \\
Q_{\Delta,1}(t, u) &= \frac{(\delta_t^2 - \delta_u^2)(t + u + d - 2 - \Sigma)}{4(\Delta - d + 1)} + (\Delta - 1)(t - u) \\
Q_{\Delta,2}(t, u) &= \frac{(d - 1)T_1}{16d(\Delta - d)} - \frac{(d - 1)T_2}{16d(\Delta - d + 1)} + \frac{T_3}{16d} + \Delta(\Delta - 1)(t - u)^2 \\
&- \frac{\delta_t^2 - \delta_u^2}{2}(t - u)(t + u + d - 2 - \Sigma) \\
&- \frac{2(\Delta^2 - \Delta + 1) - \delta_t^2 - \delta_u^2}{2d}(t + u + d - 2 - \Sigma)^2
}
where we have defined
\es{exchange-defs1}{
T_1 &= (\delta_u^2 - \delta_t^2)(t + u + d - 2 - \Sigma) \\
&\times [u(\delta_u^2 - \delta_t^2 - 8d) + t(\delta_u^2 - \delta_t^2 + 8d) - (\delta_u^2 - \delta_t^2)(\Sigma - d + 2)] \\
T_2 &= [(\delta_u - \delta_t)^2 - 4][(\delta_u + \delta_t)^2 - 4](t + u + d - 1 - \Sigma)(t + u + d - 3 - \Sigma) \\
T_3 &= [(\delta_u - \delta_t)^2 - 4][(\delta_u + \delta_t)^2 - 4] \\
&+ 8\Delta(\Delta - 1)(2\Delta^2 - 2d\Delta + 2d^2 - 6d + 2 - \delta_u^2 - \delta_t^2)
}
in terms of
\es{exchange-defs2}{
\Sigma &\equiv \Delta_1 + \Delta_2 + \Delta_3 + \Delta_4 \\
\delta_t &\equiv \Delta_1 + \Delta_4 - \Delta_2 - \Delta_3 \\
\delta_u &\equiv \Delta_2 + \Delta_4 - \Delta_1 - \Delta_3.
}

\section{Towards a proof of superconformal symmetry}
\label{app:rigor}

As discussed in the main text, bootstrap methods predict a solution for the correlator $\left < s^{I_1}_p s^{I_2}_p s^{I_3}_p s^{I_4}_p \right >$. The odd Mellin amplitude proportional to $\delta^{I_1 I_2} \delta^{I_3 I_4}$ is
\es{pppp-conj1}{
\calM^{(s,-)}_{pppp}(s, t, \alpha, \bar{\alpha}) &= \sum_{k = 2, 2 | k}^{2p - 2} \frac{p^2 k}{k^2 - 1} \mathcal{S}^{(-)}_{\sigma, k}(s, t, \alpha, \bar{\alpha}) + \sum_{k = 2, 2 | k}^{2p} \frac{p^2}{1 - k} \mathcal{S}^{(-)}_{V, k}(s, t, \alpha, \bar{\alpha}) \\
&+ \sum_{i = 0}^{\calE - 1} \left [ \tilde{a}_0^{i + 1, i} + (t + u) \tilde{a}_+^{i + 1, i} \right ] \left [ g^{0,0}_{-i - 1}(\alpha^{-1}) g^{0,0}_{-i}(\bar{\alpha}^{-1}) - (\alpha \leftrightarrow \bar{\alpha}) \right ]
}
and similarly for the even Mellin amplitude $\calM^{(s,+)}_{pppp}(s, t, \alpha, \bar{\alpha})$. This expression is conjectural because it relies on the patterns observed for small $p$ continuing to hold for all $p$. To prove this conjecture, it is enough to show that this expression can always be obtained by acting on some function with the translation of $(1 - z\alpha)(1 - \bar{z} \bar{\alpha})$ into Mellin space. At the same time, \cite{rrz19, grtw19} found that the auxiliary amplitude for this correlator is given by
\es{pppp-conj2}{
\calMt_{pppp}(s, t, \sigma, \tau) &= \sum_{0 \leq m_1 + m_2 \leq \calE} \frac{-p^2}{2 m_1!^2 m_2!^2 (p - 1 - m_1 - m_2)!^2} \frac{\sigma^{m_1} \tau^{\calE - m_1 - m_2 - 1}}{s + 2 + 2m_1 - 2p}.
}
This result is conjectural as well since it was inferred from hidden conformal symmetry. Acting on \eqref{pppp-conj2} with the difference operator in the superconformal Ward identity to find a full amplitude compatible with the expansion \eqref{after-super} has only been done for small $p$. Both gaps will be filled if we can show that \eqref{pppp-conj1} and \eqref{pppp-conj2} are compatible for all $p$. As reported in \cite{bfz21}, this has been done for the maximal supergraviton and supergluon theories in $d = 4$. The analogous exercise with $AdS_3/CFT_2$ turns out to be much more involved. In this appendix, we will prove the desired match for all residues in $s$ and the flat-space limit of the regular part in \eqref{pppp-conj1}. The former is fully due to exchange Witten diagrams and the latter is fully due to contact Witten diagrams. To complete these results, one would also need to analyze the regular part of the even amplitude and the subleading regular part of the odd amplitude --- the added complexity of such terms comes from the interplay of both exchange and contact diagrams.

A remarkable fact about the D1-D5 Mellin amplitudes (and also Mellin amplitudes in other theories that display hidden conformal symmetry) is that the complexity of each $\sigma^i \tau^j$ monomial is bounded --- the number of poles in $s$ that it has is fixed and cannot be made larger by increasing the external weight. We will therefore focus on components of the even and odd amplitudes which have the following definition.\footnote{We are intentionally avoiding the notation $\calM^{(\pm)}_{i, j}(s, t)$ since \eqref{hack-monomials} associated these with $\alpha^i \bar{\alpha}^j$ instead of $\sigma^i \tau^j$.}
\es{last-monomials}{
\calM^{(+)}_{pppp}(s, t; \alpha, \bar{\alpha}) &= \sum_{i, j = 0}^\calE \calM^{(+)}(s, t; i, j) \sigma^i \tau^j \\
\calM^{(-)}_{pppp}(s, t; \alpha, \bar{\alpha}) &= (\alpha - \bar{\alpha}) \sum_{i, j = 0}^{\calE - 1} \calM^{(-)}(s, t; i, j) \sigma^i \tau^j
}
The prediction for $\calM^{(\pm)}(s, t; i, j)$ comes from acting on \eqref{pppp-conj2} with \eqref{aux-to-full} to get
\begin{align}
\calM^{(-)}(s, t; i, j) &= \left [ \frac{p}{i! j! (p - n - 2)!} \right ]^2 \left [ (s - 2n - 2)^{-1} - \frac{1}{4} s \right ] + \dots \label{target} \\
\calM^{(+)}(s, t; i, j) &= \left [ \frac{p}{i! j! (p - n - 2)!} \right ]^2 \left [ \frac{1}{s - 2n - 2} - \frac{4i^2 + 4j^2 + 4ij + (t - 2p + 2n)(t - 2p)}{2 (p - n - 1)^2 (s - 2n)} \right. \nonumber \\
&\left. + \frac{(n - 1)(i^2 - j^2)(t + n - 2p - 5)}{(p - n)^2(p - n - 1)^2(s - 2n + 2)} \right ] + \dots \nonumber
\end{align}
where $n \equiv i + j$. Our task is to show that \eqref{pppp-conj1} also reproduces \eqref{target} when expressed in terms of $\sigma$ and $\tau$. The three parts of this calculation are residues of $\calM^{(-)}(s, t; i, j)$, residues of $\calM^{(+)}(s, t; i, j)$ and the $O(s)$ term shown in \eqref{target}.

\subsection{Equality of residues}
The $\calM^{(-)}(s, t; i, j)$ residue match works because
\es{odd-v-piecewise}{
\underset{s = 2r}{\text{Res}} \sum_{k = 2, 2 | k}^{2p} \frac{p^2}{1 - k} \mathcal{S}^{(-)}_{V, k}(s, t, i, j) = \begin{cases}
0, & r \leq n \\
\left [ \frac{p}{i! j! (p - n - 2)!} \right ]^2, & r = n + 1 \\
\left [ \frac{p}{i! j! (p - r - 1)!} \right ]^2 \frac{\Gamma \left ( n + \tfrac{1}{2} \right ) \Gamma \left ( n + \tfrac{3}{2} \right )}{\Gamma \left ( r + \tfrac{1}{2} \right ) \Gamma \left ( r - \tfrac{1}{2} \right )}, & r \geq n + 2
\end{cases}
}
and
\es{odd-sigma-piecewise}{
\underset{s = 2r}{\text{Res}} \sum_{k = 2, 2 | k}^{2p - 2} \frac{p^2 k}{k^2 - 1} \mathcal{S}^{(-)}_{\sigma, k}(s, t, i, j) = \begin{cases}
0, & r \leq n + 1 \\
-\left [ \frac{p}{i! j! (p - r - 1)!} \right ]^2 \frac{\Gamma \left ( n + \tfrac{1}{2} \right ) \Gamma \left ( n + \tfrac{3}{2} \right )}{\Gamma \left ( r + \tfrac{1}{2} \right ) \Gamma \left ( r - \tfrac{1}{2} \right )}, & r \geq n + 2
\end{cases}.
}
Proving this relies on the expressions
\es{witten-sigma-tau}{
\frac{\underset{s = k + 2 + 2m}{\text{Res}} \mathcal{S}^{(-)}_{\sigma,k}(s, t, \alpha, \bar{\alpha})}{\alpha - \bar{\alpha}} &= \sum_{n = 0}^{\calE - 1} \sigma^i \tau^j \frac{(-1)^{\frac{k}{2} - n + 1} (\tfrac{k}{2} - n)_{2n + 1}}{m! i!^2 j!^2 \Gamma[k + 1 + m] \Gamma[p - m + \tfrac{2 - k}{2}]} \\
\frac{\underset{s = k + 1 + 2m}{\text{Res}} \mathcal{S}^{(-)}_{V,k}(s, t, \alpha, \bar{\alpha})}{\alpha - \bar{\alpha}} &= \sum_{n = 0}^{\calE - 1} \sigma^i \tau^j \frac{(-1)^{\frac{k + 1}{2} - n} (\tfrac{k + 1}{2} - n)_{2n} (k^2 + 2ns - 2n + s - 1)}{2 m! i!^2 j!^2 \Gamma[k + 1 + m] \Gamma[p - m + \tfrac{3 - k}{2}]^2}.
}

Taking the residue in \eqref{odd-v-piecewise} means setting $m = r - \frac{k + 1}{2}$ and performing the sum over $k$ in steps of $2$.\footnote{If the OPE coefficients $C^{ssV}_{ppk}$ and $C^{ss\sigma}_{ppk}$ were polynomials in $p$, these sums would manifestly yield a ${}_2F_1$ function which is what happens for the maximal supergraviton and supergluon theories. We instead get ${}_3F_2$ functions because the OPE coefficients have poles with respect to $k$.} This is the same as setting $k = 2l$ and summing over $l$ in steps of $1$. Re-indexing the sum so that it runs from $l = -n$ to $l = p + 1 - n$, the left hand side of \eqref{odd-v-piecewise} is equal to
\begin{align}
\underset{s = 2r}{\text{Res}} \sum_{k = 2, 2 | k}^{2p} \frac{p^2}{1 - k} \mathcal{S}^{(-)}_{V, k}(s, t, i, j) &= -\left [ \frac{p}{i! j! (p - r - 1)!} \right ]^2 \frac{\Gamma[2n + 1]}{\Gamma[r + n + 1] \Gamma[r - n]} \label{odd-v1} \\
& \sum_{l = -n}^{p + 1 - n} \frac{(2n + 1)_l (1 + n - r)_l}{l! (1 + n + r)_l} \frac{2l^2 + 2l(2n + 1) + (n + r)(2n + 1)}{2l + 2n + 1}. \nonumber
\end{align}
This makes it easy to check the first two cases of \eqref{odd-v-piecewise}. For the third case, we can change the upper limit of the sum to $\infty$ (because the summand is only non-zero for $l \leq r - n - 1$) and the lower limit to $0$. For just the sum in \eqref{odd-v1}, this yields
\es{odd-v2}{
& \left ( r - \frac{1}{2} \right ) {}_3F_2 \left ( \hspace{-0.2cm} \begin{tabular}{c} \begin{tabular}{ccc} $n + \frac{1}{2},$ & $2n + 1,$ & $n + 1 - r$\end{tabular} \\ \begin{tabular}{cc} $n + \frac{3}{2},$ & $n + 1 + r$ \end{tabular} \end{tabular} \hspace{-0.3cm} \right ) \\
&+ \left (n + \frac{1}{2} \right ) \left [ {}_2F_1 \left ( \begin{tabular}{c} $2n + 1, n + 1 - r$ \\ $n + 1 + r$ \end{tabular} \right ) + 2 {}_2F_1 \left ( \begin{tabular}{c} $2n + 2, n + 2 - r$ \\ $n + 2 + r$ \end{tabular} \right ) \frac{n + 1 - r}{n + 1 + r} \right ]
}
where the missing hypergeometric function argument is $1$. Along with the usual ${}_2F_1$ evaluation using Gauss's theorem, the ${}_3F_2$ is fortunately of the type which can be evaluated using Dixon's theorem. The convergence conditions for both of these theorems are satisfied because $r \geq n + 2$. This leads to an expression which simplifies to \eqref{odd-v-piecewise}.

To take the residue in \eqref{odd-sigma-piecewise}, we set $m = r - \frac{k + 2}{2}$ and carry out the same re-indexing. After a partial fraction decomposition, the sum over $l$ in
\begin{align}
\underset{s = 2r}{\text{Res}} \sum_{k = 2, 2 | k}^{2p - 2} \frac{p^2 k}{k^2 - 1} \mathcal{S}^{(-)}_{\sigma, k}(s, t, i, j) &= -\left [ \frac{p}{i! j! (p - r - 1)!} \right ]^2 \frac{\Gamma[2n + 2]}{2\Gamma[r + n + 1] \Gamma[r - n - 1]} \label{odd-sigma1} \\
& \sum_{l = -n}^{p + 1 - n} \frac{(2n + 2)_l (2 + n - r)_l}{l! (1 + n + r)_l} \frac{4(l + n + 1)}{(2l + 2n + 1)(2l + 2n + 3)} \nonumber
\end{align}
gives a linear combination of the functions
\es{odd-sigma2}{
{}_3F_2  \left ( \hspace{-0.3cm} \begin{tabular}{c} \begin{tabular}{ccc} $n + \frac{1}{2},$ & $2n + 2,$ & $n + 2 - r$\end{tabular} \\ \begin{tabular}{cc} $n + \frac{3}{2},$ & $n + 1 + r$ \end{tabular} \end{tabular} \hspace{-0.3cm} \right ), \quad
{}_3F_2  \left ( \hspace{-0.3cm} \begin{tabular}{c} \begin{tabular}{ccc} $n + \frac{3}{2},$ & $2n + 2,$ & $n + 2 - r$\end{tabular} \\ \begin{tabular}{cc} $n + \frac{5}{2},$ & $n + 1 + r$ \end{tabular} \end{tabular} \hspace{-0.3cm} \right )
}
with coefficients $\frac{2}{2n + 1}$ and $\frac{2}{2n + 3}$. This time there is one extra step required to get a closed form expression matching \eqref{odd-sigma-piecewise}. Instead of applying Dixon's theorem directly, we must use the fact that the hypergeometric functions are contiguous to the ones in Dixon's theorem with parameter shifts of $\pm 1$ in the sense of \cite{c12}.

The $\calM^{(+)}(s, t; i, j)$ residue match works because
\begin{align}
&\underset{s = 2r}{\text{Res}} \sum_{k = 2, 2 | k}^{2p} \frac{p^2}{1 - k} \mathcal{S}^{(+)}_{V, k}(s, t, i, j) = \label{even-v-piecewise} \\
&\begin{cases}
0, & r \leq n - 2 \\
\left [ \frac{p}{i! j! (p - n)!} \right ]^2 (n - 1)(i^2 - j^2)(t + n - 2p - 5), & r = n - 1 \\
\left [ \frac{p}{i! j! (p - n - 1)!} \right ]^2 \frac{2}{(n + \tfrac{1}{2})(n - \tfrac{1}{2})} \left [ (1 - 4n^2)(t - 2p)^2 + 2(j - i)(t - 2p) \right. \\
\left. \times (6i^2 + 2j^2 + 8ij - 1) - 4(j^4 + 3i^4 + 6i^3j + 2ij^3 + 4i^2j^2 - i^2 - j^2 - ij) \right ], & r = n \\
\left [ \frac{p}{i! j! (p - n - 2)!} \right ]^2 \frac{16}{(n + \tfrac{1}{2})_2(n - \tfrac{1}{2})_2} \left [2(j - i)(t - 2p) + 8i^3(i + 2) + 8j^3(j + 2) \right. \\
\left. + 16ij(2i^2 + 2j^2 + 3ij + 3i + 3j + 36) + 4i^2 + 8j^2 - 4i - 1 \right ], & r = n + 1 \\
\left [ \frac{p}{i! j! (p - r - 1)!} \right ]^2 \frac{\Gamma \left ( n + \tfrac{1}{2} \right ) \Gamma \left ( n - \tfrac{1}{2} \right )}{32 \Gamma \left ( r + \tfrac{1}{2} \right ) \Gamma \left ( r + \tfrac{3}{2} \right )} [4(j - i)(t - 2p) + 8jr + 1], & r \geq n + 2
\end{cases} \nonumber
\end{align}
and
\begin{align}
&\underset{s = 2r}{\text{Res}} \sum_{k = 2, 2 | k}^{2p - 2} \frac{p^2 k}{k^2 - 1} \mathcal{S}^{(+)}_{\sigma, k}(s, t, i, j) = \label{even-sigma-piecewise} \\
&\begin{cases}
0, & r \leq n - 1 \\
\left [ \frac{p}{i! j! (p - n - 1)!} \right ]^2 \frac{8n^2}{(n + \tfrac{1}{2}) (n - \tfrac{1}{2})} [it + j(4p - 2n - t) - 2np - n^2], & r = n \\
\left [ \frac{p}{i! j! (p - n - 2)!} \right ]^2 \frac{32}{(n + \tfrac{1}{2})_2 (n - \tfrac{1}{2})_2} \left [ (j - i)(t - 2p) - 4i^3(i + 2) - 4j^3(j + 2) \right. \\
\left. - 2ij(8i^2 + 8j^2 + 8ij + 12i + 12j + 1) - 2i^2 + 6i + 1 \right ], & r = n + 1 \\
\left [ \frac{p}{i! j! (p - r - 1)!} \right ]^2 \frac{\Gamma \left ( n + \tfrac{1}{2} \right ) \Gamma \left ( n - \tfrac{1}{2} \right )}{32 \Gamma \left ( r + \tfrac{1}{2} \right ) \Gamma \left ( r + \tfrac{3}{2} \right )} [4(i - j)(t - 2p) - 8jr - 1], & r \geq n + 2
\end{cases}. \nonumber
\end{align}
This follows from the same calculations that work in the odd case except with super-Witten diagram expressions that are longer because of \eqref{r-coeffs}. Importantly, the intermediate cases are not simply analytic continuations of the residues for $r$ sufficiently large because they violate the convergence conditions for Gauss's theorem and Dixon's theorem.

\subsection{Leading contact term}

Now that the residues are known to match, we can also prove that $\tilde{a}_+^{k + 1, k}$ from \eqref{pppp-conj1} is precisely what is needed to reproduce the $O(s)$ contact term shown in \eqref{target}. Using
\es{legendre-id}{
P_{k + 1}(1 - 2\alpha) P_k(1 - 2\bar{\alpha}) - (\alpha \leftrightarrow \bar{\alpha}) = 2(\bar{\alpha} - \alpha) \sum_{n = 0}^k \sigma^i \tau^j \frac{(-1)^{k + n} (k + 1 - n)_{2n + 1}}{(k + 1)(i! j!)^2}
}
with the relation between $SU(2)$ blocks and Legendre polynomials gives
\es{3f2-derivs}{
\calM^{(c,s,-)}(s, t; i, j) &= s \frac{(-1)^{n + 1} p^2}{2 i!^2 j!^2} \frac{\Gamma(2p + 1) \Gamma(2n + 2)}{\Gamma(p - n)^2 \Gamma(p + n + 2)^2} [n + 1 - \partial_z] \\
&\times {}_3F_2 \left ( \hspace{-0.3cm} \begin{tabular}{c} \begin{tabular}{ccc} $2n + 2,$ & $n - p + 1,$ & $n - p + 1$ \end{tabular} \\ \begin{tabular}{cc} $n + p + 2,$ & $n + p + 2$ \end{tabular} \end{tabular} \hspace{-0.3cm} ; z \right ) \Biggl |_{z = -1} + O(1).
}
Leaving $z$ as a parameter above is useful because it allows us to simplify the expression with Whipple's quadratic transformation
\es{3f2-whipple}{
{}_3F_2 \left ( \hspace{-0.3cm} \begin{tabular}{c} \begin{tabular}{ccc} $a,$ & $b,$ & $c$ \end{tabular} \\ \begin{tabular}{cc} $1 + a - b,$ & $1 + a - c$ \end{tabular} \end{tabular} \hspace{-0.3cm} ; z \right ) = (1 - z)^{-a} {}_3F_2 \left ( \hspace{-0.3cm} \begin{tabular}{c} \begin{tabular}{ccc} $\frac{a}{2},$ & $\frac{a + 1}{2},$ & $1 + a - b - c$ \end{tabular} \\ \begin{tabular}{cc} $1 + a - b,$ & $1 + a - c$ \end{tabular} \end{tabular} \hspace{-0.3cm} ; -\frac{4z}{(1 - z)^2} \right ).
}
Using this on the $O(s)$ part of \eqref{3f2-derivs}, the action of the differential operator is
\begin{align}
& \frac{4(n + 1)(n + \tfrac{3}{2})(2p + 1)}{(n + p + 2)^2} \lim_{z \to -1} (1 + z)(1 - z)^{-2n - 5} {}_3F_2 \left ( \hspace{-0.3cm} \begin{tabular}{c} \begin{tabular}{ccc} $n + 2,$ & $n + \frac{5}{2},$ & $2p + 2$ \end{tabular} \\ \begin{tabular}{cc} $n + p + 3,$ & $n + p + 3$ \end{tabular} \end{tabular} \hspace{-0.3cm} ; -\frac{4z}{(1 - z)^2} \right ) \nonumber \\
& - 2(n + 1) \lim_{z \to -1} (1 + z)(1 - z)^{-2n - 3} {}_3F_2 \left ( \hspace{-0.3cm} \begin{tabular}{c} \begin{tabular}{ccc} $n + 1,$ & $n + \frac{3}{2},$ & $2p + 1$ \end{tabular} \\ \begin{tabular}{cc} $n + p + 2,$ & $n + p + 2$ \end{tabular} \end{tabular} \hspace{-0.3cm} ; -\frac{4z}{(1 - z)^2} \right ). \label{final-contact}
\end{align}
In the second limit, the rational function of $z$ vanishes at $z = -1$ and the hypergeometric function remains finite. In the first limit however, the rational function of $z$ vanishes while the hypergeometric function diverges. The value of \eqref{final-contact} is therefore computed by expanding this function around $z = -1$ to leading order. This precisely cancels the extra gamma functions in \eqref{3f2-derivs} so that the remaining contact term is the one in \eqref{target}.

\section{Three-point functions from supergravity}
\label{app:3pt}

Three-point functions between single-trace primaries, at leading order in the $1/N$ expansion, were quoted in \eqref{ope-coeffs}. Besides the supergravity spectrum, this was the main model dependent input needed for computing four-point Mellin amplitudes. This appendix will give a derivation and use position space for the only time in the paper. Calculations will be based on the quadratic and cubic parts of the $AdS_3$ effective action which were worked out in \cite{apt00}. One of our first steps in \eqref{saturation} was getting rid of R-symmetry indices by saturating them with polarizations. Here, we will instead denote them here by a collective index $r$. Without further ado, the action in \cite{apt00} is
\begin{align}
S &= \frac{N}{4\pi} \int_{AdS_3} \sqrt{-g} \; \sum_k \left [ -8k(k + 1) \nabla_\mu s^I_{r, k} \nabla^\mu s^I_{r, k} - 8k(k - 1) \nabla_\mu \sigma_{r, k} \nabla^\mu \sigma_{r, k} \right. \nonumber \\
& -\frac{1}{8} F_{\mu\nu}(A^\pm_{r, k}) F^{\mu\nu}(A^\pm_{r, k}) - \frac{(k + 1)(k - 1)}{4} A^\pm_{\mu,r,k} A^{\pm,\mu}_{r,k} \mp \frac{1}{2} P_0(A^\pm_{r, k})_\mu A^{\pm,\mu}_{r,k} \mp \frac{k}{2} P^\pm_{k - 1}(A^\pm_{r, k})_\mu A^{\pm,\mu}_{r,k} \nonumber \\
& -\frac{1}{8} F_{\mu\nu}(C^\pm_{r, k}) F^{\mu\nu}(C^\pm_{r, k}) - \frac{(k + 1)(k + 3)}{4} C^\pm_{\mu,r,k} C^{\pm,\mu}_{r,k} \pm \frac{1}{2} P_0(C^\pm_{r, k})_\mu C^{\pm,\mu}_{r,k} \mp \frac{k + 2}{2} P^\pm_{k + 3}(A^\pm_{r, k})_\mu A^{\pm,\mu}_{r,k} \nonumber \\
&\left. + \frac{1}{4} F_{\mu\nu}(A^\pm_{r, k}) F^{\mu\nu}(C^\pm_{r, k}) - \frac{(k - 1)(k + 3)}{2} A^\pm_{\mu,r,k} C^{\pm,\mu}_{r,k} \mp \frac{k + 1}{2} \left [ P_0(A^\pm_{r, k})_\mu C^{\pm,\mu}_{r,k} + P_0(C^\pm_{r, k})_\mu A^{\pm,\mu}_{r,k} \right ] \right ] \nonumber \\
& \pm \sum_{k_1, k_2, k_3} t^\pm_{r_1, r_2, r_3} W^s \left [ P^\pm_{k_3 - 1}(s^I_{r_1, k_1} \nabla s^I_{r_2, k_2})_\mu A^{\pm,\mu}_{r_3, k_3} - P^\pm_{k_3 + 3}(s^I_{r_1, k_1} \nabla s^I_{r_2, k_2})_\mu C^{\pm,\mu}_{r_3, k_3} \right ] \nonumber \\
& \pm \sum_{k_1, k_2, k_3} t^\pm_{r_1, r_2, r_3} W^\sigma \left [ P^\pm_{k_3 - 1}(\sigma_{r_1, k_1} \nabla \sigma_{r_2, k_2})_\mu A^{\pm,\mu}_{r_3, k_3} - P^\pm_{k_3 + 3}(\sigma_{r_1, k_1} \nabla \sigma_{r_2, k_2})_\mu C^{\pm,\mu}_{r_3, k_3} \right ] \nonumber \\
&+ \sum_{k_1, k_2, k_3} t^\pm_{r_1, r_2, r_3} \left [ V^{ssA} s^I_{r_1, k_1} \nabla_\mu s^I_{r_2, k_2} A^{\pm,\mu}_{r_3, k_3} + V^{ssC} s^I_{r_1, k_1} \nabla_\mu s^I_{r_2, k_2} C^{\pm,\mu}_{r_3, k_3} \right ] \nonumber \\
&+ \sum_{k_1, k_2, k_3} t^\pm_{r_1, r_2, r_3} \left [ V^{\sigma \sigma A} \sigma_{r_1, k_1} \nabla_\mu \sigma_{r_2, k_2} A^{\pm,\mu}_{r_3, k_3} + V^{\sigma \sigma C} \sigma_{r_1, k_1} \nabla_\mu \sigma_{r_2, k_2} C^{\pm,\mu}_{r_3, k_3} \right ] \nonumber \\
& + \sum_{k_1, k_2, k_3} a_{r_1, r_2, r_3} \left [ V^{ss\sigma} s^I_{r_1, k_1} s^I_{r_2, k_2} \sigma_{r_3, k_3} + V^{\sigma \sigma \sigma} \sigma_{r_1, k_1} \sigma_{r_2, k_2} \sigma_{r_3, k_3} \right ] + \dots \label{sugra-action}
\end{align}
where the suppressed terms involve other fields that are not relevant for us. The new differential operator defined in \eqref{sugra-action} is
\es{p-op}{
(P^\pm_m)_\mu^{\;\;\rho} \equiv \varepsilon_\mu^{\;\;\nu\rho} \partial_\nu \pm m \delta_\mu^\rho
}
while the cubic couplings are
\begin{align}
&W^s = 2(k_3 + 1), \quad &&W^\sigma = -2(k_3 + 1) \frac{(k_1 - 1)(k_2 - 1)}{(k_1 + 1)(k_2 + 1)} \label{cubic-couplings} \\
&V^{ssA} = -2(\Sigma^2 - 1), \quad &&V^{\sigma\sigma A} = -\frac{\Sigma^2 - 1}{(k_1 + 1)(k_2 + 1)}(k_1^2 + k_2^2 - k_3^2 - 1) \nonumber \\
&V^{ss\sigma} = -\frac{2^4 \Sigma (\Sigma^2 - 4) \alpha_1 \alpha_2 \alpha_3}{k_3 + 1}, \quad &&V^{\sigma \sigma \sigma} = -\frac{2^3 \Sigma (\Sigma^2 - 4) \alpha_1 \alpha_2 \alpha_3}{3(k_1 + 1)(k_2 + 1)(k_3 + 1)}(k_1^2 + k_2^2 + k_3^2 - 2) \nonumber \\
&V^{ssC} = 2(2\alpha_3 - 1)(2\alpha_3 - 3), \quad &&V^{\sigma\sigma C} = \frac{(2\alpha_3 + 1)(2\alpha_3 + 3)}{(k_1 + 1)(k_2 + 1)}[k_1^2 + k_2^2 - (k_3 + 2)^2 - 1] \nonumber
\end{align}
in terms of
\es{alpha-def}{
\alpha_3 \equiv \frac{k_1 + k_2 - k_3}{2}, \quad \alpha_3 \equiv \frac{k_3 + k_1 - k_2}{2}, \quad \alpha_1 \equiv \frac{k_2 + k_3 - k_1}{2}.
}
Explicit expressions for $a_{r_1, r_2, r_3}$ and $t^\pm_{r_1, r_2, r_3}$, which we will use later, may be found in \cite{c84}. The former is the $S^3$ integral of three scalar spherical harmonics while the latter integrates two scalars and one vector.

\subsection{Three scalars}

Scalar three-point functions in the supergravity limit of the D1-D5 CFT were obtained already in \cite{kst06,t07}. We will re-derive these results before moving onto the less often discussed calculation of three-point functions involving a vector. The starting point is the wave equation for a scalar bulk-bulk propagator in AdS. In Poincar\'{e} co-ordinates $x = (x_0, \vec{x})$,
\es{ads-wave-eq}{
(\Box - m^2)G_\Delta(x, y) = x_0^{d + 1} \delta(x - y), \quad m^2 = \Delta(\Delta - d).
}
The solution to \eqref{ads-wave-eq}, along with its limit as $y$ approaches the boundary, is
\es{two-propagators}{
G_\Delta(x, y) &= C_{\Delta,0} \left ( \frac{\xi}{2} \right )^\Delta {}_2F_1 \left ( \frac{\Delta}{2}, \frac{\Delta + 1}{2}; \Delta - \frac{d - 2}{2}; \xi^2 \right ), \quad \xi \equiv \frac{2 x_0 y_0}{x_0^2 + y_0^2 + |\vec{x} - \vec{y}|^2} \\
K_\Delta(x, y) &= C_{\Delta,0} \left ( \frac{x_0}{x_0^2 + |\vec{x} - \vec{y}|^2} \right )^\Delta
}
where
\es{norm-spin0}{
C_{\Delta, \ell}=\frac{(\Delta+\ell-1) \Gamma(\Delta)}{2 \pi^{d / 2}(\Delta-1) \Gamma(\Delta+1-d / 2)}
}
ensures the correct normalization.\footnote{One can see this by expanding the hypergeometric function and integrating the wave equation termwise. As explained in \cite{df98}, the bulk-bulk propagator is often expressed in a different form, related to \eqref{two-propagators} via quadratic transformation, which no longer permits exchanging the sum with the integral.} The second line of \eqref{two-propagators} is simply the large $\lambda$ limit of $\lambda^\Delta G_\Delta(x, y)$ with $y_0 = \lambda^{-1}$. We can then obtain correlators with both points on the boundary by doing the same thing to $x_0$. The holographic renormalization prescription
\es{boundary-ops}{
\mathcal{O}(x) = \frac{1}{\sqrt{C_{\Delta, 0}}} \lim_{\lambda \to \infty} \lambda^\Delta \phi(x) \bigl |_{x_0 = \lambda^{-1}}
}
therefore defines unit-normalized operators from the field $\phi$ \cite{cgp14}. It follows that the integral computing a three-point function is
\es{3pt-scalar1}{
\left < \mathcal{O}_{r_1}(x_1) \mathcal{O}_{r_2}(x_2) \mathcal{O}_{r_3}(x_3) \right > = g_{123} a_{r_1, r_2, r_3} \int \frac{dx}{x_0^{d + 1}} \frac{K_{\Delta_1}(x, x_1) K_{\Delta_2}(x, x_2) K_{\Delta_3}(x, x_3)}{\sqrt{C_{\Delta_1, 0} C_{\Delta_2, 0} C_{\Delta_3, 0}}}
}
where we have included R-symmetry for the operators as well. Rescaling fields in \eqref{sugra-action} to get canonical kinetic terms, the cubic couplings we should use for the two cases of interest are
\es{norm-cubic-couplings}{
g_{ss\sigma} &= \sqrt{\frac{4\pi}{N}} \frac{V^{ss\sigma}}{32 \sqrt{k_1 k_2 k_3 (k_1 + 1) (k_2 + 1) (k_3 - 1)}} \\
g_{\sigma \sigma \sigma} &= \sqrt{\frac{4\pi}{N}} \frac{3 V^{\sigma \sigma \sigma}}{32 \sqrt{k_1 k_2 k_3 (k_1 - 1) (k_2 - 1) (k_3 - 1)}}.
}
Note that these include symmetry factors of $2$ and $6$ respectively. A closed form expression for \eqref{3pt-scalar1} may be found in \cite{cgp14} which evaluated a family of similar integrals labelled by spin. Using this result in $d = 2$,
\es{3pt-scalar2}{
\left < \mathcal{O}_{r_1}(x_1) \mathcal{O}_{r_2}(x_2) \mathcal{O}_{r_3}(x_3) \right > = \frac{\Gamma(\tfrac{1}{2} \Sigma - 1)}{4\sqrt{2\pi}} \prod_{i = 1}^3 \frac{\Gamma(\alpha_i)}{\Gamma(\Delta_i)} \frac{g_{123} a_{r_1, r_2, r_3}}{|x_{12}|^{\Delta_1 + \Delta_2 - \Delta_3} |x_{13}|^{\Delta_1 + \Delta_3 - \Delta_2} |x_{23}|^{\Delta_2 + \Delta_3 - \Delta_1}}.
}

Although this is enough information to compute three-point functions involving the operators $s^I_{r, k}(x)$ and $\sigma_{r, k}(x)$, we are more interested in the combinations $s^I_k(x, v, \bar{v})$ and $\sigma_k(x, v, \bar{v})$. Their three-point functions depend on the polarization structure
\es{r-structure}{
(v_{12} \bar{v}_{12})^{j_1 + j_2 - j_3} (v_{23} \bar{v}_{12})^{j_2 + j_3 - j_1} (v_{31} \bar{v}_{31})^{j_3 + j_1 - j_2}
}
where $j_i$ are the equal $SU(2)_L$ and $SU(2)_R$ spins. The projection of \eqref{r-structure} onto $(m_1, m_2, m_3)$ and $(\bar{m}_1, \bar{m}_2, \bar{m}_3)$ Cartans is
\es{3j-symbol}{
& \begin{pmatrix} j_1 & j_2 & j_3 \\ \bar{m}_1 & \bar{m}_2 & \bar{m}_3 \end{pmatrix} \begin{pmatrix} j_1 & j_2 & j_3 \\ m_1 & m_2 & m_3 \end{pmatrix} \frac{(j_1 + j_2 - j_3)! (j_2 + j_3 - j_1)! (j_3 + j_1 - j_2)!}{(2j_1)! (2j_2)! (2j_3)! (j_1 + j_2 + j_3 + 1)!^{-1}} \\
&= \frac{a_{(m_1, \bar{m}_1), (m_2, \bar{m}_2), (m_3, \bar{m}_3)}}{\sqrt{(2j_1 + 1)(2j_2 + 1)(2j_3 + 1)}} \frac{(j_1 + j_2 - j_3)! (j_2 + j_3 - j_1)! (j_3 + j_1 - j_2)!}{(2j_1)! (2j_2)! (2j_3)! (j_1 + j_2 + j_3 + 1)!^{-1}}.
}
In the first line we have used the result of an explicit Taylor expansion described in \cite{b22} while in the second line we have written 3j-symbols in terms of spherical harmonic overlaps using the relations in \cite{c84}. Thanks to \eqref{3j-symbol}, the desired rewriting of \eqref{3pt-scalar2} is
\es{3pt-scalar3}{
& \left < \mathcal{O}_1(x_1, v_1, \bar{v}_1) \mathcal{O}_2(x_2, v_2, \bar{v}_2) \mathcal{O}_3(x_3, v_3, \bar{v}_3) \right > = g_{123} \frac{\Gamma(\tfrac{1}{2} \Sigma - 1)}{4\sqrt{2\pi}} \prod_{i = 1}^3 \frac{\Gamma(\alpha_i)}{\Gamma(\Delta_i)} \\
& \hspace{1cm} \frac{\sqrt{(2j_1 + 1)(2j_2 + 1)(2j_3 + 1)}(2j_1)!(2j_2)!(2j_3)!}{(j_1 + j_2 + j_3 + 1)!(j_1 + j_2 - j_3)!(j_2 + j_3 - j_1)!(j_3 + j_1 - j_2)!} \\
& \hspace{2cm} \frac{(v_{12} \bar{v}_{12})^{j_1 + j_2 - j_3} (v_{23} \bar{v}_{12})^{j_2 + j_3 - j_1} (v_{31} \bar{v}_{31})^{j_3 + j_1 - j_2}}{|x_{12}|^{\Delta_1 + \Delta_2 - \Delta_3} |x_{13}|^{\Delta_1 + \Delta_3 - \Delta_2} |x_{23}|^{\Delta_2 + \Delta_3 - \Delta_1}}.
}
It is now easily seen that we recover $C^{ss\sigma}_{k_1 k_2 k_3}$ and $C^{\sigma \sigma \sigma}_{k_1 k_2 k_3}$ from \eqref{ope-coeffs} by plugging in the cubic couplings \eqref{cubic-couplings} with $j_i = \frac{k_i}{2}$ and $\Delta_i = k_i$.

\subsection{Disentangling the gauge fields}

Since the quadratic part of the action \eqref{sugra-action} is not diagonal in the gauge fields $A^{\pm,\mu}_{r, k}$ and $C^{\pm,\mu}_{r, k}$, it is not yet clear how they are related to superconformal primaries of spin $1$. To answer this, we will identify the independent propagating degrees of freedom mostly following \cite{rrz19} but correcting some errors. This will affect the cubic terms in \eqref{sugra-action} which express how $A^{\pm,\mu}_{r, k}$ and $C^{\pm,\mu}_{r, k}$ are coupled to the currents
\es{s-sigma-currents}{
s^I_{r_1, k_1} \nabla_\mu s^I_{r_2, k_2} t^\pm_{r_1, r_2, r}, \quad \sigma_{r_1, k_1} \nabla_\mu \sigma_{r_2, k_2} t^\pm_{r_1, r_2, r}.
}
For this section, it will make sense to suppress the labels $k_1, k_2, k$ and $r$ which means we will denote the currents in \eqref{s-sigma-currents} by $J^\phi_\mu$ for $\phi \in \{ s^I, \sigma \}$. To start, the equations of motion for $A^\pm_\mu$ and $C^\pm_\mu$ are
\begin{align}
& \partial^\nu F_{\nu\mu}(A^\pm) - (k - 1)(k + 1)A^\pm_\mu \mp 2(P_0 A^\pm)_\mu \mp 2k(P^\pm_{k - 1} A^\pm)_\mu \label{a-eom} \\
& - \partial^\nu F_{\nu\mu}(C^\pm) - (k - 1)(k + 3)C^\pm_\mu \mp 2(k + 1)(P_0 C^\pm)_\mu + 2\sum_\phi (V^{\phi\phi A} \pm W^\phi P^\pm_{k - 1}) J^\phi_\mu = 0 \nonumber
\end{align}
and
\begin{align}
& \partial^\nu F_{\nu\mu}(C^\pm) - (k + 1)(k + 3)C^\pm_\mu \pm 2(P_0 C^\pm)_\mu \mp 2(k + 2)(P^\pm_{k - 1} C^\pm)_\mu \label{c-eom} \\
& - \partial^\nu F_{\nu\mu}(A^\pm) - (k - 1)(k + 3)A^\pm_\mu \mp 2(k + 1)(P_0 A^\pm)_\mu + 2\sum_\phi (V^{\phi\phi C} \mp W^\phi P^\pm_{k + 3}) J^\phi_\mu = 0 \nonumber
\end{align}
respectively. Adding these leads to the first order constraint
\es{first-order}{
4(k + 1)\left [ (P^\pm_{k - 1} A^\pm)_\mu + (P^\pm_{k + 3} C^\pm)_\mu \right ] \pm 2\sum_\phi (4W^\phi - V^{\phi \phi A} - V^{\phi \phi C}) J^\phi_\mu = 0.
}
Plugging this into \eqref{a-eom} results in
\es{second-order}{
& (P^\mp_{k + 1} P^\pm_{k - 1} A^\pm)_\mu + (P^\mp_{k + 1} P^\pm_{k + 3} C^\pm)_\mu \\
& + \sum_\phi \left [ \frac{k + 2}{k + 1} V^{\phi \phi A} - \frac{k}{k + 1} V^{\phi \phi C} \pm 2W^\phi \left ( P_0 + 2\frac{k^2 + 2k - 1}{k + 1} \right ) \right ] J^\phi_\mu = 0.
}
The second order equation \eqref{second-order} can be rewritten in terms of first order derivatives by introducing a new field which \cite{rrz19} called $L^\pm_\mu$. There is some freedom in how to define it but the most convenient choice is the one that eliminates magnetic couplings to the currents. This fixes the coefficients on the right hand side of
\es{def-l}{
L^\pm_\mu \pm \frac{1}{2} [(P_0 A^\pm)_\mu - (P_0 C^\pm)_\mu] = \sum_\phi W^\phi J^\phi_\mu.
}
We finally get
\es{last-eq}{
& \pm 2(P_0 L^\pm)_\mu \mp 2(P_0 A^\pm)_\mu \mp 2(P_0 C^\pm)_\mu - (k - 1)(k + 1)A^\pm_\mu + (k + 1)(k + 3)C^\pm_\mu \\
& + \sum_\phi \left ( \frac{k + 2}{k + 1} V^{\phi \phi A} - \frac{k}{k + 1} V^{\phi \phi C} \pm 4 \frac{k^2 + 2k - 1}{k + 1} W^\phi \right ) J^\phi_\mu = 0.
}
We can now take linear combinations of \eqref{first-order}, \eqref{def-l} and \eqref{last-eq} to to find a set of three massive Chern-Simons equations of motion with the non-diagonal mass matrix $\textbf{M}^\pm = \pm \textbf{M}$.\footnote{The fact that the mass matrices for the $+$ and $-$ cases differ by a sign is a non-trivial check of our calculations.} Explicitly,
\es{eom}{
(P_0 \textbf{1} \pm \textbf{M}) \begin{bmatrix} L^\pm_\mu \\ A^\pm_\mu \\ C^\pm_\mu \end{bmatrix} = \frac{1}{4(k + 1)} \sum_\phi Q^\pm_\phi J^\phi_\mu
}
where
\es{matrices1}{
\textbf{M} = \frac{1}{2} \begin{bmatrix}
0 & -(k - 1)^2 & (k + 3)^2 \\
-2 & k - 1 & k + 3 \\
2 & k - 1 & k + 3
\end{bmatrix}, \quad Q^\pm_\phi = \begin{bmatrix}
2(k + 1)(V^{\phi \phi A} - V^{\phi \phi C}) \pm 4(k + 1)^2 W^\phi \\ -V^{\phi \phi A} - V^{\phi \phi C} \pm 4(k + 2)W^\phi \\ - V^{\phi \phi A} - V^{\phi \phi C} \mp 4kW^\phi
\end{bmatrix}.
}
To decouple these equations, we simply diagonalize. If $\textbf{R} \textbf{M} \textbf{R}^{-1} = \Lambda$, this converts \eqref{eom} to
\es{diagonalized-eom}{
(P_0 \textbf{1} + \Lambda) \textbf{R} \begin{bmatrix} L^\pm_\mu \\ A^\pm_\mu \\ C^\pm_\mu \end{bmatrix} = \frac{1}{4(k + 1)} \sum_\phi \textbf{R} Q^\pm_\phi J^\phi_\mu
}
with
\es{matrices2}{
\Lambda = \begin{bmatrix} k - 1 & 0 & 0 \\ 0 & -k - 1 & 0 \\ 0 & 0 & k + 3 \end{bmatrix}, \quad
\textbf{R} = \zeta \begin{bmatrix}
-\frac{1}{2k} & \frac{3k + 1}{4k} & \frac{k + 3}{4k} \\
-\frac{1}{2(k + 2)} & -\frac{k - 1}{4(k + 2)} & \frac{k + 3}{4(k + 2)} \\
\frac{1}{2(k + 2)} & \frac{k - 1}{4(k + 2)} & \frac{3k + 5}{4(k + 2)}
\end{bmatrix}.
}

The steps above do not give complete information because the eigenstates $V^\pm_\mu$, $V^{\prime \pm}_\mu$ and $V^{\prime \prime \pm}_\mu$ have an inhomogeneous term proportional to $\zeta$ which may depend on $k$ and cannot be determined by equations of motion alone. The authors of \cite{rrz19} solved for $\zeta$ by demanding $\text{det} \, \textbf{R} = 1$ but this is misleading. A field redefinition is enough to rescale the inhomogeneous term by a different amount for each line of \eqref{diagonalized-eom} anyway so it is better to view these equations of motion as arising from the action
\es{eom-action}{
S_0 = \int d^3 x \; & \frac{C}{2} V^{\pm, \mu} \left [ (P^\pm_{k - 1} V^\pm)_\mu + \frac{1}{2(k + 1)} \sum_\phi (\textbf{R} Q^\pm_\phi)_1 J^\phi_\mu \right ] \\
+ & \frac{C^\prime}{2} V^{\prime \pm, \mu} \left [ (P^\pm_{-k - 1} V^{\prime \pm})_\mu + \frac{1}{2(k + 1)} \sum_\phi (\textbf{R} Q^\pm_\phi)_2 J^\phi_\mu \right ] \\
+ & \frac{C^{\prime \prime}}{2} V^{\prime \prime \pm, \mu} \left [ (P^\pm_{k + 3} V^{\prime \prime \pm})_\mu + \frac{1}{2(k + 1)} \sum_\phi (\textbf{R} Q^\pm_\phi)_3 J^\phi_\mu \right ].
}
One can now express this in terms of the original gauge fields and then eliminate $L^\pm_\mu$ using \eqref{def-l}. The resulting action has terms with three derivatives but we can demand that they cancel by tuning $C$, $C^\prime$ and $C^{\prime \prime}$. By further demanding that the terms with two derivatives match \eqref{sugra-action}, it follows that
\es{c-sol}{
C = 4 \frac{k}{\zeta}, \quad C^\prime = -8 \frac{(k + 1)(k + 2)}{\zeta k}, \quad C^{\prime \prime} = 4 \frac{k + 2}{\zeta}.
}
After plugging \eqref{c-sol} into \eqref{eom-action} and rescaling the fields to arrive at canonical kinetic terms, $V^\pm_\mu$ (which is a primary since it has the lowest scaling dimension) couples to the currents with
\es{cubic-couplings2}{
V^{ssV} &= \frac{\sqrt{k}}{2(k + 1)} (\textbf{R} Q^\pm_s)_1 = V^{ssA}/\sqrt{k} \\
V^{\sigma \sigma V} &= \frac{\sqrt{k}}{2(k + 1)} (\textbf{R} Q^\pm_\sigma)_1 = V^{\sigma \sigma A}/\sqrt{k}.
}
\subsection{Two scalars, one vector}
To properly derive three-point functions involving the vector fields we should take into account that we have massive Chern-Simons fields in the effective action (\ref{sugra-action}). In particular, the bulk-to-bulk propagators for such fields satisfy 
\es{ads-cs-eq}{
    (P^{\pm}_m)_\nu^{\;\rho}G_{\mu \rho}^{\pm}(x,y)=g_{\mu\nu} x_0^{d + 1} \delta(x-y), \quad m^2 = \Delta(\Delta - d) + d - 1.
}
To find this propagator we can use the identity
\es{ads-prop-id1}{
    (P^{-}_m P_m^{+})_{\nu}^{\;\rho}=g^{\rho\lambda}\nabla_\nu\nabla_\lambda-(\square +m^2)\delta^\rho_\nu,
}
together with the fact that $P_m^{\pm}$ are projection operators to get 
\es{ads-prop-id2}{
    G_{\mu \nu}^{\pm}(x,y)=\mp (P_m^{\mp})_\mu^{\;\rho}G_{\rho\nu}(x,y),
}
where $G_{\rho\nu}$ is the propagator for a Proca field
\es{ads-proca-eq}{
    [g^{\rho\lambda}\nabla_\nu\nabla_\lambda-(\square +m^2)\delta^\rho_\nu]G_{\mu \rho}(x,y)=g_{\mu\nu} x_0^{d + 1} \delta(x-y),
}
whose expression is known \cite{cgp14}. One can then consider the limit as $y$ approaches the boundary to get the bulk-to-boundary propagator
\es{vector-prop}{
        K^{\mu i}_\Delta(x,y ) =C_{\Delta,1}\left(\frac{x_0}{x_0^2 + |\vec{x}-\vec{y}|^2}\right)^{\Delta} \partial^\mu\left[\frac{(x-y)^i}{x_0^2 + |\vec{x}-\vec{y}|^2}\right].
}
Then, we can get the desired three-point function between two scalars and a self-dual vector by integrating the three bulk-to-boundary propagators
\es{ssV3point}{
\left < \mathcal{O}_{r_1}(x_1) \mathcal{O}_{r_2}(x_2) \mathcal{O}^{+i}_{r_3}(x_3) \right > = -g_{123} t^+_{r_1, r_2, r_3} \int \frac{d^3 x}{x_0^3} \frac{K_{\Delta_1}(x, x_1) \nabla_\rho K_{\Delta_2}(x, x_2) (P^{-}_{\Delta_3-1})_\mu^{\;\;\rho} K^{\mu i}_{\Delta_3}(x, x_3)}{\sqrt{C_{\Delta_1, 0} C_{\Delta_2, 0} C_{\Delta_3, 1}}}.
}
Expanding the projector makes \eqref{ssV3point} a sum of two integrals. The first one has already been computed in the literature \cite{cgp14} and will give a prefactor similar to the one in \eqref{3pt-scalar2} times the position dependence of a $\left < \mathcal{O} \mathcal{O} \mathcal{O}^i \right >$ three-point function. Conformal symmetry is then enough to infer the result of the second integral. It must yield the same prefactor but the position dependence of $i \varepsilon^{ij} \left < \mathcal{O} \mathcal{O} \mathcal{O}_j \right >$ so that we get something self-dual.\footnote{The identity we have in mind is
\begin{align}
\frac{\bar{z}_{12} z_{23} z_{31}}{|z_{12}|^{2\Delta_\phi - \Delta + 1} |z_{23}|^{\Delta + 1} |z_{31}|^{\Delta + 1}} = \frac{|x_{23}|^2 x_{13}^\parallel - |x_{13}|^2 x_{23}^\parallel}{|x_{12}|^{2\Delta_\phi - \Delta + 1} |x_{23}|^{\Delta + 1} |x_{31}|^{\Delta + 1}} + i \frac{|x_{23}|^2 x_{13}^\perp - |x_{13}|^2 x_{23}^\perp}{|x_{12}|^{2\Delta_\phi - \Delta + 1} |x_{23}|^{\Delta + 1} |x_{31}|^{\Delta + 1}}
\end{align}
for $z = x^\parallel + i x^\perp$.
} All in all,
\es{3pt-vector}{
\left < \mathcal{O}_{r_1}(x_1) \mathcal{O}_{r_2}(x_2) \mathcal{O}^{+}_{r_3}(x_3) \right > &= \frac{\Gamma(\tfrac{\Sigma - 1}{2})}{2\sqrt{2\pi \Delta_3}} \prod_{i = 1}^3 \frac{\Gamma(\alpha_i + \tfrac{1}{2})}{\Gamma(\Delta_i)} \frac{g_{123} t^+_{r_1, r_2, r_3} \bar{z}_{12} z_{23} z_{31}}{|z_{12}|^{\Delta_1 + \Delta_2 - \Delta_3 + 1} |z_{23}|^{\Delta_3 + \Delta_{21} + 1} |z_{31}|^{\Delta_3 + \Delta_{12} + 1}}.
}
The fact that we contract the $R$-symmetry indices with polarization spinors must again be taken into account. Since $V^+$ has $\bar{\jmath} = j + 1$, the polarization structure analogous to \eqref{r-structure} we should consider is
\es{r-structure2}{
(v_{12} \bar{v}_{12})^{j_1 + j_2 - j_3 - 1} (v_{23} \bar{v}_{12})^{j_2 + j_3 - j_1} (v_{31} \bar{v}_{31})^{j_3 + j_1 - j_2} v_{12} \bar{v}_{23} \bar{v}_{31}.
}
Its projection onto the $SU(2)_L$ Cartans $(m_1, m_2, m_3)$ and $SU(2)_R$ Cartans $(\bar{m}_1, \bar{m}_2, \bar{m}_3)$ is
\begin{align}
& \begin{pmatrix} j_1 & j_2 & j_3 + 1 \\ \bar{m}_1 & \bar{m}_2 & \bar{m}_3 \end{pmatrix} \begin{pmatrix} j_1 & j_2 & j_3 \\ m_1 & m_2 & m_3 \end{pmatrix} \frac{(j_1 + j_2 - j_3 - 1)! (j_2 + j_3 - j_1)! (j_3 + j_1 - j_2)!}{(2j_1)! (2j_2)! (2j_3)! (j_1 + j_2 + j_3 + 1)!^{-1}} \nonumber \\
&\times \sqrt{\frac{(j_1 + j_2 - j_3 - 1)(j_2 + j_3 - j_1 + 1)(j_3 + j_1 - j_2 + 1)}{2j_3(2j_3 + 1) (j_1 + j_2 + j_3 + 2)^{-1}}} \label{3j-symbol2} \\
&= \frac{t^+_{(m_1, \bar{m}_1), (m_2, \bar{m}_2), (m_3, \bar{m}_3)}}{\sqrt{(2j_1 + 1)(2j_2 + 1)(2j_3 + 1)}} \frac{(j_1 + j_2 - j_3 - 1)! (j_2 + j_3 - j_1)! (j_3 + j_1 - j_2)!}{(2j_1)! (2j_2)! (2j_3)! (j_1 + j_2 + j_3 + 1)!^{-1}} \sqrt{\frac{j_3 + 1}{j_3}} \nonumber
\end{align}
where we have used the results of \cite{c84,b22}. Plugging this into \eqref{3pt-vector} therefore gives
\es{3pt-vector2}{
& \left < \mathcal{O}_1(x_1, v_1, \bar{v}_1) \mathcal{O}_2(x_2, v_2, \bar{v}_2) \mathcal{O}^+_3(x_3, v_3, \bar{v}_3) \right > = g_{123} \frac{\Gamma(\tfrac{\Sigma - 1}{2})}{2\sqrt{2\pi \Delta_3}} \prod_{i = 1}^3 \frac{\Gamma(\alpha_i + \tfrac{1}{2})}{\Gamma(\Delta_i)} \\
& \frac{\sqrt{(2j_1 + 1)(2j_2 + 1)(2j_3 + 1)}(2j_1)!(2j_2)!(2j_3)!}{(j_1 + j_2 + j_3 + 1)!(j_1 + j_2 - j_3 - 1)!(j_2 + j_3 - j_1)!(j_3 + j_1 - j_2)!} \sqrt{\frac{j_3}{j_3 + 1}} \\
& \frac{(v_{12} \bar{v}_{12})^{j_1 + j_2 - j_3 - 1} (v_{23} \bar{v}_{12})^{j_2 + j_3 - j_1} (v_{31} \bar{v}_{31})^{j_3 + j_1 - j_2}}{|x_{12}|^{\Delta_1 + \Delta_2 - \Delta_3 - 1} |x_{13}|^{\Delta_1 + \Delta_3 - \Delta_2 - 1} |x_{23}|^{\Delta_2 + \Delta_3 - \Delta_1 - 1}} \frac{v_{12} \bar{v}_{23} \bar{v}_{31}}{z_{12} \bar{z}_{23} \bar{z}_{31}}.
}
To retrieve $C^{ssV}_{k_1 k_2 k_3}$ and $C^{\sigma \sigma V}_{k_1 k_2 k_3}$ from \eqref{ope-coeffs}, we use the cubic couplings \eqref{cubic-couplings2} from the last subsection and set $j_1 = \frac{k_1}{2}$, $j_2 = \frac{k_2}{2}$, $j_3 = \frac{k_3 - 1}{2}$ and $\Delta_i = k_i$.

\bibliographystyle{utphys}
\bibliography{refs}
\end{document}